\documentclass{article}
\usepackage{graphicx,epsfig,epsf,cite}
\usepackage{mathptmx}      
\usepackage{amsmath}
\usepackage{amssymb}
\usepackage{subfigure}
\usepackage{color}
\usepackage{appendix}
\usepackage{soul}

\usepackage{mathtools}

\usepackage{sidecap}

\usepackage{setspace}

\graphicspath{{FIGURES/}}

\begin{document}
\title{\large\bf
A contact model for sticking of adhesive meso-particles 
}
\author{
\normalsize A. Singh, V. Magnanimo and S. Luding\\
\normalsize Multi Scale Mechanics, CTW, MESA+, UTwente, \\
\normalsize    P.O.Box 217, 7500 AE Enschede, Netherlands,\\
\normalsize    e-mail: {\tt a.singh-1@utwente.nl, v.magnanimo@utwente.nl, s.luding@utwente.nl} \\
}
\date{}


\maketitle
\begin{abstract}{

The interaction between visco-elasto-plastic and adhesive particles is the subject 
of this study, where \textquotedblleft meso-particles\textquotedblright\ are introduced,
i.e., 
simplified 
particles, whose contact mechanics 
is not taken into account in all details. A few examples of meso-particles 
include agglomerates or groups of primary particles, or inhomogeneous particles with 
micro-structures of the scale of the contact deformation, such as core-shell materials.

A simple, flexible contact model for meso-particles is proposed, which allows to model 
the bulk behavior of assemblies of many particles in both rapid and slow, quasi-static 
flow situations. 
An attempt is made to categorize existing contact models 
for the normal force, discuss all the essential mechanical ingredients
that must enter the model (qualitatively) and finally solve it analytically.

The model combines a short-ranged, non-contact part 
(resembling either dry or wet materials) with an elaborate, 
visco-elasto-plastic and adhesive contact law.
Using energy conservation arguments, 
an analytical expression for the coefficient of restitution is
derived in terms of the impact velocity (for pair interactions or, equivalently, 
without loss of generality,  for quasi-static situations
in terms of the maximum overlap or confining stress).

Adhesive particles (or meso-particles) stick to each other at very low impact 
velocity, while they rebound less dissipatively with increasing velocity, 
in agreement with previous 
studies.
For even higher impact velocities an interesting second sticking and 
rebound regime is reported.
The low velocity sticking is due to non-contact adhesive forces, 
the first rebound regime is due to stronger elastic and kinetic energies
with little dissipation,
while the high velocity sticking is generated by the non-linearly 
increasing, history dependent plastic dissipation and adhesive contact force.
As the model allows also for a stiff, more elastic core material,
this causes the second rebound regime at even higher velocities. 

}\end{abstract}

\clearpage
\clearpage
\section*{Nomenclature}

\begin{tabular}{ccp{15cm}}

$m_i$ & : & mass of $i^{th}$ particle. \\
$a_i$ & : & Radius of $i^{th}$ particle. \\
$m_r$ & : & Reduced mass of a pair of particles. \\
$\delta$ & : & Contact overlap between particles.\\
$v_i$ & : & Relative velocity before collision. \\
$v_f$ & : & Relative velocity after collision. \\
${v_i}^{\infty}$ & : & Relative velocity before collision at infinite separation. \\
${v_f}^{\infty}$ & : & Relative velocity after collision at infinite separation. \\
$v^n$ & : & Normal component of relative velocity. \\
$e$ & : & Coefficient of restitution. \\
$e_n$ & : & Normal coefficient of restitution. \\
$\epsilon_{i}$ & : & Pull-in coefficient of restitution. \\
$\epsilon_{o}$ & : & Pull-off coefficient of restitution. \\
$k$ & : &Spring stiffness. \\
$k_1$ & : & Slope of loading plastic branch.\\
$k_2$ & : & Slope of unloading and re-loading elastic branch.\\
$k_c$ & : & Slope of irreversible, tensile adhesive branch.\\
$k_p$ & : & Slope of unloading and re-loading limit branch; end of plastic regime.\\
$v_{p}$ & :& Relative velocity before collision for which the limit case is reached. \\
$\phi_f$ & :& Dimensionless plasticity depth. \\
$\delta_{\rm max}$ & :& Maximum overlap between particles during a collision. \\
$\delta^p_{\rm max}$ & :& Maximum overlap between particles for the limit case. \\
$\delta_{\rm 0}$ & :& Force free overlap $\cong$ plastic contact deformation. \\
$\delta_{\rm min}$ & :& Overlap between particles at the maximum (negative) attractive force. \\
$\delta_{\rm c}$ & :& Kinetic energy free overlap between particles. \\
$W_{\rm diss}$ & : & Amount of energy dissipated during collision. \\
$\eta$ & : & Dimensionless plasticity of the contact. \\
$\beta$ & : & Adhesivity: dimensionless adhesive strength of the contact. \\
$\chi$ & : & Scaled initial velocity relative to $v_p$. \\
$f_a$ & : & Non-contact adhesive force at zero overlap. \\
$\delta_{\rm a}$ & :& Non-contact separation between particles at which attractive force becomes active. \\
$k_c^a$ & : & Strength of non-contact adhesive force.\\
\end{tabular}

\clearpage

\section{Introduction}
\label{sec:Intro}

Granular materials and powders are ubiquitous in industry and nature. For this reason, the past decades 
have witnessed a strong interest in research aiming for better understanding and predicting their behavior
in all regimes from flow to static as well as the transitions between these states. 
Especially, the impact of fine particles with other particles or surfaces is of fundamental importance. 
The interaction force between two particles is a combination of elasto-plastic deformation, 
viscous dissipation, and adhesion -- due to both mechanical contact-
and long ranged non-contact forces. Pair interactions that can be used in bulk simulations with many particles
and multiple contacts per particle are the focus here, and we use the special, elementary case of pair interactions to 
understand them analytically.

Different regimes can be observed for collisions between two particles: 
For example, a particle can either stick to another particle/surface or it rebounds, 
depending upon the relative strength of adhesion and impact velocity, size and material
various material parameters \cite{Paulick2015}. 
This problem needs to be well understood, as it forms the basis for understanding 
rather complex, many-particle flows in realistic systems, related to e.g.\ astrophysics 
(dust agglomeration, Saturn's rings, planet formation) or industrial processes 
(handling of fine powders, granulation, filling and discharging of silos). 
Particularly interesting are the interaction mechanisms for adhesive materials such as
 asphalt, ice particles or clusters/agglomerates of fine powders (often made of even smaller
 primary particles). Some of these materials can 
be physically visualized as having a plastic outer shell with a stronger and more elastic 
inner core. 
Understanding this can then be applied to particle-surface collisions in kinetic spraying,
where the solid micro-sized powder particle is accelerated towards a substrate. In cold spray,
 bonding occurs when impact velocities of particles exceed a critical value, which depends
 on various material parameters \cite{Van99,Zhang05,Schmidt06,Paulick2015}.
However, for even higher velocities particles rebound from the surface \cite{Wu06,Wu11}.
Due to the inhomogeneity of most realistic materials, their non-sphericity and their surface
irregularity, one can not include all these details -- but rather has to focus on the essential phenomena 
and ingredients, finding a compromise between simplicity and realistic contact mechanics. 

\subsection{Contact Models Review}\label{sec:Model_review}
Computer simulations have turned out to be a powerful tool to investigate the physics of 
particulate systems, especially valuable as experimental difficulties are considerable and since
there is no generally accepted theory of granular flows. A very popular simulation scheme is an adaptation 
of the classical Molecular Dynamics technique called Discrete Element Method (DEM)
(for details see 
Refs.\ \cite{cundall79,bashir91,herrmann98,thornton00,thornton01pg,vermeer01,latzel03,luding06c,Luding08}).
It involves integrating Newton's equations of motion for a system of ``soft'', deformable grains, 
starting from a given initial configuration. DEM can be successfully applied to adhesive particles, 
if a proper force-overlap law (contact model) is used.

The JKR model \cite{JKR71} is a widely accepted contact model for adhesive elastic spheres 
and gives an expression for the normal force in terms of the normal deformation. 
Derjaguin {\em et~al.}\ \cite{Derjaguin1975} 
suggested that the attractive forces act only just outside the contact zone, 
where surface separation is small, and is referred to as DMT model.
An interesting approach for dry adhesive particles was proposed by 
Molerus \cite{Molerus75, Molerus78}, who explained consolidation and non-rapid 
flow of adhesive particles in terms of adhesive forces at particle contacts. 
Thornton and Yin \cite{thornton91} compared the results of elastic spheres with 
and without adhesion, a work that was later extended to adhesive 
elasto-plastic spheres \cite{Thornton98}. 
Molerus's model was further developed by Tomas, who introduced a complex contact 
model \cite{tomas00, Tomas01-1, Tomas01-2} by coupling elasto-plastic contact 
behavior with non-linear adhesion and hysteresis involving dissipation and a history 
(compression) dependent adhesive force.
The contact model subsequently proposed by Luding \cite{Luding2005Discrete,Luding08} 
works in the same spirit as that of Tomas \cite{Tomas01-1}, only reducing complexity by 
using piece-wise linear branches in an otherwise non-linear contact model in spirit
(as explained later in this study). In the original version \cite{Luding08}, a short-ranged force 
beyond contact was mentioned, but not specified, which is one of the issues
tackled in the present study.
Contact details, such as a possible non-linear Hertzian law for small deformation, and 
non-linear loading-unloading hysteresis are over-simplified in Ludings model, 
as compared to the model proposed by Tomas \cite{Tomas01-1}.
This is partly due to the lack of the experimental reference data or theories, 
but also motivated by the wish to keep the model as simple as possible.  
The model consists of several basic mechanisms, i.e., non-linear elasticity, plasticity and adhesion 
as relevant for, e.g.\ core-shell materials or agglomerates of fine, dry primary powder particles
\cite{dominik1997physics,Boltachev2015nanopowders}.
A possible connection between the microscopic contact model and the macroscopic, continuum 
description for adhesive particles was recently proposed by Luding {\em et~al.}\ \cite{Luding11},
as further explored by Singh {\em et~al.}\ \cite{Singh2014effect,Singh2014micro} for dry adhesion, 
by studying the force anisotropy and force distributions in steady state bulk shear in the
\footnote{The details on the geometry are explained in Refs.\ \cite{singh2015role,Singh2014micro,singh2013effect}.},
which is further generalized to wet adhesion by Roy {\em et al.}\ \cite{Roy2015micromacro_wet}, 
or studied under shear-reversal \cite{Kumar2014memory,Rojas2015}.

Jiang {\em et~al.}\ \cite{Jiang12} experimentally investigated the force-displacement 
behavior of idealized bonded granules. This was later used to study the mechanical 
behavior of loose cemented granular materials using DEM simulations \cite{Jiang13}. 
Kempton {\em et~al.}\ \cite{Kempton2012} proposed a meso-scale contact model combining 
linear hysteretic, simplified JKR and linear bonding force models, 
to simulate agglomerates of sub-particles. The phenomenology of such particles
is nicely described by Dominik and Tielens \cite{dominik1997physics}.
Walton {\em et~al.}\ {\cite{Walton04,walton2009}} also proposed contact models in similar
spirit as that of Luding {\cite{Luding08}} and Tomas {\cite{Tomas01-1}}, separating the 
pull-off force from the slope of the tensile attractive force as independent mechanisms.
Most recently two contact models were proposed by Thakur {\em et~al.}\ {\cite{Thakur2013}}
and by Pasha {\em et~al.}\ \cite{pasha2014linear}, which work in the same spirit as Luding's model, 
but treat loading and un/re-loading behaviors differently. The former excludes the non-linear 
elastic stiffness in the plastic regime, and both deal with a more brittle, abrupt reduction of the adhesive
contact force. The authors further used their models to study the scaling and effect of DEM parameters in an 
uniaxial compression test \cite{Thakur13DEM}, and compared part of their results with other models
\cite{pasha2014linear}.

When two particles interact, their behavior is intermediate between the extremes of 
perfectly elastic and fully inelastic, possibly even fragmenting, where the latter is not considered
in this study. Considering a dynamic collision is our choice here, but without loss of generality,
most of our results can also be applied to a slow, quasi-static loading-unloading cycle that
activates the plastic loss of energy, by replacing kinetic with potential energies. 
Rozenblat {\em et~al.}\ \ \cite{Tomas13} have recently proposed an empirical relation between 
impact velocity and static compression force. 

The amount of energy dissipated during a collision can be best quantified by the coefficient of restitution, 
which is the ratio of magnitude of post-collision and pre-collision normal relative velocities of the particles. 
It quantifies the amount of energy that is not dissipated during the collision.  
For the case of plastic and viscoelastic collisions, it was suggested that dissipation 
depends on impact velocity \cite{johnson89, Kuwabara87,walton86b}; this can be realized by viscoelastic forces 
\cite{Kuwabara87,luding94d,brilliantov96,luding98c} and follows from plastic deformations too \cite{Zhang02}.

Early experimental studies \cite{Dahneke73, Dahneke75} on adhesive polystyrene latex spheres 
of micrometer size showed sticking of particles for velocities below a threshold and an increasing 
coefficient of restitution for velocities increasing above the threshold.
Wall {\em et~al.}\ \cite{Wall90} further confirmed these findings for highly mono-disperse ammonium 
particles. Thornton {\em et~al.}\ \cite{Thornton98} and Brilliantov {\em et~al.}\ \cite{Brilliantov07} 
proposed an adhesive visco-elasto-plastic contact model in agreement with these experiments.
Work by Sorace {\em et~al.}\ \cite{Sorace09} also confirms the sticking at low velocities 
for particle sizes of the order of a few mm. 
Li {\em et~al.}\ \cite{Li11a} proposed a dynamical model based on JKR 
 for the impact of micro-sized spheres with a flat surface, whereas realisitc
particle contacts are usually not flat \cite{Hanaor2015contact}.
 Recently, Saitoh {\em et~al.}\ \cite{Brilliantov10} even reported
 negative coefficients of restitution in nanocluster simulations, which is an artefact of the
 wrong definition of the coefficient of restitution; one has to relate the relative velocities
 to the normal directions before and after collision and not just in the frame before collision,
 which is especially a serious effect for softer particles \cite{Poschel11}. 
Jasevi\u{c}ius {\em et~al.}\ \cite{Jasevicius09,Jasevicius11} have recently studied the 
rebound behavior of ultrafine silica particles using the contact model by 
Tomas \cite{tomas00, Tomas01-1, Tomas01-2, Tomas07}.
They found that energy absorption due to attractive forces is the main source of 
energy dissipation at lower impact velocities or compression, 
while plastic deformation-induced dissipation becomes more important with
increasing impact velocity. They found some discrepancies between numerical
and experimental observations and concluded that these might be due to the lack of  
knowledge of particle- and contact-parameters, including surface
roughness, adsorption layers on particle surfaces, and microscopic material property
distributions (inhomogeneities), which in essence are features of the meso-particles 
that we aim to study. 

In a more recent study, Shinbrot {\em et~al.}\ \cite{siu2015nonlinear} studied
charged primary particles with interesting single particle dynamics in the 
electromagnetic field. They found ensembles of attractive (charged) particles 
can forming collective contacts or even fingers, extending the concepts 
of ``contact'' well beyond the idealized picture of perfect spheres,
as shown also in the appendix of the present study. 

Finally, Rathbone {\em et~al.}\ \cite{rathbone2015accurate} presented a new 
force-displacement law for elasto-plastic materials and compare it to their FEM 
results that resolve the deformations in the particle contact zone. This was complemented
by an experimental study comparing various models and their influence on the
bulk flow behavior \cite{Paulick2015}.


\subsection{Model classification}\label{sec:Model_classification}
Since our main focus is on dry particles, here we do not review the diverse works that involve
liquid \cite{Herminghaus05} or strong solid bridges \cite{Brendel11}.
Even though oblique collisions between two particles are of practical relevance and have 
been studied in detail by Thornton {\em et~al.}\ \cite{Thornton11,Thornton13}, here we focus 
on central normal collisions without loss of generality. Finally, we also disregard
many minute details of non-contact forces, as, e.g.\ due to van der Waals forces, for
the sake of brevity, but will propose a very simple mesoscale non-contact 
force model in section \ref{sec:non-contact}.
 
Based on our review of adhesive, elasto-visco-plastic contact models, here we propose a 
{\em possible} classification, by dividing them into three groups 
(based on their complexity and aim): \\
 (1) {\bf Academic} contact models, \\
 (2) {\bf Mesoscopic} contact models, and \\
 (3) {\bf Realistic}, fully detailed contact models. \\
Here we focus on adhesive elastic, and elasto-plastic contact models mainly, while
the effect of various forces on adhesion of fine particles is reviewed in Ref.\ \cite{Walton08review},
and some of the more complex models are reviewed and compared in Ref.\ \cite{Thornton13}.

\begin{enumerate}
  \item \textbf{Academic contact models} 
  allow for easy analytical solution,
  as for example the linear spring-dashpot model \cite{luding98c},
  or piece-wise linear models with constant unloading stiffness
  (see e.g.\ Walton and Braun \cite{walton86}), 
  which feature a constant coefficient of restitution
  (independent of impact velocity). 
Also the Hertzian visco-elastic models belong to this
  class, even though they provide a velocity dependent coefficient of 
  restitution, for a summary see Ref.\ \cite{luding98c} and references therein,
  while for a recent comparison see Ref.\ \cite{Nasato2015}. 
  However, no academic model can fully describe realistic, practically relevant contacts. 
  Either the material or the geometry/mechanics is too idealized; 
  in application, there is hardly any contact that is perfectly linear or Hertzian
  visco-elastic. Academic models thus miss most details of real contacts, but
  can be treated analytically.
  \item \textbf{Mesoscopic contact models} 
  (or, with other words, contact models for meso-particles) are a compromise, 
  (i) still rather easy to implement, (ii) aimed for fast ensemble/bulk-simulations
  with many particles and various materials, and (iii) contain most relevant
  mechanisms, but not all the minute details of every primary particle and every single contact.
  They are often piece-wise linear, e.g.\ with a variable unloading stiffness or with an extended 
  adhesive force, leading to a variable coefficient of restitution, etc., see
  Refs.\ \cite{walton86,Luding08,walton2009,Ooi12,Thakur2013}).
  \item \textbf{Realistic, full-detail contact models} have (i) the most realistic, 
  but often rather complicated formulation, (ii) can reproduce with similar precision the pair 
  interaction and the bulk behavior, but (iii) are valid only for the limited class of materials
  they are particularly designed for, since they do include all the minute details of these 
  interactions. A few examples include:
  \begin{enumerate}
    \item \textbf{visco-elastic models:} 
    Walton \cite{walton93}, Brilliantov \cite{Brilliantov05,Brilliantov07}, Haiat \cite{Haiat03};
    \item \textbf{adhesive elastic models:} 
    JKR \cite{JKR71}, Dahneke \cite{Dahneke72}, 
    DMT \cite{Derjaguin1975}, Thornton and Yin \cite{thornton91};
    \item \textbf{adhesive elasto-plastic models:} 
    Molerus\cite{Molerus75}, Thornton and Ning \cite{Thornton98}, Tomas \cite{tomas00, Tomas01-1, Tomas01-2, Tomas07},
    Pasha {\em et~al.}\ \cite{pasha2014linear}.
  \end{enumerate}
\end{enumerate}
While the realistic models are designed for a special particulate material in mind,
our main goal is to define and apply mesoscopic contact models to simulate the bulk behavior 
of a variety of assemblies of many particles (for which no valid realistic model is available), 
we focus on the second class: mesoscopic contact models. 

\subsection{Focus and Overview of this study}
\label{sec:Overview}

In particular, we study the dependence of the
coefficient of restitution for two meso-particles 
on impact velocity and contact/material parameters, for a wide 
range of impact velocities, using the complete version of the 
contact model by Luding \cite{Luding08}, with a specific piece-wise linear
non-contact force term. We observe sticking of particles at low velocity, 
which is consistent with previous theoretical and 
experimental works \cite{Wall90,Thornton98,Sorace09}.
Pasha {\em et~al.}\ \cite{pasha2014linear} recently also reproduced the low 
velocity sticking using an extension of the similar, but simpler model \cite{luding01e}. 
Above a certain small velocity, dissipation is not strong enough 
to dissipate all relative kinetic energy and the coefficient of restitution 
begins to increase. We want to understand the full regime of relative velocities, and
thus focus also on the less explored intermediate and high velocity regimes, 
as easily accessible in numerical simulations. 
In the intermediate regime, we observe a decrease in the coefficient
of restitution, as observed previously for idealized/homogeneous particles \cite{Thornton98,Brilliantov07},
however the functional behavior is different compared to the predictions by Thornton \cite{Thornton98}.
In Appendix\ \ref{sec:tuning_model}, 
we show that this property can be tuned by simple modifications to our model.
Tanaka {\em et~al.}\ \cite{Tanaka12} have recently reported similar results, 
when simulating the collision of more realistic dust aggregates, consisting of
many thousands of nanoparticles that interact via the JKR model. 
With further increase in impact velocity, we find a second sticking regime due
to the non-linearly increasing adhesive and plastic dissipation.  For even higher
velocities, the second, intermediate sticking regime is terminated by a 
second rebound regime due to the elastic core that can be specified in
the model.
Finally, since the physical systems under consideration also are viscous in 
nature, we conclude with some simulations with added viscous damping, which
is always added on top of the other model ingredients, but sometimes neglected
in order to allow for analytical solutions. 

An exemplary application of our model that shows the unexpected high velocity 
sticking and rebound regime (which might not be observed in the case of 
homogeneous granular materials) is, the coating process
in cold sprays. In these studies, the researchers are interested in 
 analyzing the deposition efficiency of the powder on a substrate as a function of the 
 impact velocity. Bonding/coating happens when the impact velocity
 of the particles exceeds a ``critical velocity'', with values of the order 
 of $10^2$\,m/s \cite{Schmidt06,Wu06,Wu11}. 
 Interestingly, when the velocity is further increased the
 particles do not bond (stick) to the substrate anymore, and a decrease
 in the deposition efficiency (inverse of the coefficient of restitution) is observed \cite{Wu06}.
 Schmidt {\em et~al.}\ \cite{Schmidt06} have used numerical simulations to explore 
 the effect of various material properties on the critical velocity, 
 while Zhou {\em et~al.}\ \cite{Wu11} studied the effect of impact velocity and
 material properties on the coating process, showing that properties of both particle and substrate 
 influence the rebound.
 Using our model, one could explore the dependence of the deposition efficiency 
 on the impact velocity, leading to the synergy between different communities.

The paper is arranged as follows:
In section \ref{sec:mdsoft}, we introduce the DEM simulation method and the
basic contact models for the normal direction;  one type of meso-models is further elaborated on in the following section \ref{sec:split_COR}, where the coefficient of restitution is computed analytically, and
dimensionless contact parameters are proposed in section \ref{sec:dimless}.
The limit of negligible non-contact forces is considered in section \ref{sec:fa0},
where various special cases are discussed, the contact model parameters are 
studied, and also asymptotic solutions and limit values are given, 
before the study is concluded in section\ \ref{sec:con}.

\section{Discrete Element Method} 
\label{sec:mdsoft}

The elementary units of particulate systems as granular materials or powders
are grains that deform under applied stress. Since the realistic and detailed modeling 
of real particles in contact is too complicated, it is necessary 
to relate the interaction force to the overlap $\delta$ between two particles
in contact. Note that the evaluation of the inter-particle forces based on the
overlap may not be sufficient to account for the inhomogeneous stress
distribution inside the particles, for internal re-arrangements \cite{dominik1997physics},
and for possible multi-contact effects \cite{johnson89}. 
However, this price has to be paid in order to simulate large samples of 
particles with a minimal complexity and still taking various 
physical contact properties such as non-linear contact elasticity, 
plastic deformation or load-dependent adhesion into account.

\subsection{Equations of Motion} \label{sec:eqom}

If all forces acting on a spherical particle $p$, either from other
particles, from boundaries or externally, are known -- let
their vector sum be $\vec f_p$ -- then the problem is reduced to the 
integration of Newton's equations of motion for the translational
degrees of freedom (the rotational degrees are not considered here
since we focus only on normal forces) for each particle:
$m_p \frac{{\rm d}^2}{{\rm d} t^2} \vec{r_p} = \vec f_p + m_p \vec g$
where, $m_p$ is the mass of particle $p$, 
$\vec r_p$ its position, $\vec f_p = \sum_c \vec f^c_p$
is the total force due to all contacts $c$,
and $\vec g$ is the acceleration due to volume forces like gravity.
With tools as nicely described in textbooks as \cite{allen87,rapaport95,Poschel05}, 
the integration over many time-steps is a straightforward exercise.  
The typically short-ranged interactions in granular media allow for 
further optimization by using linked-cell (LC) or alternative methods
in order to make the neighborhood search more efficient 
\cite{Ogarko2011algorithm,Krijgsman2015CPM}. 
However, such optimization issues are not of concern in this study, 
since only normal pair collisions are considered.

\subsection{Normal Contact Force Laws}\label{sec:forcelaws}

Two spherical particles $i$ and $j$, with radii $a_i$ and $a_j$,
 $r_i$ and $r_j$ being the position vectors
respectively, interact if their overlap,
\begin{equation}
\delta = (a_i + a_j) - (\vec r_i - \vec r_j) \cdot \vec n ~,
\end{equation}
is either positive, $\delta > 0$, for mechanical contact,
or smaller than a cut-off, $0 \ge \delta > \delta_a$, for 
non-contact interactions, with the unit vector
$\vec n = \vec n_{ij}= (\vec r_i - \vec r_j) / |\vec r_i - \vec r_j|$
pointing from $j$ to $i$.  The force on particle $i$, from particle 
$j$, at contact $c$, can be decomposed into a normal and a tangential part
as $\vec{f}^{c} := \vec{f}^{c}_{i} = f^n \vec{n} + f^t \vec{t}$,
where $\vec{n} \cdot \vec{t} = 0$, $n$ and $t$ being normal and tangential parts respectively. 
In this paper, we focus on frictionless particles, i.e., only normal 
forces will be considered, for tangential forces and torques, 
see e.g.\ Ref.\ \cite{Luding08} and references therein.

In the following, we discuss various normal contact force models, 
as shown schematically in Fig.\ 1.
We start with the linear contact model (Fig.\ 1(a)) for non-adhesive particles,
before we introduce a more complex contact model that is able to
describe the realistic interaction between adhesive, inhomogeneous
\footnote{Examples of inhomogeneous particles include core-shell materials,
clusters of fine primary particles or randomly micro-porous particles.},
slightly non-spherical particles (Fig.\ 1(b)).


\begin{figure*}
  \centering
    \mbox{
\subfigure[]{\includegraphics[scale=0.3]{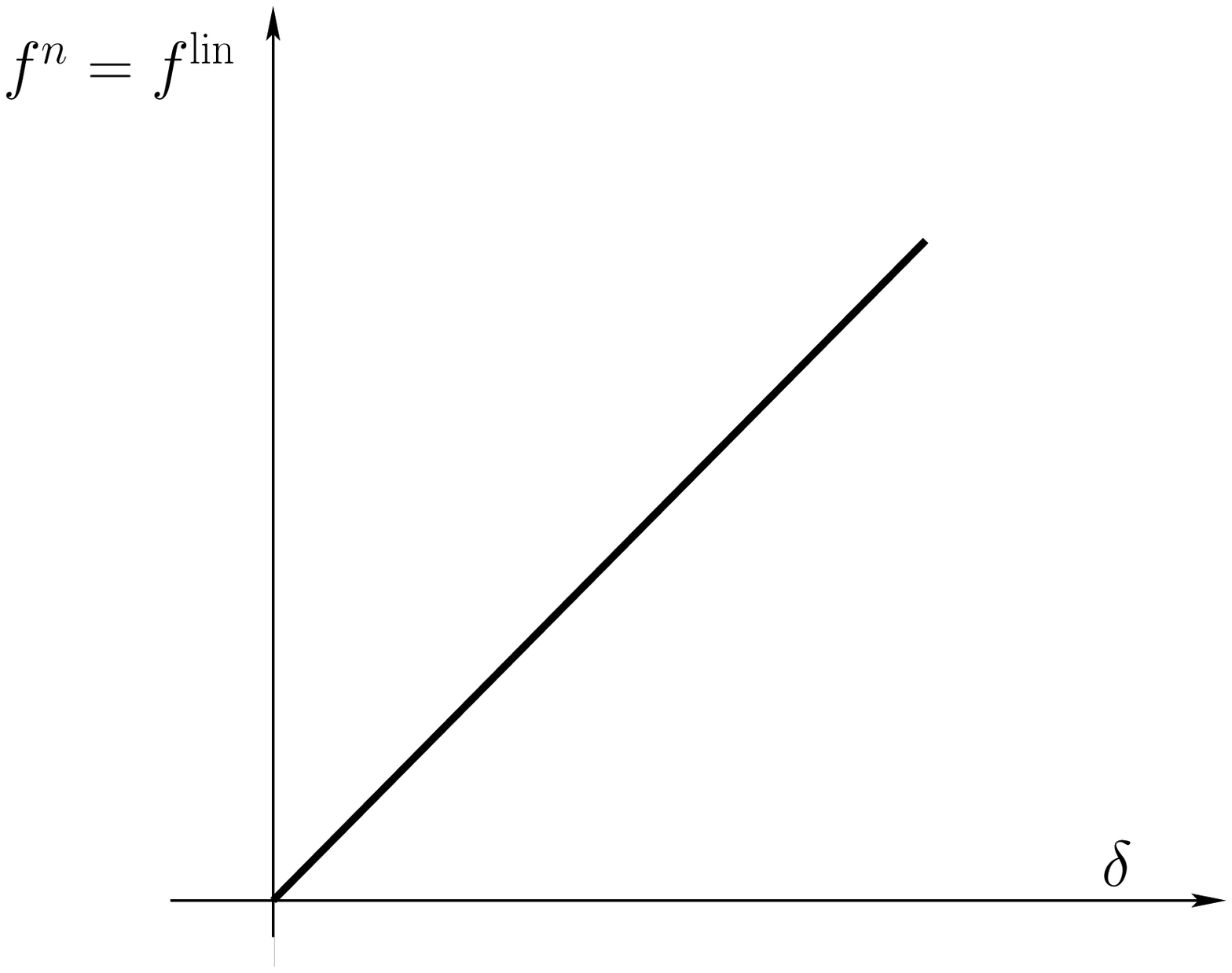}\label{fig:flin}}\quad
\subfigure[]{\includegraphics[scale=0.3]{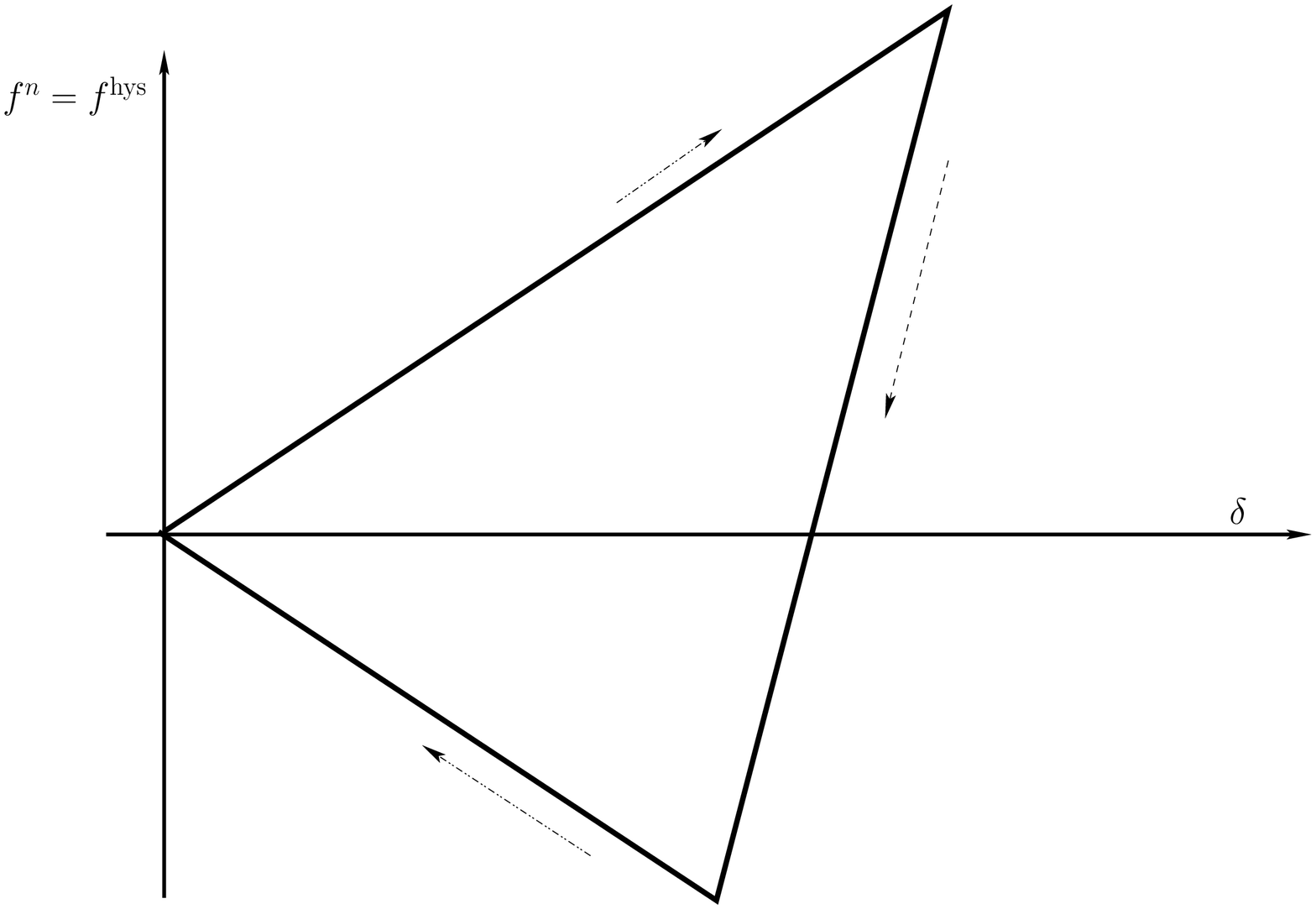}\label{fig:fnonadh-sch}}
}
\caption{
Schematic plots of contact forces for
(a) the linear normal model for a perfectly elastic collision, and
(b) the force-overlap relation for an elasto-plastic adhesive collision}
\label{fig:sch_for}
\end{figure*}


\subsubsection{Linear Normal Contact Model}\label{sec:Linearconmodel}

Modelling a force that leads to an inelastic collision requires at least two ingredients:
repulsion and some sort of dissipation. The simplest (but academic) normal force law 
with the desired properties is the damped harmonic oscillator
\begin{equation}
f^n = k \delta + \gamma_0 v^n ~,
\label{eq:fn_LSD}
\end{equation}
with spring stiffness $k$, viscous damping $\gamma_0$, 
and normal relative velocity 
$v^n=-{\vec{v}}_{ij} \cdot {\vec{n}}
    =-(\vec{v}_i - \vec{v}_j) \cdot \vec{n} = \dot \delta$.
This model (also called linear spring dashpot (LSD) model) 
has the advantage that its analytical solution (with initial 
conditions $\delta(0)=0$ and $\dot \delta(0)=v^n_0$)
allows easy calculations of important quantities \cite{luding98c}. 
For the non-viscous case, the linear normal contact model is 
given schematically in Fig.\ 1.
 
The typical response time (contact duration) and the
eigenfrequency of the contact are related as
\begin{equation}
t_c = {\frac{\pi}{\omega}} ~ {\rm ~~~and~~~}
\omega=\sqrt{({k}/{m_{r}})-\eta_0^{2}} ~
\label{eq:tc12}
\end{equation}
with the rescaled damping coefficient $\eta_0=\gamma_0/(2 m_{r})$,
and the reduced mass $m_{r}=m_i m_j / (m_i+m_j)$, where the
$\eta_0$ is defined such that it has the same units as $\omega$, 
i.e., frequency. From the solution of the equation of a half-period of the 
oscillation, one also obtains the coefficient of restitution
\begin{equation}
e^{\rm LSD}_n = v_f/v_i
  = \exp \left ( -\pi \eta_0 / \omega \right )
  = \exp \left ( - \eta_0 t_c         \right ) ~,
\label{eq:rnLSD}
\end{equation}
which quantifies the ratio of normal relative velocities after 
($v_f$) and before ($v_i$) the collision. 
Note that in this model $e_n$ is independent of $v_i$.
For a more detailed review on this and other, more realistic, 
non-linear contact models, see \cite{luding98c,Luding08}
and references therein.

The contact duration in Eq.\ \eqref{eq:tc12} is also of practical and
technical importance, since the integration of the equations of motion
is stable only if the integration time-step $\Delta t$
is much smaller than $t_c$.
Note that $t_c$ depends on the magnitude of dissipation: In the 
extreme case of an over-damped spring (high dissipation), 
$t_c$ can become very large
(which renders the contact behavior artificial \cite{luding94d}).
Therefore, the use of neither too weak nor too strong viscous
dissipation is recommended, so that some artificial effects are not observed by the use of viscous damping.


\subsubsection{Adhesive Elasto-Plastic Contacts}\label{sec:plasadh}

\begin{figure*}
\centering
\includegraphics[scale=0.5]{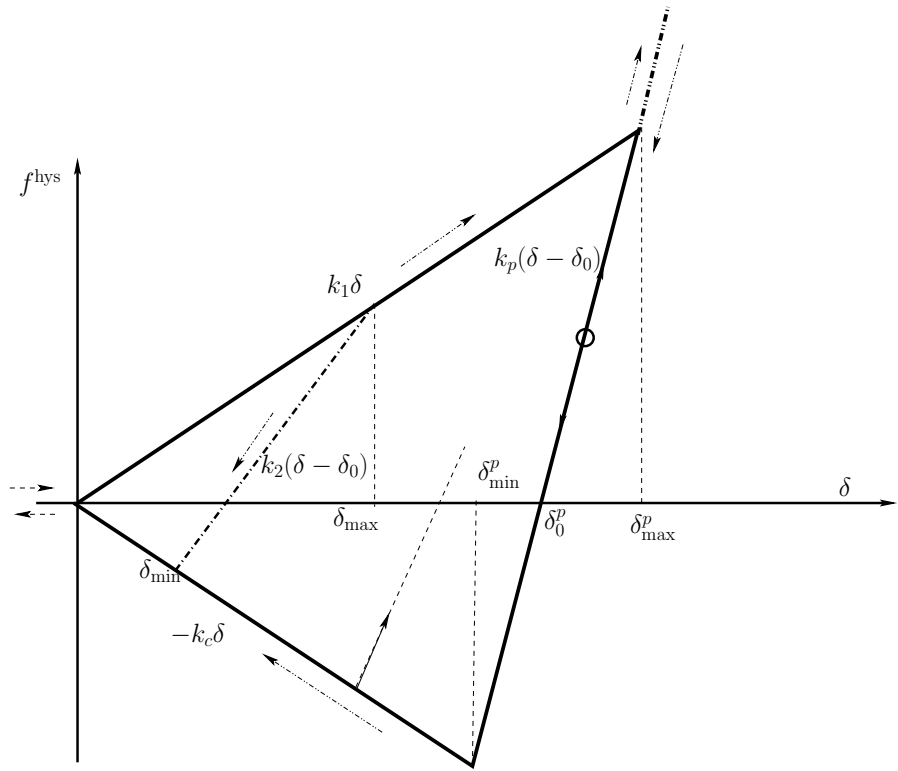}
\caption{Schematic graph of the piece-wise linear, hysteretic, 
         and adhesive force-displacement model in normal direction from Ref.\ \cite{Luding08}.
}
\label{fig:fnonadh}
\end{figure*}

For completeness, we re-introduce the piece-wise linear hysteretic 
model \cite{Luding08} as an alternative to non-linear spring-dashpot models 
or more complex hysteretic models 
\cite{tomas00,Tomas01-1,Tomas01-2,Thornton98,vu-quoc99b,zhang00}. 
It reflects permanent plastic deformation 
\footnote{After a contact is opened, the pair forgets its previous contact, 
since we assume that the contact points at a future re-contact of the same two
particles are not the same anymore.}, 
which takes place at the contact, and the non-linear increase of both elastic stiffness
and attractive (adhesive) forces with the maximal compression force.

In Fig.\ 2, the normal force at contact is plotted against
the overlap $\delta$ between two particles.
The force law can be written as
\begin{equation}
f^{\rm hys} = \left \{
\begin{array}{lll}
k_1  \delta           & 
            & {\rm if~~} k_2 (\delta-\delta_0) \ge k_1 \delta \\
k_2 (\delta-\delta_0) & 
            & {\rm if~~} k_1 \delta > k_2 (\delta-\delta_0) > -k_c \delta \\
-k_c \delta           & 
            & {\rm if~~} -k_c \delta \ge k_2 (\delta-\delta_0)
\end{array}
\right .
\label{fhys}
\end{equation}
with $k_1 \le k_2 \le k_p$, respectively the initial loading stiffness, 
the un-/re-loading stiffness and the elastic limit stiffness. 
The latter defines the limit force branch $k_p(\delta-\delta^p_0)$,
as will be motivated next in more detail, and
$k_2$ interpolates between $k_1$ and $k_p$, see 
Eq.\ \eqref{eq:k2_del}.
For $k_c=0$,
the above contact model 
reduces to that proposed by Walton and Braun \cite{walton86},
with coefficient of restitution 
\begin{equation}
e^{\rm WB}_n = \sqrt{k_1/k_2} ~.
\label{eq:rnWB}
\end{equation}
 
During the initial loading the force increases linearly with
overlap $\delta$ along $k_1$, until the maximum overlap 
$\delta_{\rm max} = v_i\sqrt{{m_r}/{k_1}}$ (for 
binary collisions) is reached, which is a history parameter
for each contact.
During unloading the force decreases along $k_2$, 
see Eq.\ \eqref{eq:k2_del}, from its maximum value 
$k_1\delta_{\rm max} $ at $\delta_{\rm max}$ 
down to zero at overlap 
\begin{equation}\label{eq:del}
 \delta_0=(1-k_1/k_2)\delta_{\rm max} ~,
\end{equation}
where $\delta_0$ resembles the {\em permanent plastic contact deformation}.
Re-loading at any instant 
leads to an increase of the force along the (elastic)
branch with slope $k_2$, until the maximum overlap $\delta_{\rm max}$ 
(which was stored in memory) is reached; 
for still increasing overlap $\delta$, the force again increases with slope 
$k_1$ and the history parameter $\delta_{\rm max}$ has to be updated.

Unloading below $\delta_0$ leads to a negative, {\em attractive} 
(adhesive) force, which follows the line with slope $k_2$, until 
the extreme adhesive force $-k_c \delta_{\rm min}$ is reached. 
The corresponding overlap is 
\begin{equation}
 \delta_{\rm min}=\frac{({k_2-k_1})}{({k_2+k_c})} \delta_{\rm max} ~.
\end{equation}
Further unloading follows the irreversible tensile limit branch, with slope 
$-k_c$, with the attractive force $f^{\rm hys}=-k_c \delta$.

The lines with slope $k_1$ and $-k_c$ define the range of possible
positive and negative forces.
  Between these two extremes,
unloading and/or re-loading follow the line with slope $k_2$. 
A non-linear un-/re-loading behavior would be more realistic, however,
due to a lack of detailed experimental informations, the piece-wise 
linear model is used as a compromise, besides that it is easier to implement. 
The elastic $k_2$ branch becomes non-linear and ellipsoidal if
a moderate normal viscous damping force is active at the contact,
as in the LSD model. 

In order to account for realistic load-dependent contact behavior, the $k_2$ 
value is chosen to depend on the maximum overlap  $\delta_{\rm max}$, 
i.e.\ {\em contacts are stiffer and more strongly plastically deformed 
for larger previous deformations} so that the dissipation depends on the previous 
deformation history. 
The dependence of $k_2$ on overlap $\delta_{\rm max}$ is chosen 
empirically as linear interpolation: 
\begin{equation}
k_2(\delta_{\rm max}) = \left \{
\begin{array}{lll}
k_{p}\,{\rm~~~~~~~~~~~~~~~~~~~~~~~~if~~} \delta_{\rm max} / \delta^{p}_{\rm max} \ge 1 \\
k_1 + (k_{p}-k_1)  {\delta_{\rm max} } / { \delta^{p}_{\rm max} } & \\
~~~~~~~~~~{\rm~~~~~~~~~~~~~~~~~~~if~~} \delta_{\rm max} / \delta^{p}_{\rm max} < 1
\end{array}
\label{eq:k2_del}
\right . 
\end{equation}
where $k_{p}$ is the maximal (elastic) stiffness, and 
\begin{equation}
\delta^p_{\rm max} 
 = \frac{k_{p}}{k_{p}-k_1} \phi_f \frac{2{a_1 a_2}}{a_1+a_2}~, 
\label{eq:kpdelp}
\end{equation}
is the plastic flow limit overlap, 
with $\phi_f$ the dimensionless plasticity depth, $a_1$ and $a_2$ being the radii 
of the two particles.  This can be further simplified to
\begin{equation}\label{eq:kpdel0}
\delta^p_{0} = \phi_f a_{12}, 
\end{equation}
where $\delta^p_{0}$ represents the plastic contact deformation at the limit overlap,
and $a_{12}=\frac{2{a_1 a_2}}{a_1+a_2}$ is the reduced radius. 
In the range $\delta_{\rm max}< \delta^{p}_{\rm max}$, the stiffness $k_2$ can also 
be written as:
\begin{equation}\label{eq:kpdelpWB}
 k_2=k_1+\frac{(k_{p}-k_1)}{k_1 \delta^{p}_{\rm max}} f^{\rm max}, 
\end{equation}
where $f^{\rm max}=k_1 \delta_{\rm max}$ is the same as Eq.\ (4) 
in \cite{walton86} with prefactor $S=\frac{(k_{p}-k_1)}{k_1 \delta^{p}_{\rm max}}$.

From energy balance considerations, one can define the ``plastic'' limit velocity
\begin{equation}
v_p = \sqrt{k_1/m_r} \, \delta^p_{\rm max} ~,
\label{eq:vpdef}
\end{equation}
below which the contact behavior is elasto-plastic, and above which 
the perfectly elastic limit-branch is reached. Impact velocities larger than
$v_p$ have consequences, as discussed next (see Sec.\ \ref{sec:shortcomings}).

In summary, the adhesive, elasto-plastic, hysteretic normal contact model 
is defined by the four parameters $k_1$, $k_p$, $k_c$ and $\phi_f$
that, respectively, account for the initial plastic loading stiffness,
the maximal, plastic limit (elastic) stiffness, the adhesion strength, and 
the plastic overlap-range of the model.
This involves an empirical choice 
for the load-dependent, intermediate elastic branch stiffness $k_2$,
which renders the model non-linear in its behavior (i.e.\ higher confinging
stress leads to stiffer contacts like in the Hertz model), even though the present
model is piece-wise linear.

\subsubsection{Motivation of the original contact model}\label{sec:disc}

To study a collision between two ideal, homogeneous spheres, one should
refer to realistic, full-detail contact models with a solid experimental and 
theoretical foundation \cite{JKR71,Thornton98, tomas00}. 
These contact models feature a small elastic regime and 
the particles increasingly deform plastically with increasing,
not too large deformation (overlap).
During unloading, their contacts end at finite overlap due to flattening. 
An alternative model was recently proposed, see Ref.\ \cite{pasha2014linear},
that follows the philosophy of {\em plastically flattened contacts} with instantaneous detachment 
at positive overlaps.

However, one has to also consider the possibility of rougher contacts
\cite{Hanaor2015contact}, and possible
non-contact forces that are usually neglected for very large particles,
but can become dominant and hysteretic as well as long-ranged for rather 
small spheres \cite{dominik1997physics,tomas00}. 

The mesoscopic contact model used here, as originally developed for sintering 
\cite{Luding2005Discrete}, and later defined in a temperature-independent form 
\cite{Luding08}, follows a different approach in two respects: 
(i) it introduces a limit to the plastic deformation of the particles/material
for various reasons as summarized below and also in subsection \ref{sec:shortcomings}, 
and (ii) the contacts are {\em not idealized as perfectly flat}, and thus do not have to lose 
mechanical contact immediately at un-loading, as will be detailed in subsection \ref{sec:tensile}.

 Note that a limit to the slope $k_p$ resembles a simplification of
 {\em different contact behavior at large deformations}:
 \\*
 (i) for low compression, due to the wide probability distribution of forces in bulk
 granular matter, only few contacts should reach the limit, which 
 would not affect much the collective, bulk behavior;
 \\*
 (ii) for strong compression, in many particle systems, i.e., for large deformations, 
 the particles cannot be assumed to be spherical anymore, and they deform plastically 
 or could even break;
 \\*
 (iii) from the macroscopic point of view, too large deformations would 
 lead to volume fractions larger than unity, which for most materials
 (except highly micro-porous, fractal ones) would be unaccountable;
 \\*
 (iv) at small deformation, contacts are due to surface roughness realized by multiple 
 surface asperities and at large deformation, the single pair point-contact 
 argument breaks down and multiple contacts of a single particle can 
 not be assumed to be independent anymore;
 \\*
 (v) finally, (larger) meso-particles have a lower stiffness than (smaller) primary particles
 \cite{Thakur2013}, which is also numerically relevant, since the time step has to be chosen 
 such that it is well below the minimal contact duration of all the contacts. 
 If $k_2$ is not limited the time-step could become prohibitively small,
 only because of a few extreme (large compression) contact situations.
\\*
The following two subsections discuss the two major differences of the present 
 piece-wise linear (yet non-linear) model as compared to other existing models: 
 (i) the elastic limit branch, and (ii) the elastic re-loading
 or non-contact-loss, as well as their reasons, relevance
 and possible changes/tuning -- in case needed.
\subsubsection{Shortcomings, physical relevance and possible tuning}\label{sec:shortcomings}

 In the context of collisions between perfect homogeneous elasto-plastic spheres, 
 a purely elastic threshold/limit and enduring elastic behavior after a
 sharply defined contact-loss are indeed
 questionable, as the plastic deformation of the single particle 
 cannot become reversible/elastic.
 Nevertheless, there are many materials that support the idea of a more elastic behavior
 at large compression (due to either very high impact velocity or multiple strong contact 
 forces), as discussed further in the paragraphs below.

\paragraph{\textbf{Mesoscopic contact model applied to real materials:}}
\label{sec:mesoscopic_model}
 First we want to recall that the present model is mainly aimed
 to reproduce the behavior of multi-particle systems of realistic fine
 and ultra-fine powders, which are typically non-spherical and often 
 mesoscopic in size with internal micro-structure and micro-porosity
 on the scale of typical contact deformation. 
 For example, think of clusters/agglomerates of primary nano-particles that 
 form fine micron-sized secondary powder particles, or
 other fluffy materials \cite{dominik1997physics}. 
 The primary particles are possibly better described by other contact models,
 but in order to simulate a reasonable number of secondary (meso) particles one 
 cannot rely on this bottom-up approach and hence a mesoscopic contact model
 needs to be used.
 During the bulk compression of such a system, the material deforms plastically and 
 both the bulk and particles' internal porosity reduces \cite{dominik1997physics}.  
 Plastic deformation diminishes if the material becomes so dense, 
 with minimal porosity, such that the elastic/stiff primary particles dominate.
 Beyond this point the system deforms {\em more} elastically, i.e. the stiffness becomes 
 high and the (irrecoverable) plastic deformations are much smaller than at 
 weaker compression.

 In their compression experiments of granular beds with
 micrometer sized granules of micro-crystalline cellulose, 
 Persson {\em et~al.}\ \cite{Persson12} found that a contact model 
 where a limit on plastic deformation is introduced  
 can very well describe the bulk behavior.
 Experimentally they observe a strong elasto-plastic bulk-behavior for the assembly at low 
 compression strain/stress. In this phase the height of the bed decreases, irreversibly with the 
 applied load. It becomes strongly non-linear beyond a certain strain/stress,
 which is accompanied by a dramatic increase of the stiffness of the aggregate. 
 They associate this change in the behavior to the loss of porosity and the subsequent 
 more elastic bulk response to the particles that are now closely in touch with each 
 other. In this new, re-structured, very compacted configurations, further
 void reduction is not allowed anymore and thus the behavior gets more elastic.
 While the elastic limit in the contact model does not affect the description of the bulk behavior
 in the first part, the threshold is found to play a key role in order to reproduce the material 
 stiffening (see Fig.\ 8 in Ref.\ \cite{Persson12}).

 Note that in an assembly of particles, not all the contacts will
 reach the limit branch and deform elastically simultaneously. 
 That is, even if few contacts are in the elastic limit, the system
 will always retain some plasticity, hence \textit{the assembly will never 
 be fully elastic}.
 
\paragraph{\textbf{Application to pair interactions:}}\label{para:pair}

Interestingly, the contact model in Sec.\ \ref{sec:plasadh} 
is suitable to describe the collision between pairs of particles, 
when special classes of materials are considered,
such that the behavior at high velocity and thus 
large deformation drastically changes. 

(i) \textit{Core-shell materials.}
The model is perfectly suited for plastic
core-shell materials, such as asphalt or ice particles,
having a ``soft'' plastic outer shell and a rather stiff, elastic inner core. 
For such materials the stiffness increases with the load due to
an increasing contact surface. For higher deformations, the inner cores can 
come in contact, which turns out to be almost elastic 
when compared to the behavior of the external shell. The model was
successfully applied to model asphalt, where the elastic inner core
is surrounded by a plastic oil or bitumen layer \cite{Tom12}.
Alternatively,
the plastic shell can be seen as the range of overlaps, where the 
surface roughness and inhomogeneities lead to a different contact
mechanics as for the more homogeneous inner core. 
 
(ii) \textit{Cold spray.}
 An other interesting system that can be effectively reproduced
 by introducing an elastic limit in the contact model is cold spray. 
 Researchers have experimentally and numerically shown
 that spray-particles rebound from the substrate at low velocities, 
 while they stick at intermediate impact energy
 \cite{Van99,Zhang05,Schmidt06,Moridi13}.  
 Wu {\em et~al.}\ \cite{Wu06} experimentally found that 
 rebound re-appears with a further increase in velocity (Fig.\ 3 in Ref.\ \cite{Wu06}). 
 Schmidt {\em et~al.}\ \cite{Schmidt06} relate the decrease of the deposition efficiency 
 (inverse of coefficient of restitution) to a transition from a plastic impact to hydrodynamic 
 penetration (Fig.\ 16 in Ref.\ \cite{Schmidt06}). Recently,
 Moridi {\em et~al.}\ \cite{Moridi13} numerically studied the sticking and rebound processes, 
 by using the adhesive elasto-plastic contact model of 
 Luding \cite{Luding08}, and their prediction of the velocity dependent behavior
 is in good agreement with experiments.

(iii) \textit{Sintering.}
 As an additional example, we want to recall that the present mesoscopic contact model has already 
 been applied to the case of sintering, see Ref.\ \cite{Luding2005Discrete,luding2011discrete}.
 For large deformations, large stresses, or high temperatures, the material goes 
 to a fluid-like state rather than being solid. Hence, the elasticity of the
 system (nearly incompressible melt) determines its limit stiffness, while $\phi_f$ 
 determines the maximal volume fraction that can be reached.

All the realistic situations described above clearly hint at a modification in 
the contact phenomenology that can not be described solely by an elasto-plastic 
model beyond some threshold in the overlap/force. 
The limit stiffness $k_p$ and the plastic layer depth $\phi_f$ in our model allow 
the transition of the material to a new state.
Dissipation on
the limit branch -- which otherwise would be perfectly elastic --
can be taken care of, by adding a viscous damping force
(as the simplest option). Due to viscous damping, 
the unloading and re-loading will follow different paths,
so that the collision will never be perfectly elastic, 
which is in agreement with the description in 
Jasevi\u{c}ius {\em et~al.}\ \cite{Jasevicius09,Jasevicius11} 
and will be shown later in Appendix\ \ref{sec:viseffect}.

Finally, note that an elastic limit branch is surely not the ultimate 
solution, but a simple first model attempt -- possibly requiring 
material- and problem-adapted improvements in the future.

\paragraph{\textbf{Tuning of the contact model:}}\label{sec:tuning_model}
 The change in behavior at large contact deformations is thus a feature
 of the contact model which allows us to describe many special types of materials.
 Nevertheless, if desired (without changing the model), the parameters can be tuned
 in order to reproduce the behavior of materials where the plasticity
 increases with deformation without limits, i.e., the elastic core feature can be removed.
 The limit-branch where plastic deformation ends is defined by the dimensionless 
 parameters plasticity depth, $\phi_f$, and maximal (elastic) stiffness, $k_p$. 
 Owing to the flexibility of the model, it can be tuned such that the limit overlap 
 is set to a much higher value that is never reached by the contacts. 
 When the new value of ${\phi_f}^{'}$ is chosen, a new ${k_p}^{'}$ can be calculated
 to describe the behavior at higher overlap (as detailed in Appendix\ \ref{sec:mod}).
 In this way the model with the extended ${\phi_f}^{'}$ exhibits elasto-plastic behavior for a higher
 velocity/compression-force range, while keeping the physics of the system for smaller
 overlap intact.

\subsubsection{Irreversibility of the tensile branch}\label{sec:tensile}

 Finally we discuss a feature of the contact model in \cite{Luding08}, 
 that postulates the irreversibility, i.e. partial elasticity, of the tensile 
 $k_c$ branch, as discussed in Sec.\ \ref{sec:plasadh}. While this is 
 unphysical in some situations, e.g.\ for homogeneous plastic spheres, 
 we once again emphasize that we are interested 
 in non-homogeneous, non-spherical meso-particles, as e.g.\ clusters/agglomerates 
 of primary particles in contact with internal structures of the order of typical
 contact deformation.  
 The perfectly flat surface detachment due to plasticity happens
 only in the case of ideal, elasto-plastic adhesive, perfectly spherical particles 
 (which experience a large enough tensile force).
 In almost all other cases, the shape of the detaching surfaces and the hence the subsequent
 unloading behavior depends on the relative strengths of plastic dissipation, attractive forces, 
 and various other contact mechanisms. 
In the case of meso-particles such as the core-shell materials \cite{Tom12}, 
 assemblies of micro-porous fine powders \cite{dominik1997physics,Persson12}, 
 or atomic nanoparticles \cite{Tanaka12}, other details such as rotations can be important.
 We first briefly discuss the case of ideal elasto-plastic adhesive particles and later 
 describe the behavior of many particle systems, which is the main focus of this work.
 

 Ideal homogeneous millimeter sized particles detach with a permanently flattened surface 
 created during deformation
 are well described using contact models presented in \cite{Thornton98,pasha2014linear}.
 This flattened surface is of the order of micrometers and the plastic
 dissipation during mechanical contact is dominant over the van der Waals force.
 During unloading, when the particles detach, the force suddenly drops to zero from the tensile branch.  
 When there is no contact, further un- and re-loading involves no force. Even when the contact is
 re-established, the contact is still assumed to be elastic, i.e., it follows the previous contact-unloading path.
 This leads to very little or practically no plastic deformation at the re-established contact, 
 until the (previously reached) maximum overlap is reached again and the plasticity kicks in.
 

 On the other hand for ultra-fine ideal spherical particles of the order of 
 macro-meters  \cite{tomas00,Tomas07,Tomas09ch}, the van der Waals force is much
 stronger and unloading adhesion is due to purely non-contact forces. 
 Therefore, the non-contact forces do not vanish and even extend beyond
 the mechanical first contact distance. The contact model of Tomas
 \cite{tomas00,Tomas07} is reversible for non-contact and features a strong
 plastic deformation for the re-established contact -- in contrast to the 
 previous case of large particles.

 The contact model by Luding \cite{Luding08} follows similar considerations as others,
 except for the fact that the mechanical contact does {\em not} detach (for details see the next section). 
 The irreversible, elastic re-loading before complete detachment can be seen 
 as a compromise between small and large particle mechanics, i.e. between
 weak and strong attractive forces. It also could be interpreted as a premature
 re-establishment of mechanical contact, e.g.\ due to a rotation of the deformed, 
 non-spherical particles. Detachment and remaining non-contact is only then valid if the
 particles do not rotate relative to each other; in case of rotations, 
 both sliding and rolling degrees of freedom can lead to a mechanical
 contact much earlier than in the ideal case of a perfect normal collision
 of ideal particles.
 In the spirit of a mesoscopic model, the irreversible contact model is 
 due to the ensemble of possible contacts, where some behave as imagined 
 in the ideal case, whereas some behave strongly different, e.g.\ due to  
 relative rotation.  However, there are several other reasons to
 consider an irreversible unloading branch, as summarized in the following.

In the case of asphalt (core-shell material with a stone core and bitumen-shell), 
depending on the composition of the bitumen (outer shell), which can contain a considerable 
amount of fine solid, when the outer shells collide the collision is plastic. In contrast, 
 the collision between the inner cores is rather elastic (even though the inner cores
 collide when the contact deformation is very large).
 Hence, such a material will behave softly for loading, but will be rather stiff for re-loading 
 (elastic $k_2$ branch), since the cores can then be in contact.
 A more detailed study of this class of materials goes beyond the scope of this study 
 and the interested reader is referred to Ref.\ \cite{Tom12}

For atomistic nano-particles and for porous materials, one thing in common
is the fact that the {\em scale of a typical deformation} can be much larger than the inhomogeneities
 of the particles and that the adhesion between primary particles is strong enough to keep them
 agglomerated during their re-arrangements (see Fig. 5 in Ref.\ \cite{Tanaka12} and the
phenomenology in Ref.\ \cite{dominik1997physics}, as well as recent results for different
deformation modes \cite{Boltachev2015nanopowders}).
Thus the deformation of the bulk material will be plastic (irreversible), 
even if the primary particles would be perfectly elastic.

For agglomerates or other mesoscopic particles, 
we can not assume permanent ideal flattening and complete, instantaneous
loss of mechanical contact during unloading \cite{dominik1997physics}. 
In average, many contacts between particles might be lost, 
but -- due to their strong attraction -- many others will still 
remain in contact. 
Strong clusters of primary particles will remain intact and can form 
threads, a bridge or clumps during unloading -- which either keeps the two surfaces in 
contact beyond the (idealized) detachment point \cite{dominik1997physics}
or can even lead to an additional elastic repulsion due to a clump of particles 
between the surfaces 
(see Fig.\ 3 in Ref.\ \cite{Luding08} and Appendix\ \ref{sec:agg}).

During re-loading, the (elastic) connecting elements influence the bulk response.  
At the same time, the re-arrangements of the primary particles (and clusters)
can happen both inside and on the surface, which leads to reshaping, very
likely leaving a non-flat contact surface 
\cite{dominik1997physics,Paulick2015,Hanaor2015contact}. 
As often mentioned for granular systems, the interaction of several
elastic particles does not imply bulk elasticity of the granular assembly, 
due to (irreversible) re-arrangements in the bulk material -- especially
under reversal of direction \cite{Rojas2015}. 
Thus, in the present model an irreversible tensile branch is assumed, 
without distinction between the behavior before and after the first contact-loss-point
other than the intrinsic non-linearity in the model: The elastic stiffness for
re-loading $k_2$ decreases the closer it comes to $\delta=0$; in the present
version of the contact model, $k_2$ for unloading from the $k_1$ branch and 
for re-loading from the $k_c$ branch are exactly matched (for the sake of simplicity).

It is also important to mention that large deformation, and hence large forces are rare,
 thanks to the exponential distribution of the deformation and thus forces, 
 as shown by our studies using this contact model \cite{Luding2005Discrete,Singh2014effect,Singh2014friction}. 
 Hence, such large deformations are rare and do not strongly affect the 
 bulk behavior, as long as compression is not too strong. 

As a final remark, for almost all models on the market
 -- due to convenience and numerical simplicity, in case of complete detachment $\delta < 0$ 
 -- the contact is set to its initial state, 
 since it is very unlikely that the two particles will touch again at exactly the same contact 
 point as before.
On the other hand in the present model a long-range interaction is introduced, in the same
 spirit as \cite{Tomas01-1,Tomas07}, which could be used to extend the contact memory to
 much larger separation distances.
Re-loading from a non-contact situation ($\delta<0$)  is, however, assumed to be 
starting from a ``new'' contact, since contact model and non-contact
forces are considered as distinct mechanisms, for the sake of simplicity. 
Non-contact forces will be detailed in the next subsection.

\begin{figure*}
  \centering
\mbox{
\subfigure[]{\includegraphics[scale=0.4]{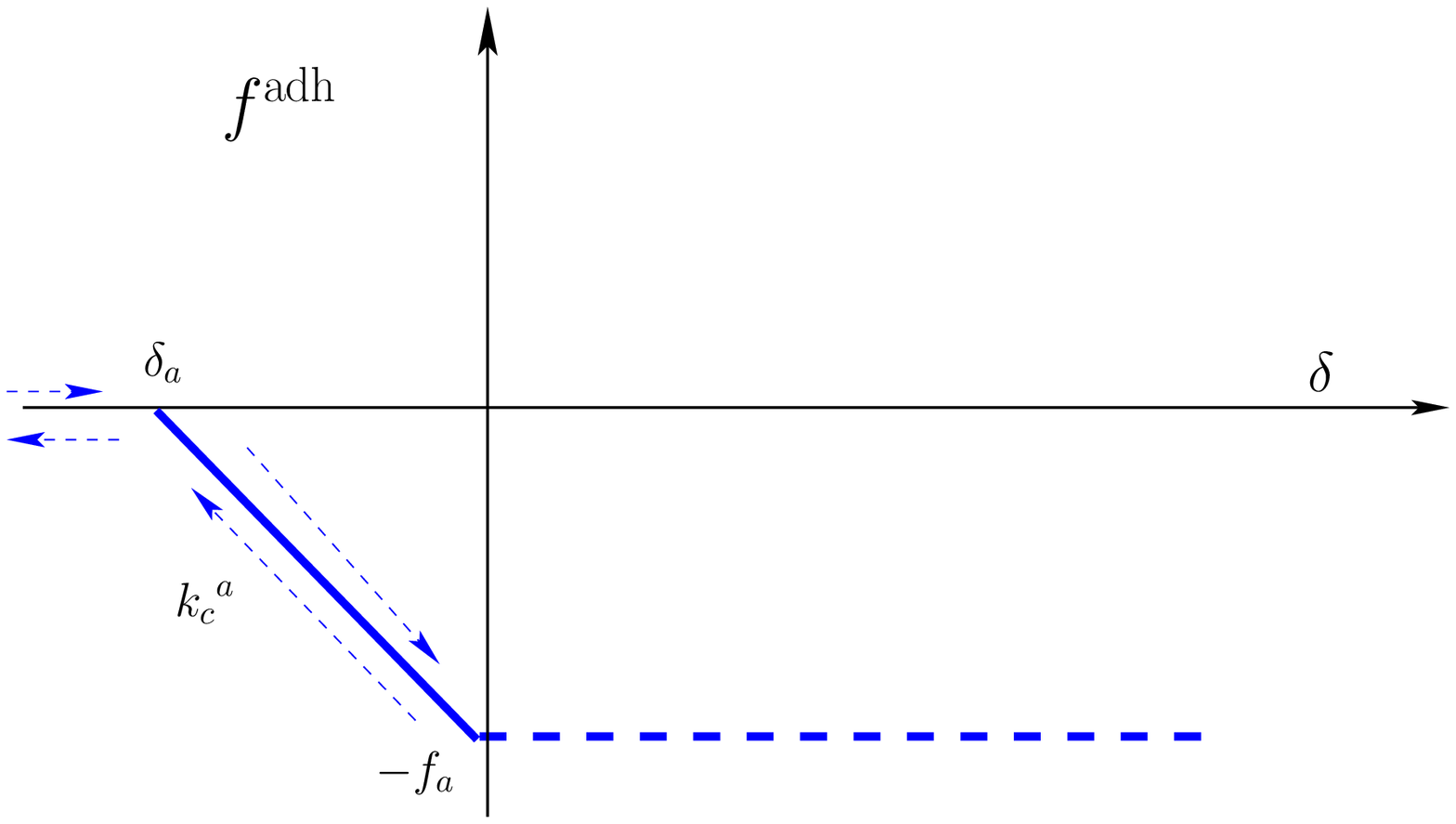}\label{fig:fadh-sch-off}}\quad
\subfigure[]{\includegraphics[scale=0.4]{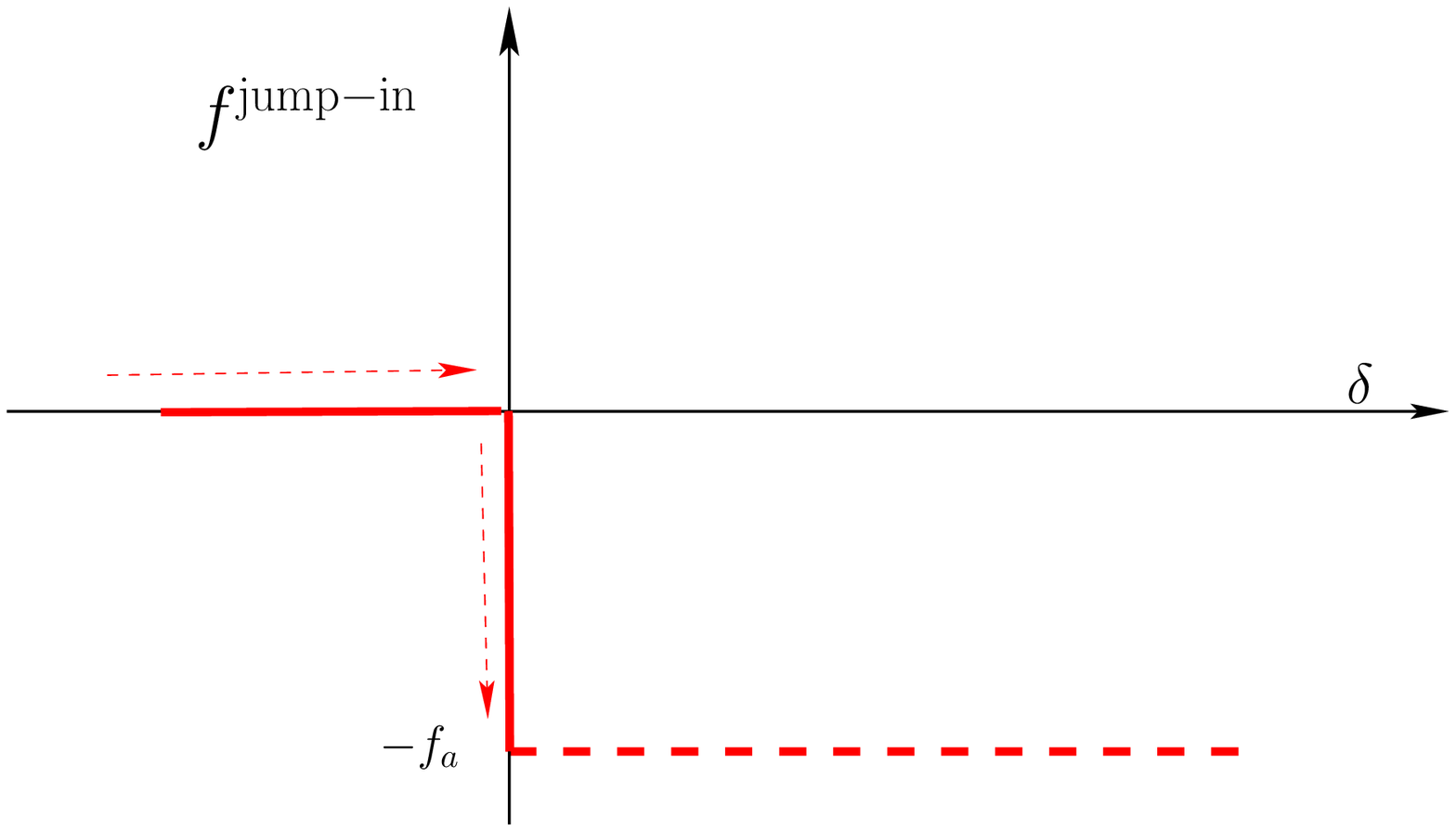}\label{fig:fadh-sch-in}}
}
\caption{
Schematic plots of 
(a) the non-contact adhesive force-overlap relation and
(b) the non-contact jump-in force-overlap relation.}
\label{fig:sch_noncontact}
\end{figure*}

\subsection{Non-contact normal force}
\label{sec:non-contact}
It has been shown in many studies that long-range interactions are 
present when dry adhesive particles collide, 
i.e.\ non-contact forces are present for negative overlap $\delta$ 
\cite{Thornton98,Luding08,Tomas07,muller2014simulation,mader2015microscopic}.
In the previous section, we have studied the force laws for contact 
overlap \ $\delta>0$. In this section we introduce a description
for non-contact, long range, adhesive forces, focusing on the two
non-contact models schematically shown in Fig.\ 3
-- both piece-wise linear in the spirit of the mesoscopic model --
namely the reversible model and the jump-in (irreversible) non-contact
models (where the latter could be seen as an idealized, mesoscopic
representation of a liquid bridge, just for completeness).
Later, in the next section, we will combine non-contact and contact forces.

\subsubsection{Reversible Adhesive force}
\label{para:revadheps_i}

In Fig.\ 3(a) we consider 
the reversible attractive case, where a (linear) van der Waals type 
long-range adhesive force is assumed. 
The force law can be written as
\begin{equation}
f^{\rm adh} =  \left \{
\begin{array}{lll}
-f_a  & & {\rm if~~}    {\delta}>0 \\
-k^a_c \delta - f_a        & 
        & {\rm if~~}    0\geq{\delta}>{\delta_a} \\
0                        &
        & {\rm if~~}    {\delta_a}>{\delta} \\
\end{array}
\label{eq:fadh}
\right . 
\end{equation}
with the range of interaction $\delta_a=-{f_a}/{k^a_c} < 0$, 
where $k^a_c>0$ is the adhesive ``stiffness'' of the material
\footnote{Since the $k_c$-branch has a negative slope, this
parameter does not represent a true stiffness of the material, 
which must have a positive sign.} and
$f_a>0$ is the (constant) adhesive force magnitude, active 
also for overlap $\delta>0$, in addition to the contact force.
%
When $\delta=0$ the force is $-f_a$.
The adhesive force $f^{\rm adh}$ is active when particles 
are closer than $\delta_a$, when it starts to increase/decrease linearly 
along $-k_c^{a}$, for approach/separation, respectively. 
In the results and theory part of the paper, for the sake 
of simplicity and without loss of generality, the adhesive stiffness can 
be either chosen as infinite, which corresponds to zero range non-contact
force ($\delta_a=0$), or as coincident with the contact adhesive stiffness,
e.g.\ in Sec.\ \ref{sec:plasadh}, that is $k^a_c=k_c$.

\subsubsection{Jump-in (Irreversible) Adhesive force}
\label{para:irradheps_i}
In Fig.\ 3(b) we report the behavior of the non-contact force versus 
overlap when the approach between particles is described by a discontinuous (irreversible) 
attractive law.
The jump-in force can be simply written as
\begin{equation}
f^{\rm jump-in} =  \left \{
\begin{array}{lll}
0            & 
            & {\rm if~~}    {\delta} < 0 \\
-f_a                            &
            & {\rm if~~}    {\delta} \ge 0 \\
\end{array}
\label{fatt_irradh}
\right . ~.
\end{equation}
As suggested in previous 
studies \cite{JKR71,Thornton98,Brilliantov07}, there is no attractive 
force before the particles come into contact; the 
adhesive force becomes active and suddenly drops to 
a negative value, $-f_{a}$, at contact, when $\delta=0$. 
The jump-in force resembles
the limit case $k^a_c \rightarrow \infty$ of Eq.\ \eqref{eq:fadh}.
Note that the behavior is defined here only for approach of the particles. 
We assume the model to be irreversible, as in the unloading stage, during separation, 
the particles will not follow this same path (details will be discussed below).

\section{Coefficient of Restitution}
\label{sec:split_COR}
 
The amount of dissipated energy relative to the incident kinetic energy 
is quantified by $1-e^2$, in terms of the coefficient of restitution $e$.
Considering a pair collision, 
with particles approaching from infinite distance,
the coefficient of restitution is defined as
\begin{subequations}
\begin{equation}
e=\frac{{v_f}^{\infty}}{{v_i}^{\infty}} ~
\end{equation}
where ${v_f}^{\infty}$ and ${v_i}^{\infty}$ are final and initial
velocities, respectively, at infinite separations (distance beyond which 
there is no long range interaction). Assuming superposition of the 
non-contact and contact forces, the restitution coefficient can be 
further decomposed including terms of final and initial velocities, 
$v_f$ and $v_i$, at overlap $\delta=0$, where the mechanical 
contact-force becomes active:
\begin{equation}
e = \frac{{v_f}^{\infty}}{v_f} \frac{v_f}{v_i} \frac{v_i}{{v_i}^{\infty}}
  = {\epsilon_o}{e_n}{\epsilon_i} ~, 
\label{eq:splitCOR}
\end{equation}
\end{subequations}
and ${\epsilon_i}$ and ${\epsilon_o}$ are the pull-in and pull-off 
coefficients of restitution, that describe the non-contact parts of
the interaction ($\delta < 0$), 
for approach and separation of particles, respectively.
The coefficient of restitution for particles in mechanical contact 
($\delta > 0$) is $e_n$, 
as analytically computed in subsection \ref{sec:e_n}.
\\*
In the following, we will analyze each term in
Eq.\ \eqref{eq:splitCOR} separately, based on energy considerations.
This provides the coefficient of restitution for a wide, general class 
of meso interaction models with
superposed non-contact and contact components, as defined in 
sections \ref{sec:forcelaws}-\ref{sec:non-contact}. 

For the middle term, $e_n$, different contact models with their
respective coefficients of restitution can be used, 
e.g.\ $e_n^{\rm LSD}$ from Eq.\ (\ref{eq:rnLSD}),
$e_n^{\rm  WB}$ from Eq.\ (\ref{eq:rnWB}), or
$e_n^{\rm HYS}$ as calculated below in subsection \ref{sec:e_n}.
Prior to this, we specify ${\epsilon_i}$ in subsection \ref{sec:pull-in}
and then ${\epsilon_o}$ in subsection \ref{sec:pull-off}, for the 
simplest piece-wise linear non-contact models. \footnote{
If other, possibly non-linear non-contact forces such as
square-well, van der Waals or Coulomb are used, see 
Refs.\ \cite{muller08t,muller11,gonzalez13,gonzalez14},
the respective coefficient of restitution has to be computed, 
and also the long-range nature has to be accounted for,
which goes far beyond the scope of this paper.}

\subsection{Pull-in coefficient of restitution}
\label{sec:pull-in}

In order to describe the pull-in coefficient of restitution $\epsilon_i$,
we focus on the two non-contact models proposed in Sec.\ \ref{sec:non-contact}, 
as simple interpretations of the adhesive force during the approach of the 
particles. 

When the \textit{reversible adhesive} contact model is used, 
energy conservation leads to an increase in velocity due to the 
attractive branch from $\delta_a$ ($<0$) to contact:
\begin{subequations}
 \begin{equation}
   \frac{1}{2}m_r{{v_i}^{\infty}}^2
 = \frac{1}{2}f_a\delta_a + \frac{1}{2}m_r{v_i}^2 ~,
 \end{equation}
which yields
\begin{equation}\label{eq:epsirev}
 {\epsilon_i^{\rm rev-adh}}
     =\frac{v_i}{{v_i}^{\infty}}
     =\sqrt{1-\frac{f_a\delta_a}{m_r{{v_i}^{\infty}}^2}} 
     =\sqrt{1+\frac{f_a^2/k_c^a}{m_r{{v_i}^{\infty}}^2}} ~.
\end{equation}
\end{subequations}
The pull-in coefficient of restitution is thus larger than unity;
it increases with increasing adhesive force magnitude $f_{a}$
and decreases with the adhesive strength of the material $k_{c}^a$
(which leads to a smaller cutoff distance).

On the other hand, 
if the \textit{irreversible adhesive jump-in} model is implemented, 
a constant value $\epsilon_i^{\rm jump-in}=1$ is obtained for first 
approach of two particles, before contact, 
as $f^{\rm jump-in}=0$ for $\delta<0$ and 
the velocity remains constant $v_i={v_i}^{\infty}$.

%

\subsection{Pull-off coefficient of restitution}
\label{sec:pull-off}

The pull-off coefficient of restitution
is defined for particles that lose contact and separate. 
Using the \textit{adhesive reversible} model, 
as described in section \ref{para:revadheps_i},
energy balance leads to a reduction in velocity during
separation:
\begin{subequations}
 \begin{equation}
  \frac{1}{2}m_r{{v_f}^{\infty}}^2
    = \frac{1}{2}f_a\delta_a + \frac{1}{2}m_r{v_f}^{2} ~, 
 \end{equation}
which yields
\begin{equation}\label{eq:epsfrev}
 {\epsilon_{o}}=\frac{{v_f}^{\infty}}{{v_f}}
       =\sqrt{1+\frac{f_a\delta_a}{m_r{v_f}^2}} 
       =\sqrt{1-\frac{f_a^2/k_c^a}{m_r{v_f}^2}} ~,
\end{equation}
\end{subequations}
due to the negative overlap $\delta_a$ at which the contact ends.
Similar to Eq.\ (\ref{eq:epsirev}), the pull-off coefficient of restitution 
depends on both the adhesive force magnitude $f_{a}$ and stiffness $k_{c}$, 
given the separation velocity $v_{f}$ at the end of the mechanical contact.

It is worthwhile to note that the force-overlap picture described above, 
with $\epsilon_{o}<1$ defined as in Eq.\ (\ref{eq:epsfrev}) 
refers to a system with sufficiently high impact velocity, 
so that the particles can separate with a finite kinetic energy 
at the end of collision, i.e., 
${v_f}^2 > f_a^2/(m_r k_c^a) =: (v_f^a)^2$
or, equivalently, 
$v_i^\infty > v_f^a/(e_n \epsilon_i)$, where $v_f^a$ denotes the 
minimal relative velocity at the end of the contact, for which 
particles can still separate.
If the kinetic energy reaches zero before the separation, 
e.g.\ the particles start re-loading along the adhesive 
branch until the value $\delta=0$ is reached and the contact
model kicks in.

\begin{figure*}
\centering
\subfigure[]{
\includegraphics[width=0.4\textwidth]{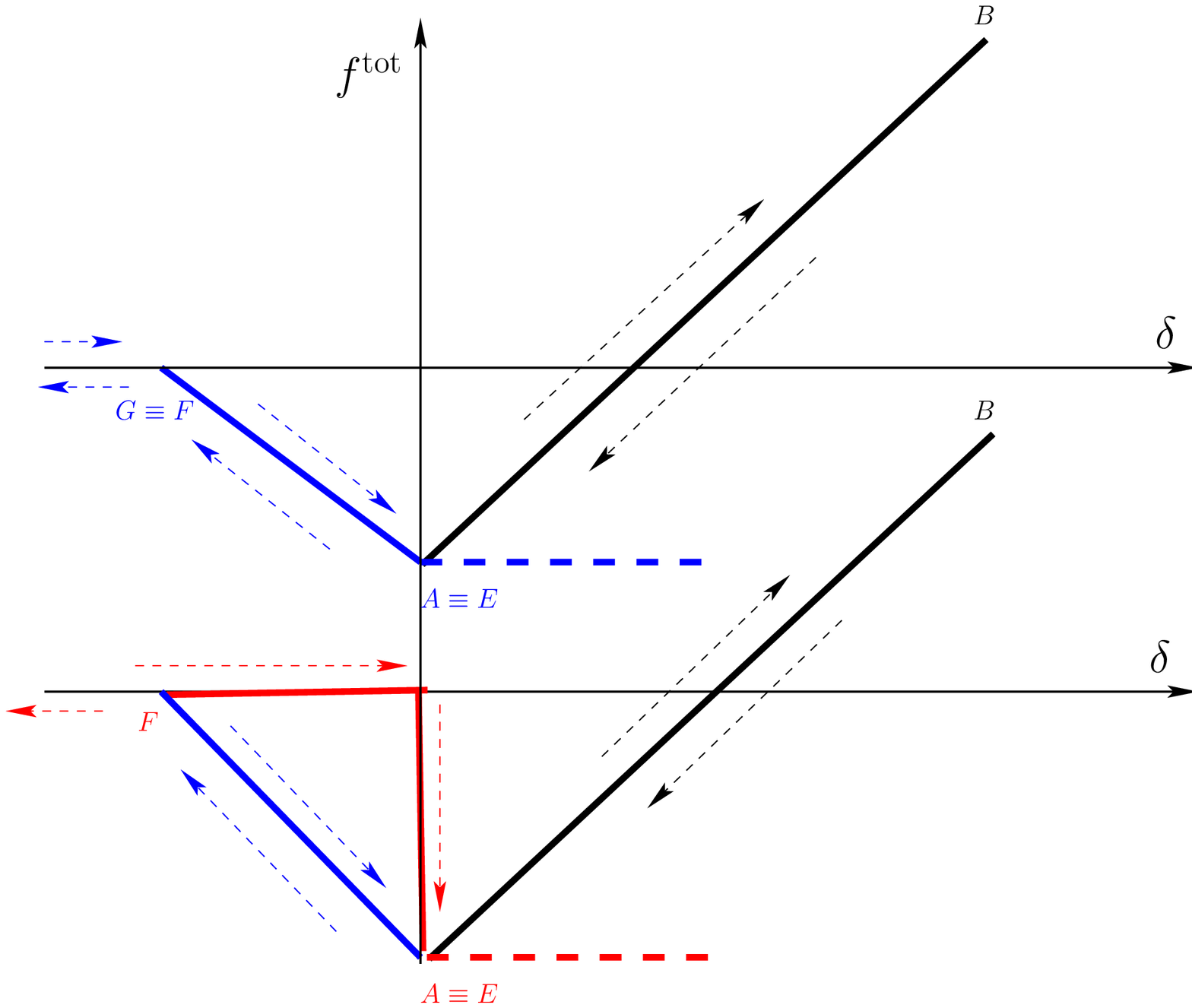}
\label{fig:flinadh}
}
\subfigure[]{
\includegraphics[width=0.5\textwidth]{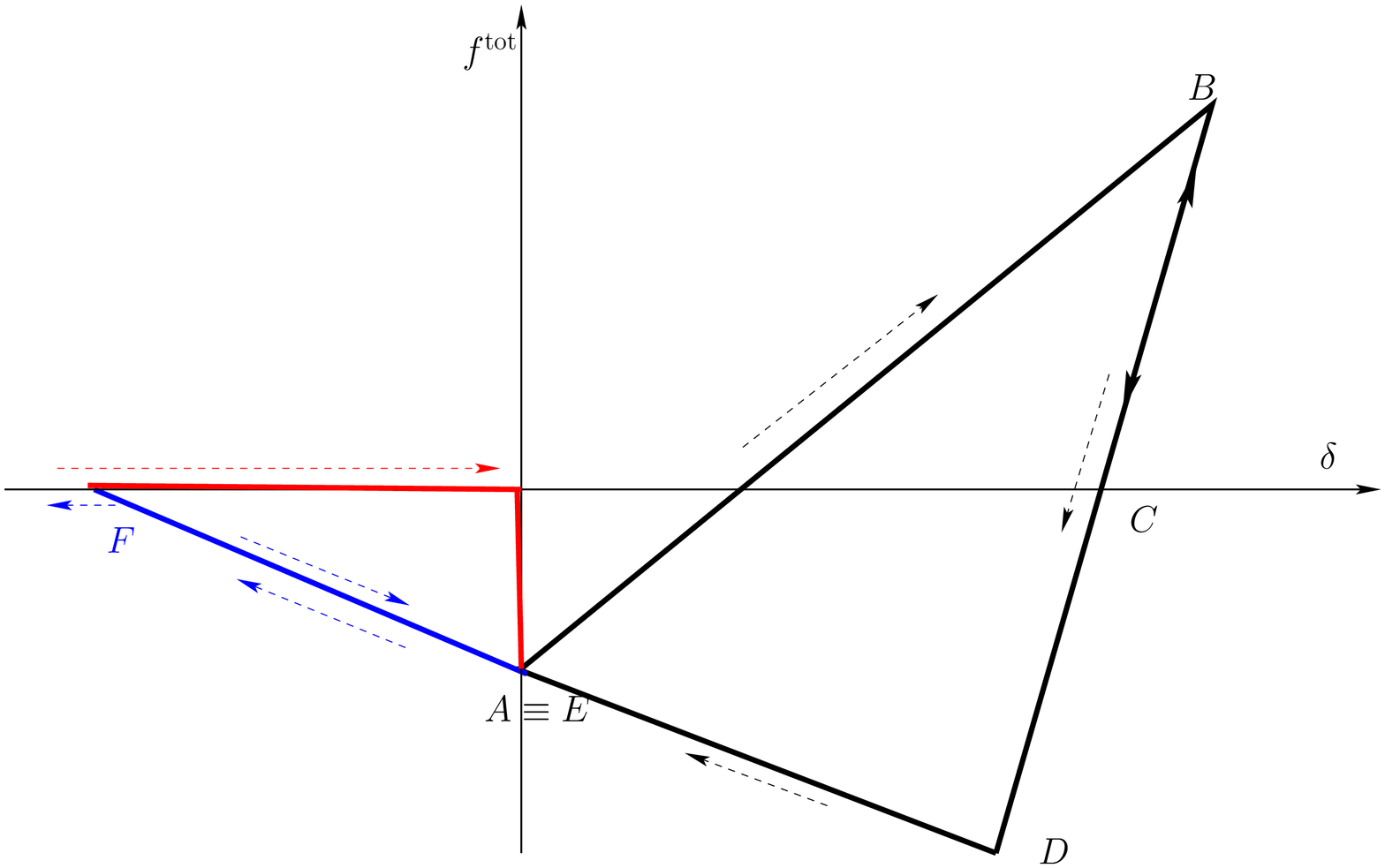}
\label{fig:fnonahd_mod}
}
\centering
\caption{(a) Reversible and irreversible non-contact forces,
where the top blue line (for negative overlap) represents the former 
and the bottom red line (for negative overlap) the latter. The black 
line for positive overlap represents the linear contact force as  
superimposed on the non-contact force.
(b) Force-displacement law for elasto-plastic, adhesive contacts 
superimposed on the irreversible non-contact adhesive force.}
\label{fig:force_mod}
\end{figure*}


\subsection{Elasto-plastic coefficient of restitution}
\label{sec:e_n}

The key result of this paper is the analytical study of the 
coefficient of restitution as function of the impact velocity,
for the model presented in Fig.\ 4(b),
disregarding viscous forces in order to allow for a 
closed analytical treatment.  The impact velocity $v_i$ 
is considered for two cases $v_i \le v_p$ and $v_i > v_p$,
with the plastic-limit velocity $v_p$ (needed to reach the 
elastic branch), defined as:
\begin{eqnarray}
v_p & = & \sqrt{ \frac{k_1}{m_r} \left [ (\delta^p_{\rm max}-f_a/k_1)^2
            -({f_a}/{k_1})^2 \right ] }  \nonumber \\
    & = & \sqrt{ \frac{1}{m_r}  \delta^p_{\rm max} \left [
                                     k_1 \delta^p_{\rm max} - 2 f_a 
                             \right ] } ~, 
\label{eq:vp}
\end{eqnarray}
where the term(s) with $f_a$ represent the energy gained or lost by this
(attractive, negative) constant force, with zero reached at overlap 
$\delta_a^{(1)}=f_a/k_1$, and $\delta^p_{\rm max}$ defined in 
Eq.\ (\ref{eq:kpdelp}). 
The velocity $v_p$ needed to reach the limit branch thus decays 
with increasing non-contact attraction force $f_a$.

\subsubsection{Plastic contact with initial relative velocity $v_i< v_p$}
\label{sec:smallvel}

When $v_i<v_p$ the particles after loading to $\delta_{\rm max}$, unload with slope $k_2$ and the system
deforms along the path $ 0 \rightarrow \delta_{\rm max} \rightarrow \delta_{\rm 0}^a \rightarrow \delta_{\rm min} \rightarrow 0 $, corresponding to
 $ A \rightarrow B \rightarrow C \rightarrow D \rightarrow E$ in Fig.\ 4(b).

The initial kinetic energy (at $\delta=0$ overlap, with adhesive force $f_a$ and with initial velocity $v_i<v_p$) is completely transformed to 
potential energy at the maximum overlap $\delta_{\rm max}$ where energy-balance provides:
\begin{subequations}\label{eq:10}
\begin{equation}\label{eq:10a}
E_i := \frac{1}{2}m_rv^2_{i}
 =\frac{1}{2}(k_1\delta_{\rm max}-f_a)\left(\delta_{\rm max}-\frac{f_a}{k_1}\right)-\frac{1}{2}\frac{f^2_a}{k_1}
 =\frac{1}{2} \delta_{\rm max} ( k_1 \delta_{\rm max} - 2 f_a )~,
\end{equation}
so that the physical (positive) solution yields:
\begin{equation}\label{eq:Delmax}
 \delta_{\rm max}
 = \frac{f_a+\sqrt{f^2_a+k_1m_rv^2_i}}{k_1}
 = \delta^{(1)}_a+\sqrt{\left (\delta^{(1)}_a \right )^2+m_rv^2_i/k_1} ~,
\end{equation}
with zero force during loading at $\delta^{(1)}_a=f_a/k_1$. 
The relative velocity is reversed at $\delta_{\rm max}$, and
unloading proceeds from point $B$ along the slope 
$k_2=k_2(\delta_{\rm max})$. 
Part of the potential energy is dissipated, the rest is converted 
to kinetic energy at point $C$, the force-free overlap $\delta^a_{\rm 0}$,
in the presence of the attractive force $f_a$:
\begin{equation}\label{eq:10b}
\frac{1}{2}m_rv^2_{0}
 = \frac{1}{2}(k_1\delta_{\rm max}-f_a)(\delta_{\rm max}-\delta^a_{\rm 0})
 = \frac{k_1}{2 k_2}\left(m_rv^2_{i}+\frac{f^2_a}{k_1}\right) 
 = \frac{1}{2k_2} (k_1 \delta_{\rm max}-f_a)^2 ~,
\end{equation}
where $\delta^a_{\rm 0}=[(k_2-k_1)\delta_{\rm max}+f_a]/{k_2}
=:\delta_{\rm 0}+\delta^{(2)}_a$, with $\delta^{(2)}_a=f_a/k_2$, 
and the second identity follows from Eq.\ \eqref{eq:10a},
using the force balance at the point of reversal 
$k_2(\delta_{\rm max}-\delta_0^a) = k_1 \delta_{\rm max}-f_a$.
\footnote{
From this point, we can derive the coefficient of restitution for the special case of
$k_c=0$ final energy, using the final energy $E_f^{(1)}(k_c=0) := (f_a^2 + 2 k_1 E_i)/(2 k_2)$.}

Further unloading, below $\delta^a_{\rm 0}$, leads to attractive forces. 
The kinetic energy at $\delta^a_{\rm 0}$ is partly converted to 
potential energy at point $D$, with overlap $\delta_{\rm min}$, where the 
minimal (maximally attractive) force is reached. 
Energy balance provides:
\begin{equation}\label{eq:10c}
\frac{1}{2}m_rv^2_{\rm min}
 = \frac{1}{2}m_rv^2_{0}-\frac{1}{2}k_2(\delta^a_{\rm 0}-\delta_{\rm min})^2 
 = \frac{1}{2}m_rv^2_{0}-  \frac{1}{2 k_2} (k_c \delta_{\rm min} + f_a)^2
 ~,
\end{equation}
where 
$\delta_{\rm min}=\frac{k_2 \delta_0^a-f_a}{k_2+k_c}
                           =\frac{(k_2-k_1)\delta_{\rm max}}{k_2+k_c}$, 
and the second identity follows from inserting
$\delta_0^a = (1+k_c/k_2) \delta_{\rm min} + f_a/k_2$.

The total energy is finally converted exclusively to kinetic energy at 
point $E$, the end of the collision cycle (with overlap $\delta=0$):
\begin{equation}\label{eq:10d}
 \frac{1}{2}m_rv^2_{f}=\frac{1}{2}m_rv^2_{\rm min}-\frac{1}{2}k_c\delta^2_{\rm min}-f_a\delta_{\rm min} ~.
\end{equation}
\end{subequations}

Using Eqs.\ \eqref{eq:10b}, \eqref{eq:10c}, and \eqref{eq:10d} with the definition of $\delta_{\rm min}$, 
and combining terms proportional to powers of $f_a$ and $\delta_{\rm max}$ yields the final kinetic energy 
after contact:
\begin{equation}\label{eq:cor1}
E_f^{(1)} :=
\frac{1}{2}m_rv^2_{f}=\left[ \frac{k_1}{k_2}-\frac{k_c}{k_1 k_2}\frac{(k_2-k_1)^2}{(k_2+k_c)} \right] 
                                    \frac{1}{2}k_1\delta^2_{\rm max} - f_a\delta_{\rm max} ~,
\end{equation}
with $\delta_{\rm max}$ as defined in Eq.\ \eqref{eq:Delmax}. 
Note that the quadratic terms proportional to $f_a^2$ have cancelled each other, 
and that the special cases of non-cohesive ($k_c=0$ and/or $f_a=0$) are 
simple to obtain from this
analytical form. Finally, dividing the final by the initial kinetic energy, Eq.\ \eqref{eq:10a},
we have expressed the coefficient of restitution 
\begin{equation}\label{eq:cor11}
 {e^{(1)}_n} = \sqrt{E_f^{(1)}/E_i} 
\end{equation}
as a function of maximal overlap reached, $\delta_{\rm max}$, non-contact adhesive force, $f_a$, 
elastic unloading stiffness, $k_2=k_2(\delta_{\rm max})$, and the constants plastic stiffness, $k_1$,
and cohesive ``stiffness'', $k_c$.

\subsubsection{Plastic-elastic contact with initial relative velocity $v_i \geqslant v_p$}
\label{sec:largevel}

When the initial relative velocity $v_i$ is large enough such that 
$v_i\geqslant v_p$, the estimated maximum overlap 
$\delta_{\rm max}$ as defined in Eq.\ \eqref{eq:Delmax} is greater than $ \delta^p_{\rm max}$.
Let $v_1$ be the velocity at overlap $\delta^p_{\rm max}$.
The system deforms along the path
$0 \rightarrow \delta^p_{\rm max} \rightarrow \delta_{\rm max} 
   \rightarrow \delta_{\rm 0}^a \rightarrow \delta_{\rm min} \rightarrow 0$.

The initial kinetic energy (at $\delta=0$ overlap, with adhesive force $f_a$ and with initial velocity $v_i\ge v_p$)  is not completely converted to potential energy at $\delta=\delta^p_{\rm max}$, where energy balance provides:
\begin{subequations}
\begin{equation}\label{eq:13a}
\frac{1}{2}m_rv^2_{1}
 = \frac{1}{2}m_rv^2_{i} - \frac{1}{2}m_rv^2_{p}
 = \frac{1}{2}m_rv^2_{i}
   - \frac{1}{2} \delta^p_{\rm max} ( k_1 \delta^p_{\rm max} - 2 f_a )
  ~,
\end{equation}
using the definition of $v_p$ in Eq.\ \eqref{eq:vp}.

From this point the loading continues along the elastic limit branch with slope 
$k_p$ until all kinetic energy is transferred to potential energy at overlap 
$\delta_{\rm max}>\delta^p_{\rm max}$, where the relative velocity changes 
sign, i.e., the contact starts to unload with slope $k_p$. Since there is no
energy disspated on the $k_p$-branch (in the absence of viscosity), 
the potential energy is completely converted to kinetic energy at the 
force-free overlap $\delta^{ap}_{0}$, on the plastic limit branch
\begin{equation}\label{eq:13b}
\frac{1}{2}m_rv^2_{0}
 = \frac{1}{2k_p} (k_1 \delta^p_{\rm max}-f_a)^2 + \frac{1}{2}m_rv^2_{1}~,
\end{equation}
with the first term taken from Eq.\ \eqref{eq:10b}, but replacing 
$\delta_{\rm max}$ with $\delta^p_{\rm max}$ and $k_2$ by $k_p$.

Further unloading, still with slope $k_p$, leads to attractive forces.
The kinetic energy at $\delta^{ap}_{\rm 0}$ is partly converted to potential energy 
at $\delta^p_{\rm min}$, where energy balance yields:
\begin{equation}\label{eq:13c}
\frac{1}{2}m_rv^2_{\rm min} 
 = \frac{1}{2}m_rv^2_{0} - \frac{1}{2}k_p(\delta^{ap}_{\rm 0}-\delta^p_{\rm min})^2
 = \frac{1}{2}m_rv^2_{0} - \frac{1}{2 k_p} (k_c \delta^{p}_{\rm min}-f_a)^2 ~.
\end{equation}

Some of the remaining potential energy is converted to kinetic energy so that 
at the end of collision cycle (with overlap $\delta=0$) one has
\begin{equation}\label{eq:13d}
 \frac{1}{2}m_rv^2_{f}=\frac{1}{2}m_rv^2_{\rm min}
      -\frac{1}{2}k_c \left( \delta^p_{\rm min}\right)^2 - f_a \delta^p_{\rm min} ~,
\end{equation}
analogously to Eq.\ \eqref{eq:10d}

When Eq.\ \eqref{eq:13d} is combined with Eqs.\ \eqref{eq:13b} and \eqref{eq:13c}, 
and inserting the definitions
$\delta^p_{\rm min}=\frac{k_p \delta_0^{ap}-f_a}{k_p+k_c}
                               =\frac{(k_p-k_1)\delta^p_{\rm max}}{k_p+k_c}$, 
and
$\delta_0^{ap} = (1+k_c/k_p) \delta^p_{\rm min} + f_a/k_p$,
one obtains (similar to the previous subsection):
%
%
%
\begin{equation}\label{eq:13f}
E_f^{(2)}
 =\frac{1}{2}m_rv^2_{f}
 =\frac{1}{2}m_rv^2_{i} 
    - \left[1 - \frac{k_1}{k_p} + \frac{k_c}{k_1 k_p} \frac{(k_p-k_1)^2}{(k_p+k_c)} \right]
     \frac{k_1}{2} \left({\delta^p_{\rm max}}\right)^2 
\end{equation}
\end{subequations}
Dividing the final by the initial kinetic energy, we obtain the coefficient of restitution 
\begin{equation}\label{eq:cor21}
 {e^{(2)}_n} = \sqrt{E_f^{(2)}/E_i} =: \sqrt{1-E_{\rm diss}/E_i} ~,
\end{equation}
with constants $k_1$, $k_p$, $k_c$, $f_a$, and $\delta^p_{\rm max}$.
Note that $e_n^{(2)}$, interestingly, does not depend on $f_a$ at all,
since the constant energy $E_{\rm diss}$ is lost exclusively in the hysteretic loop
(not affected by $f_a$).
Thus, even though $E_{\rm diss}$ does not depend on the impact velocity, 
the coefficient of restitution does, because of its definition.

As final note, when the elastic limit regime is not used, or modified towards
larger $\delta_{\rm max}^p$, as defined in appendix \ref{sec:mod}, the
limit velocity, $v_p$, increases, and the energy lost, $E_{\rm diss}$, 
increases as well (faster than linear), so that 
the coefficient of restitution just becomes $e^{(2)}_n=0$, 
due to complete loss of the initial kinetic energy, i.e., sticking,
for all $v \le v_p$.


%

 
\subsection{Combined coefficient of restitution}\label{sec:com_COR}

The results from previous subsections can now be combined 
in Eq.\ \eqref{eq:splitCOR} to compute the coefficient of restitution 
as a function of impact velocity for the irreversible elasto-plastic 
contact model presented in Fig.\ 4:
\begin{equation}
e = {\epsilon_{o}} e_n {\epsilon_{i}}=  \left \{
\begin{array}{lll}
{\epsilon_{o}}e^{(1)}_n{\epsilon_{i}}            & 
            & {\rm for~~}    v_i<v_p \\
{\epsilon_{o}}e^{(2)}_n{\epsilon_{i}}                            &
            & {\rm for~~}    v_i \ge v_p \\
\end{array}
\label{eq:fullCOR1}
\right . ~,
\end{equation}
 with $v_p$ from Eq.\ (\ref{eq:vp})
and $\epsilon_{o}=1$ or $<0$ for reversible and irreversible 
non-contact forces, respectively. Note that, without loss of generality, also
other shapes of non-contact, possibly long-range interactions can be used
here to compute $\epsilon_{o}$ and $\epsilon_{i}$, however, going into
these details goes beyond the scope of this paper, which only covers the most
simple, linear non-contact force.

\section{Dimensionless parameters}
\label{sec:dimless}

In order to define the dimensionless parameters of the problem, 
we first introduce the relevant energy scales, before we use their ratios 
further on:
\begin{subequations}
\begin{align}
  \mathrm{Intial\ kinetic\ energy}: 
     ~ & {E_i} = \displaystyle\frac{1}{2}m_rv^2_i ~,\\
\mathrm{Potential\ energy\ stored\ at\ \delta^{\it p}_{\rm max}}:  
     ~ & {E_p} = \displaystyle\frac{1}{2}k_1{\delta^p_{\rm max}}^2 ~,\\
\mathrm{Attractive\ non-contact\ potential\ energy}: 
     ~ & {E_a} = \displaystyle\frac{1}{2}\frac{f^2_a}{k_1} ~ ~.
\end{align}
\end{subequations}
%
%
The first two dimensionless parameters are simply given by ratios of
material parameters, while last two (independent) are scaled energies:
\begin{subequations}
\begin{align}
\label{eq:plas}
  \mathrm{Plasticity}: ~ & \eta = \displaystyle\frac{k_p-k_1}{k_1} ~,\\
\label{eq:coh}
 \mathrm{Plastic\ (contact)\ adhesivity}:
      ~ & \beta = \displaystyle\frac{k_c}{k_1} ~,\\
\label{eq:coh2}
 \mathrm{Non-contact\ adhesivity}:
     ~ & \alpha=\sqrt{\frac{E_a}{E_i}}=\sqrt{\frac{f^2_a}{k_1m_rv^2_{i}}} ~, \\
\label{eq:coh21}
 \mathrm{Dimensionless\ (inverse)\ impact\ velocity}:
    ~ & {\psi}=\sqrt{\frac{E_p}{E_i}}
                             =\frac{\delta^p_{\rm max}}{v_i}\sqrt{\frac{k_1}{m_r}} ~. 
\end{align}
\end{subequations}
from which one can derive the dependent abbreviations:
\begin{subequations}
\begin{align}
\label{eq:p_deg}
\mathrm{Scaled\ maximal\ deformation}: 
~ & \chi = \displaystyle\frac{\delta_{\rm max}}{\delta^p_{\rm max}} 
        = \frac{1}{\psi} \left( \sqrt{1+\alpha^2} + \alpha \right ) ~, \\
\label{eq:vp_dimless}
 \mathrm{Dimensionless\ impact\ velocity}:
   ~ & {\zeta}=\frac{v_i}{v_p}= \frac{1}{\sqrt{\psi^2-2\alpha\psi}}  ~.
\end{align}
\end{subequations}

Using Eqs.\ \eqref{eq:plas}, \eqref{eq:coh} and \eqref{eq:p_deg}, $k_2$ can be 
rewritten in non-dimensional form
\begin{equation}
\frac{k_2(\chi)}{k_1} = \left \{
\begin{array}{lll}
1 + \eta\chi, & {\rm ~if~}\chi < 1\\
1 + \eta, & {\rm ~if~} \chi \ge 1 
\end{array} 
\label{eq:k2_vel}
\right . ~,
\end{equation}
so that Eq.\ \eqref{eq:cor11} becomes:
\begin{equation}\label{eq:cor111}
  e^{(1)}_n = \sqrt{\left(\frac{1}{1+\eta\chi}-\frac{\beta\eta^2\chi^2}
                               {(1+\eta\chi)(1+\beta+\eta\chi)}\right)\psi^2\chi^2 
              - 2 \alpha \psi \chi } ~,
\end{equation}
and, similarly, Eq.\ \eqref{eq:cor21} in non-dimensional form reads:
\begin{equation}\label{eq:cor211}
  e^{(2)}_n 
   = \sqrt{1-\left(1-\frac{1}{1+\eta}
                +\frac{\beta\eta^2}{(1+\eta)(1+\beta+\eta)}\right)\psi^2
      }  ~,
\end{equation}
where the abbreviation $\psi \chi = \left( \sqrt{1+\alpha^2} + \alpha \right )$ was used.

To validate our analytical results, we confront our theoretical predictions with the 
results of two-particle DEM simulations in Fig.\ 5,
which shows $e$ plotted against the dimensionless impact velocity 
$\zeta$, for elasto-plastic adhesive spheres with different 
non-contact adhesion strength $f_a$ (and thus $\alpha$).
The lines are the analytical solutions for the coefficient of restitution, 
see Eqs.\ \eqref{eq:cor111} and \eqref{eq:cor211},
and the symbols are simulations, with perfect agreement, 
validating our theoretical predictions.
\begin{figure*}
\centering
\includegraphics[scale=0.5,angle=-90]{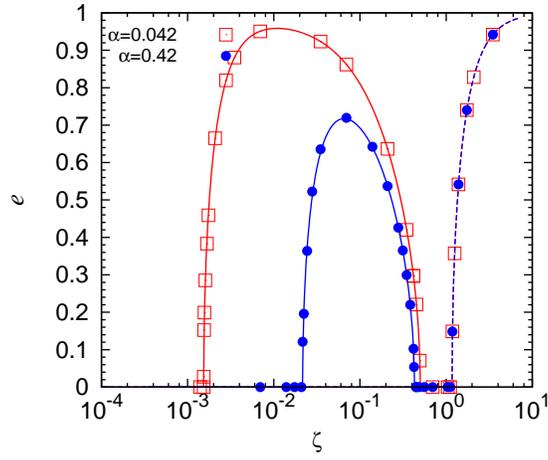}
 \caption{Restitution coefficient $e$ plotted as a function of the dimensionless
impact velocity $\zeta$, see Eq.\ (\ref{eq:vp_dimless}), for elasto-plastic 
adhesive spheres with the irreversible non-contact branch. 
Solid and dashed lines correspond to the analytical
expressions in Eqs.\ \eqref{eq:cor111} and \eqref{eq:cor211}, 
respectively, and the squares and circles are results of DEM simulations
for values of $f_a$ as given in the legend in terms of
$\alpha=0.042$ and $0.42$ (for impact velocity $v_i=0.01$\,ms$^{-1}$).
Simulation parameters used here are 
$k_1=10^2$\,Nm$^{-1}$, $k_p=5\times 10^2$\,Nm$^{-1}$, ($\eta=4$),
$k_c=10^2$\,$\mathrm{Nm^{-1}}$, ($\beta=1$), with $\phi_f=0.1$,
for particles with radius $1.1\times 10^{-3}$\,m, density $2000$\,kg/m$^{3}$, 
and thus mass $m_r = 5.6 \times 10^{-6}$\,kg.
}
\label{fig:coeff_vali}
\end{figure*}

For low velocity, the coefficient of restitution $e$ is zero, i.e.,
 the particles stick to each other. This behavior is in qualitative agreement 
 with previous experimental and numerical results 
 \cite{Wall90, Thornton98, Sorace09}.
With increasing impact velocity, $e$ begins to increase and then decreases 
again, displaying a second sticking regime (for the parameters used here). For 
even higher impact velocity, $v > v_p$ (and thus $\zeta>1$), another increase 
is observed, which will be explained in more detail in the next section.

Besides the onset of the plastic-limit regime at $\zeta=1$, we observe 
three further velocities $\zeta_c^{(a)}$, $\zeta_c^{(b)}$  and $\zeta_c^{(c)}$, 
for the end of the first sticking regime, i.e.\ $0 \le \zeta \le \zeta_c^{(a)}$, 
and the second sticking regime, i.e.\ $\zeta_c^{(b)} \le \zeta \le \zeta_c^{(c)}$.
While $\zeta_c^{(c)}$ is constant,
the critical velocity, $\zeta_c^{(a)}$, required to separate the particles
increases, whereas $\zeta_c^{(b)}$, required to enter the high-velocity sticking 
regime, decreases with the non-contact adhesion $f_a$. 


The further study of these critical velocities and the comparison to 
existing literature (theories and experiments) goes beyond the scope of the present study,
since there are just too many possibilities for materials and particle 
sizes.
We only refer to one example, where the end of the low-velocity sticking regime
was predicted as a non-linear function of the surface energy/adhesion
\cite{dominik1997physics} and hope that our proposal of using dimensonless
numbers will in future facilitate calibration of contact models with both
theories and experiments.

\section{Results using the mesoscopic contact model only ($f_a=0$)}
\label{sec:fa0}

Having understood the results for the contact model with finite non-contact 
force $f_a$, we will restrict our analytical study to the special case of 
negligible non-contact adhesive forces $f_a=0$, in the following,
which corresponds to the range of moderate to large impact velocity
or weak $f_a$, i.e.\ $\alpha \ll 1$.

For this special case, the dimensionless parameters reduce to 
$\alpha=0$, and $\chi=\frac{1}{\psi}=\zeta=\frac{v_i}{v_p}$.
 The expressions for the coefficients of restitution presented in 
 Eqs.\ \eqref{eq:cor111} and \eqref{eq:cor211} reduce to
\begin{equation}\label{eq:fine1}
e^{(1)}_n(\eta,\beta,\chi \le 1)
 = \sqrt{\frac{1}{1+\eta\chi}-\frac{\beta\eta^2\chi^2}{(1+\eta\chi)(1+\beta+\eta\chi)}} 
\end{equation}
and
\begin{equation}\label{eq:fine2}
e^{(2)}_n(\eta,\beta,\chi > 1)
 = \sqrt{1 - \left[ 1 - e^{(1)}_n(\eta,\beta,\chi=1)^2 \right]\frac{1}{\chi^2}} ~.
\end{equation}
 
\begin{figure*}
\centering
\includegraphics[scale=0.5,angle=-90]{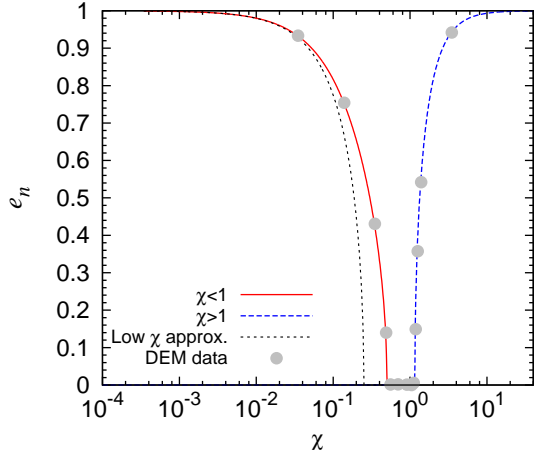}
 \caption{Restitution coefficient plotted as a function of 
the scaled initial velocity $\chi$ for a collision without viscous and 
non-contact forces ($\alpha=0$). 
The solid red line corresponds to the analytical expression in Eq.\ \eqref{eq:fine1},
the dashed blue line to Eq.\ \eqref{eq:fine2}, the thin black line represents the 
low velocity approximation, and the circles are DEM simulation data.
The material parameters are as in Fig.\ 5,
i.e.\ $\eta=4$ and $\beta=1$.
}
\label{fig:rest_vel}
\end{figure*}

\subsection{Qualitative Description}\label{sec:quandesc}
In Fig.\ 6, the analytical prediction for the
coefficient of restitution, from Eqs.\ \eqref{eq:fine1} 
and \eqref{eq:fine2}, is compared to the numerical integration
of the contact model, for different scaled initial velocities $\chi$.
We confirm the validity of the theoretical
prediction for the coefficient of restitution in the whole range.

For very small $\chi < 10^{-2}$, the approximation
$e^{(1)}_n \approx {1-\frac{\eta\chi}{2}}$ predicts the data very well.
With increasing initial relative velocity, dissipation increases 
non-linearly with the initial kinetic energy, leading to a 
convex decrease of $e^{(1)}_n$ (due to the log-scale plot). 
The coefficient of restitution $e^{(1)}_n$ becomes zero when 
a critical scaled initial velocity $\chi^{(b)}_{c}$ (see Eq.\ \eqref{eq:vel_1} below) 
is reached. At this point, the amount of dissipated energy becomes equal to the 
initial kinetic energy, leading to sticking of particles. 
The coefficient of restitution remains zero until a second critical 
scaled initial velocity $\chi^{(c)}_c$ is reached, i.e.\ sticking 
 is observed for $\chi^{(b)}_{c} \le \chi \le \chi^{(c)}_{c}$.
Finally, for $\chi>\chi^{(c)}_c$, the dissipated energy 
remains constant (the elastic limit branch is reached), while the initial kinetic energy 
increases. As a result, the kinetic energy after collision increases and so does the 
coefficient of restitution $e_n$.  Existence of sticking at such
 high velocities was recently reported by Kothe {\em et~al.}\ \cite{Kothe2013}, 
 who studied the outcome of collisions between sub-mm-sized dust agglomerates 
 in micro-gravity \cite{dominik1997physics}
\footnote{Note that $\xi \gg 1$ is the regime where the physics of the 
contact changes, dependent on the material and other considerations; 
modifications to the contact model could/should then be applied, however, 
this goes beyond the scope of this paper, where we use the elastic limit branch
or the generalized fully plastic model without it. Beyond the limits of the model, 
at such large deformations, the particles cannot be assumed to be spherical 
anymore and neither are contacts isolated from each other.
}.
Note that an increase of $e_n$ for high velocity is a familiar observation
 in studies focused on the cold-spray technique \cite{Van99,Zhang05,Schmidt06}.
 Above a certain (critical) velocity the spray particles adhere to the substrate, and 
 they do so for a range of impact velocities; increasing the impact velocity further leads 
 to unsuccessful deposition, i.e.\ the particles will bounce from the substrate. 
 The sticking and non-sticking phenomenology of such materials has been 
 extensively studied experimentally and numerically in 
 Refs.\ \cite{Van99,Zhang05,Schmidt06,Wu06,Wu11}. 

In Fig.\ 7, we compare the variation of the force 
with overlap in the various regimes of $\chi$ as discussed above,
but here for $\phi_f=0.05$.
For very small $\chi$, the unloading slope $k_2 \approx k_1$, 
(see Fig.\ 7(a) for a moderate $\chi=0.34$), 
and the amount of dissipated energy is small, increasing with $\chi$. 
The kinetic energy after collision is almost equal to the 
initial kinetic energy, i.e.\ $e_n \sim 1$, see Fig.\ 6.
In Figs.\ 7(b) and 7(c), the 
force-overlap variation is shown for sticking particles, 
for the cases $\chi^{(b)}_c<\chi<1$ and $1<\chi<\chi^{(c)}_c$, 
respectively (more details will be given in the following subsection). 
Finally, in Fig.\ 7(d), the case $\chi>\chi^{(c)}_c$ 
is displayed, for which the initial kinetic energy is larger than the 
dissipation, resulting in separation of the particles. 
The corresponding energy variation is described in detail in 
appendix \ref{sec:viseffect}.


 
\begin{figure*}
  \centering
    \mbox{
\subfigure[]{\includegraphics[scale=0.45,angle=-90]{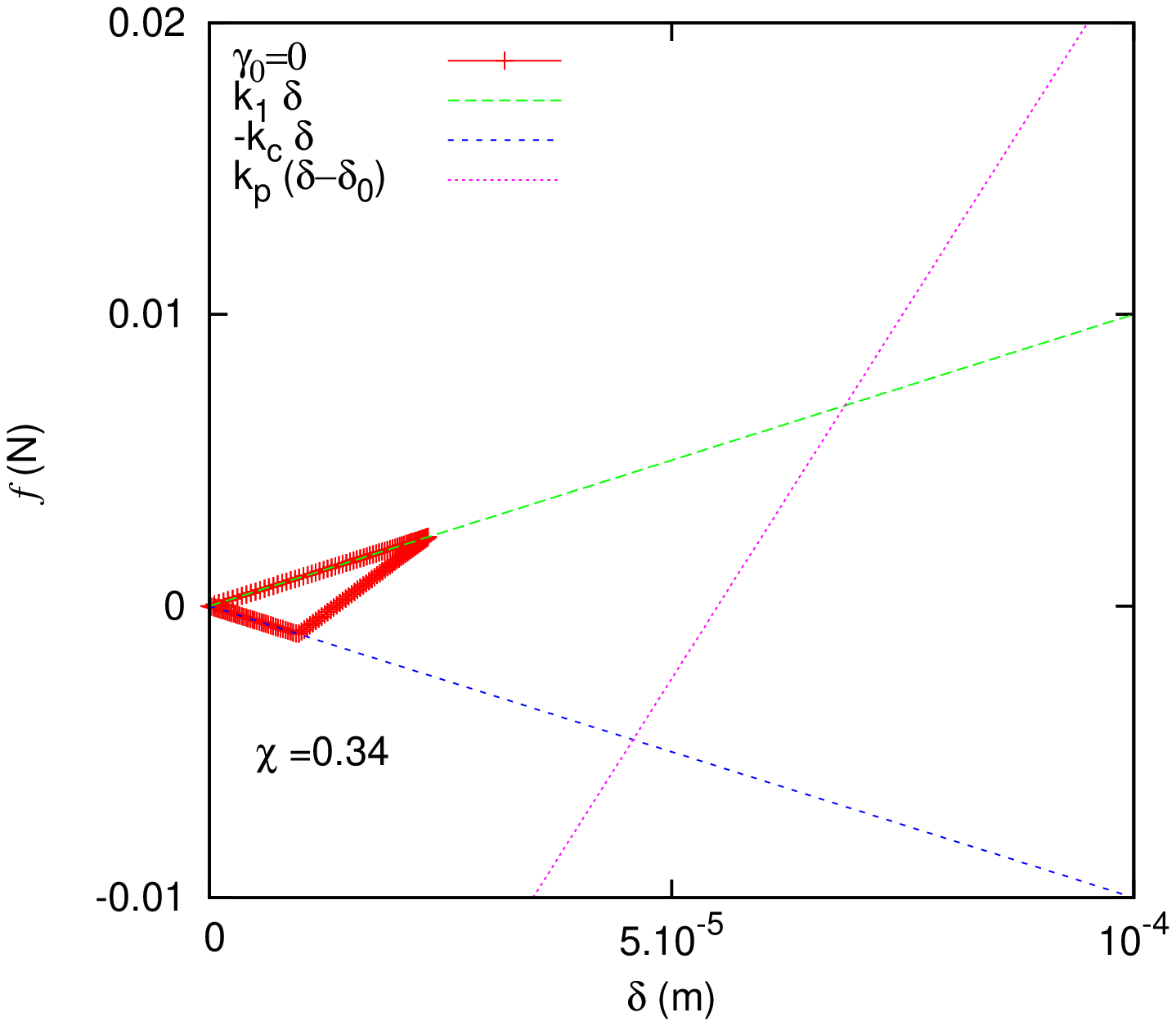}\label{fig:low_vel_force}}\quad
\subfigure[]{\includegraphics[scale=0.45,angle=-90]{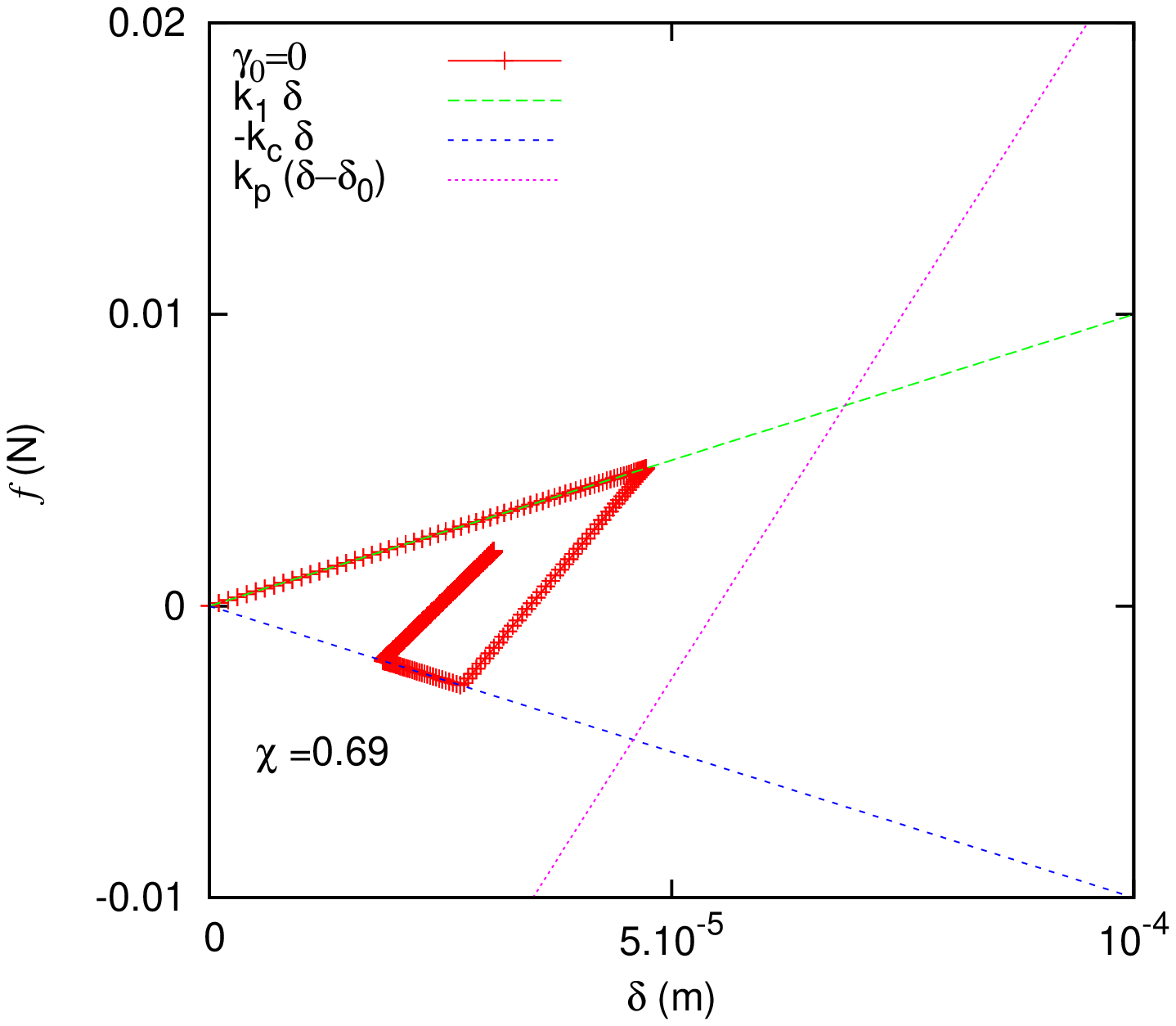}\label{fig:low_vel_stick}}
}
\mbox{
\subfigure[]{\includegraphics[scale=0.45,angle=-90]{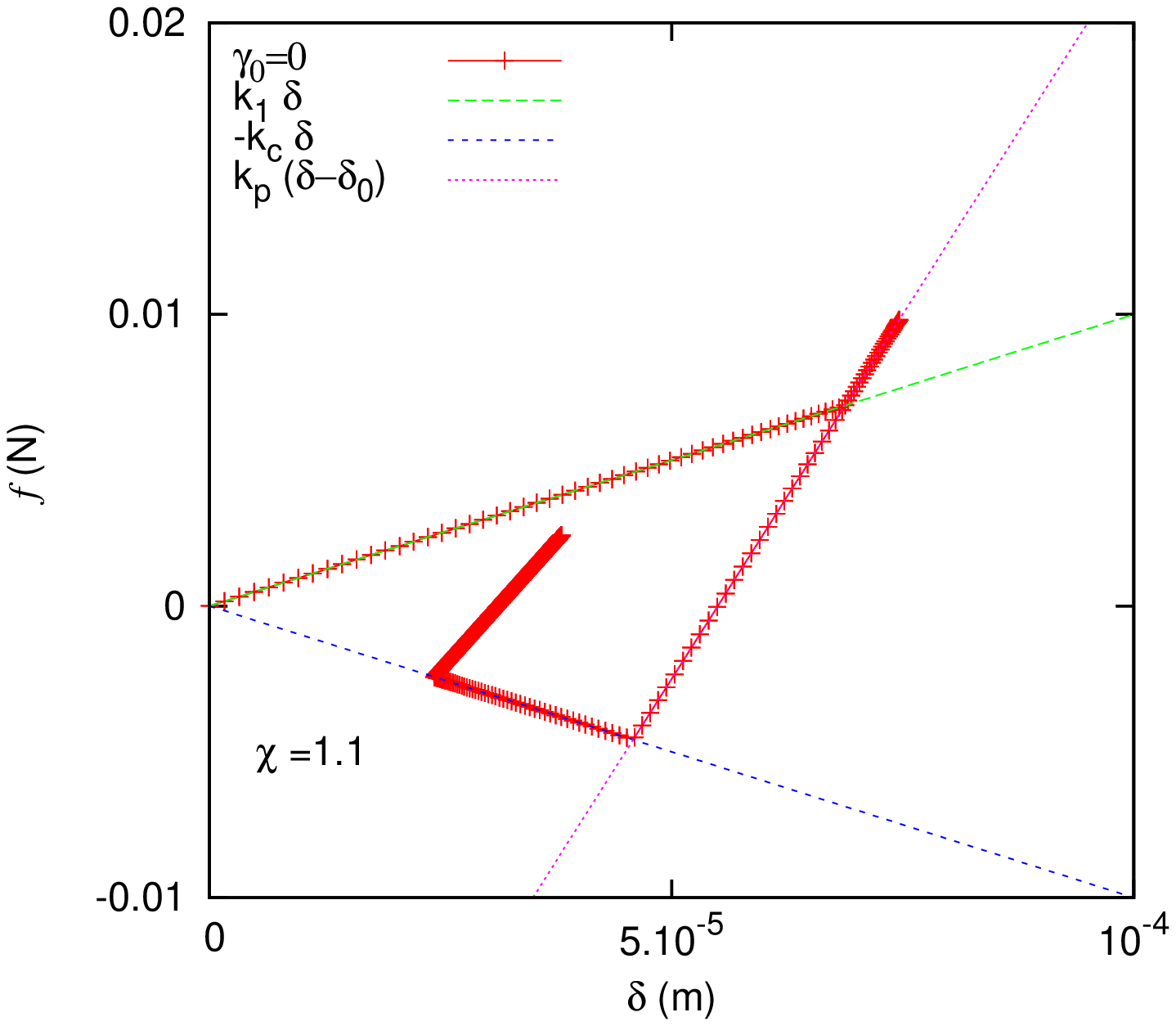}\label{fig:large_vel_stick}}\quad
\subfigure[]{\includegraphics[scale=0.45,angle=-90]{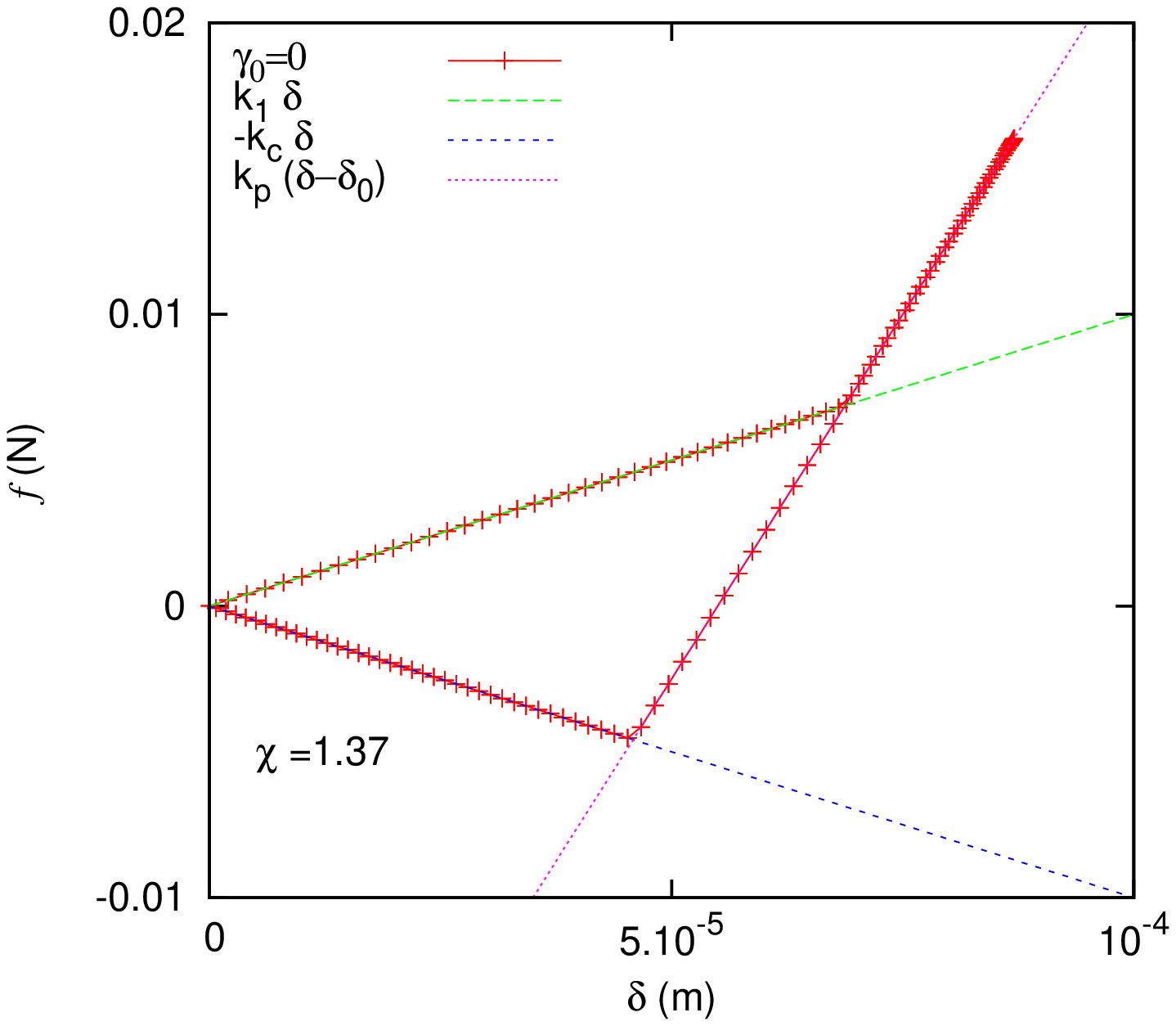}\label{fig:large_vel_force}}
}
\caption{Contact force during one collision, plotted against the 
overlap for different scaled initial velocities $\chi=0.34$, 0.69, 1.1, and 1.37, 
respectively. The three straight lines represent the plastic branch, 
with slope $k_1$, the adhesive branch, with slope $-k_c$, and the 
limit branch with slope $k_p$, for $k_1=10^2$\,Nm$^{-1}$, 
$k_p=5\times 10^2$\,Nm$^{-1}$, $k_c=10^2$ $\mathrm{\,Nm^{-1}}$,
i.e.\ $\eta=4$ and $\beta=1$, and $\phi_f=0.05$.}
\label{fig:vel_f1}
\end{figure*}

\subsection{Sticking regime limits and overlaps}
\label{sec:sticking}

In this section we focus on the range of 
$\chi^{(b)}_c<\chi\le \chi^{(c)}_c$, where the particles stick 
to each other (implying that $\beta$ is large enough $\beta\ge\beta^*$
with minimal $\beta^*$ for sticking) 
and calculation of the critical values $\chi^{(b)}_c$ and $\chi^{(c)}_c$. 
When $\chi = \chi^{(b)}_c$ all of the initial kinetic energy of the 
particles is just dissipated during the collision. 
Hence the particles stick and 
$ e^{(1)}_n(\eta,\beta, \chi^{(b)}_c)=0 $,
which leads to
$ 1+\beta+\eta\chi-\beta\eta^2\chi^2=0 $.
Only the positive solution is physically possible, 
as particles with negative initial relative velocity cannot collide, 
so that
\begin{equation}\label{eq:vel_1}
 \chi^{(b)}_c=\frac{1}{2\beta\eta}\left[1+\sqrt{1+4\beta(1+\beta)}\right] ~
\end{equation}
is the lower limit of the sticking regime.
For larger $\chi>\chi^{(b)}_c$, the dissipation is strong enough to
consume all the initial kinetic energy, hence the particles loose their
kinetic energy at a positive, finite overlap $\delta_{\rm c}$, see
Fig.\ 7(b).
The contact deforms along the path 
$0 \rightarrow \delta_{\rm max} \rightarrow \delta_{\rm 0} 
   \rightarrow \delta_{\rm min} \rightarrow \delta_{\rm c}$.
Thereafter, in the absence of other sources of dissipation, particles keep 
oscillating along the same slope $k_2$. In order to compute $\delta_{\rm c}$, 
we use the energy balance relations in Eqs.\ \eqref{eq:10}, and
conservation of energy along $\delta_{\rm min} \rightarrow \delta_{\rm c}$, 
similar to Eq.\ \eqref{eq:10d},
\begin{subequations}\label{eq:2.4.3}
\begin{equation}\label{2.4.3c}
 \frac{1}{2}m_rv^2_{\rm min} 
   - \frac{1}{2}k_c \left \{ \delta_{\rm min}^2
                                   -  \delta^2_{\rm c} \right \} = 0 ~,
\end{equation}
with vanishing velocity $v_c=0$ at overlap $\delta_c$.
Using the definitions around Eqs.\ \eqref{eq:10} and
re-writing in terms of $k_c$ and $\delta_{\rm max}$ leads to
\begin{equation}
k_c\delta^2_{\rm c}+
 \left \{\frac{k^2_1}{k_2}-\frac{k_c(k_2-k_1)^2}{k_2(k_2+k_c)}\right \} \delta^2_{\rm max}=0
\end{equation}
and thus to the sticking overlap in regime (1), for $\chi^{(b)}_c v_p < v_i <v_p$:
\begin{equation}
\frac{\delta^{(1)}_{\rm c}}{\delta^p_{\rm max}} 
  =  \frac{\delta_{\rm max}}{\delta^p_{\rm max}}
           \sqrt{\frac{(k_2-k_1)^2}{k_2 (k_2+k_c)}-\frac{k^2_1}{k_2k_c}} ~.
\end{equation}
\end{subequations}
In terms of dimensionless parameters, as defined earlier, this can be written as:
\begin{equation}\label{delc_fine1}
\frac{\delta^{(1)}_{\rm c}}{\delta^p_{\rm max}} 
 = \chi \sqrt{\frac{\eta^2\chi^2}{(1+\eta\chi)(1+\beta+\eta\chi)}-\frac{1}{\beta(1+\eta\chi)}} 
 = \frac{\chi}{\sqrt{\beta}} \hat{e}_n^{(1)} ~,
\end{equation}
where $\hat{e}_n^{(1)}$ denotes the result from Eq.\ \eqref{eq:fine1}
with positive argument under the root.

\begin{figure*}
\centering
\includegraphics[scale=0.5]{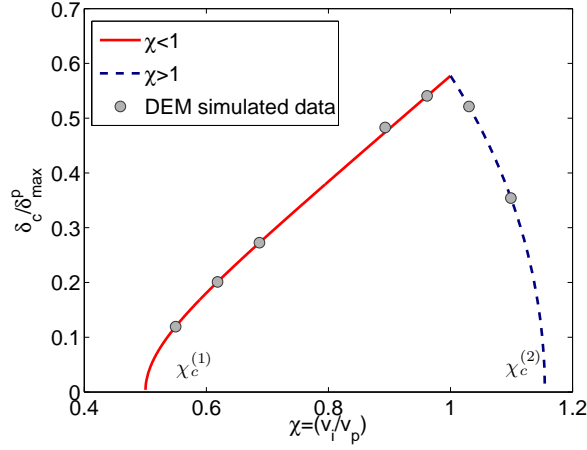}
\caption{Kinetic energy-free contact overlap $\delta_{\rm_{c}}$ 
plotted as a function of the scaled initial velocity 
$\chi={v_i}/{v_p}$; the increasing branch corresponds to $\chi<1$, 
while the decreasing branch corresponds to $\chi>1$. The dots are simulations 
for $\eta=4$ and $\beta=1$, as in Fig.\ 7, which yields
$\delta^{\rm max}_{\rm_c}/{\delta^p_{\rm max}}=(1/3)^{1/2}$
in Eq.\ \eqref{eq:deltacmax}.}
\label{fig:sticking}
\end{figure*}

For larger initial relative velocities, $\chi\ge1$, the coefficient of 
restitution is given by Eq.\ \eqref{eq:fine2}, so that
the upper limit of the sticking regime $\chi^{(c)}_c>1$ 
can be computed by setting $ e^{(2)}_n(\eta,\beta,\chi^{(c)}_c)=0 $.
Again, only the positive solution is physically meaningful, so that
\begin{equation}\label{eq:vel_2}
\chi^{(c)}_c=\sqrt{1-\frac{1}{1+\eta}+\frac{\beta\eta^2}{(1+\eta)(1+\beta+\eta)}}
\end{equation}
is the maximum value of $\chi$ for which particles stick to each other. 
For $\chi\leqslant \chi^{(c)}_{c}$ particles deform along the path 
$0 \rightarrow \delta^p_{\rm max}  \rightarrow \delta_{\rm max}
   \rightarrow \delta_{\rm 0} \rightarrow \delta_{\rm min} \rightarrow \delta_{\rm c}$
and then keep oscillating on the branch with stiffness $k_2$, with $\delta_{\rm c}$ 
being one of the extrema of the oscillation, see Fig.\ 7(c).
Similar to the considerations above, we compute
the sticking overlap in regime (2), for $v_p <v_i < \chi^{(c)}_c v_p$:
%
in dimensionless parameters:
\begin{equation}\label{delc_fine2}
 \frac{\delta^{(2)}_{\rm c}}{\delta^{p}_{\rm max}}
  = \sqrt{\frac{\eta^2}{(1+\eta)(1+\beta+\eta)}+\frac{\eta}{\beta(1+\eta)}-\frac{\chi^2}{\beta}}
  = \frac{\chi}{\sqrt{\beta}} \hat{e}_n^{(2)} ~,
\end{equation}
where $\hat{e}_n^{(2)}$ 
denotes the result from Eq.\ \eqref{eq:fine2}
with positive argument under the root.

In Fig.\ 8, the scaled sticking overlap 
${\delta_{\rm_c}}/{\delta^p_{\rm max}}$ 
is plotted for different $\chi$, showing perfect agreement of the analytical 
expressions in Eqs.\ \eqref{delc_fine1} and \eqref{delc_fine2},
with the numerical solution for a pair-collision.
In the sticking regime, the stopping overlap increases with $\chi$, 
and reaches a maximum at $\chi=1$, 
\begin{equation}
\frac{\delta_{\rm_c}({\chi=1})}{\delta^p_{\rm max}}
 = \sqrt{\frac{\beta \eta^2 - (1+\eta+\beta)}
        {\beta (1+\eta)(1+\eta+\beta)}}
\label{eq:deltacmax}
\end{equation}
which depends on the the adhesivity $\beta$ and the plasticity $\eta$
only. For $\chi>1$, dissipation gets weaker, relatively to the increasing 
initial kinetic energy,
and $\delta^{(2)}_{\rm_c}/{\delta^p_{\rm max}}$ decreases
until it reaches $0$ for $\chi=\chi^{(c)}_c$.

\subsection{Contacts for different adhesivity $\beta$}\label{sec:betadep}

In the previous subsections, we studied the dependence of the 
coefficient of restitution $e_n$ on the scaled initial velocity 
$\chi$ for fixed adhesivity $\beta$, whereas here the 
dependence of $e_n$ on $\beta$ is analyzed.

A special adhesivity $\beta^*$ can be calculated such that $e_n=0$ 
for $\chi=1$, which is the case of maximum dissipation, and leads
to sticking only at exactly $\chi=1$, i.e.\ there is no sticking for $\beta<\beta^*$.
Using Eq.\ \eqref{eq:fine1}, we get
\begin{subequations}
\begin{equation}
1+\beta^*+\eta-\beta^*\eta^2=0 ~,
\end{equation}
so that
\begin{equation}\label{eq:critical_kc}
 \beta^*=\frac{1}{\eta-1} ~.
\end{equation}
\end{subequations}
In Fig.\ 9, we plot the coefficient of restitution $e$ as
a function of the scaled initial velocity $\chi$ for different values of 
adhesivity $\beta$. For $\beta<\beta^*$, in Fig.\ 9,  
the coefficient of restitution $e_n$ decreases with increasing $\chi<1$, 
reaches its positive minimum at $\chi=1$, and increases for $\chi>1$.
In this range, the particles (after collision) always have a non-zero 
relative separation velocity $v_f$.
When $\beta=\beta^*$, $e_n$ follows a similar trend, becomes zero 
at $\chi=1$, and increases with increasing scaled initial velocity for $\chi>1$. 
This is the minimum value of adhesivity for which $e_n$ can become zero
and particles start to stick to each other.
For $\beta=\beta^*$, the sticking regime upper and lower limits coincide, 
$\chi^{(b)}_c=\chi^{(c)}_c=1$.  
If $\beta>\beta^*$, $e_n$ decreases and becomes zero at 
$\chi=\chi^{(b)}_c<1$, it remains zero until $\chi=\chi^{(c)}_c>1$,
and increases with increasing relative initial velocity thereafter.
Hence, the range of velocities for which sticking happens 
is determined by the material properties of the particles. 
Indeed Zhou {\em et~al.}\ \cite{Wu11} presented similar 
conclusions about the deposition efficiency in cold spray.
Simulations with viscous forces change the value of $\beta^*$ 
and are not shown here, see Appendix\ \ref{sec:viseffect}.
\begin{figure*}
\centering
\includegraphics[scale=0.5]{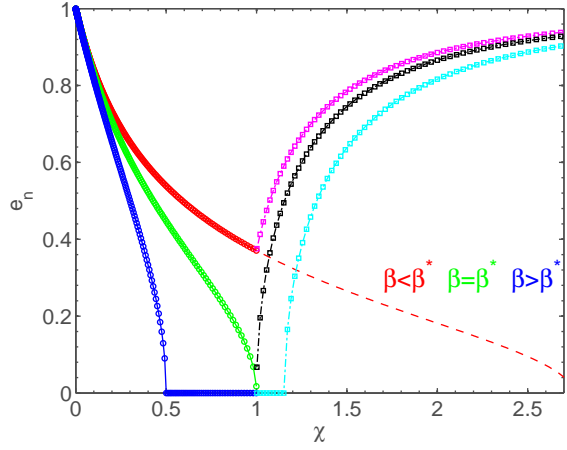}
\caption{Coefficient of restitution $e_n$ plotted against the scaled initial velocity 
$\chi$. Circles with different colors correspond to different adhesivity $\beta$
(red for $\beta<\beta^*$, green for $\beta=\beta^*$ and blue for $\beta>\beta^*$) for $\chi \le 1$ , while magenta, black
 and cyan squares correspond to the respective values of $\beta$ for $\chi>1$.
Other parameters used are $k_1=10^2$, $k_p=5\times 10^2$, and different $k_c$
(all in units of $\mathrm{Nm^{-1}}$), i.e.\ $\eta=4$ and $\beta/\beta^*=1/3$, 
$1$, and $3$, with $\beta^*=1/3$. The dashed red line represents the solution with the tuned
 fully plastic model with a new ${\phi_f}^{'}=0.5$ and newly calculated ${k_p}^{'}$, see Appendix\ \ref{sec:mod} .
}
\label{fig:res_kc}
\end{figure*}

\section{Conclusions and outlook}\label{sec:con}

Various classes of contact models for 
non-linear elastic, adhesive and visco-elasto-plastic particles are reviewed.
Instead of focusing on the well understood models for perfect spheres of 
homogeneous (visco) elastic or elasto-plastic materials, 
a special mesoscopic adhesive (visco) elasto-plastic contact model is considered, 
aimed at describing the macroscale behavior of assemblies of realistic 
fine particles (different from perfectly homogeneous spheres).
An analytical solution for the coefficient of restitution of pair-contacts
is given as reference, for validation, and to understand the role of 
the contact model parameters.

\paragraph{Mesoscopic Contact Model}
The contact model by Luding \cite{Luding08},
including short-ranged (non-contact) interactions, is critically discussed and
 compared to alternative approaches in subsection \ref{sec:Model_classification}.  
 The model introduced in section \ref{sec:mdsoft}
 is simple (piece-wise linear), yet it catches the important
 features of particle interactions that affect the bulk behavior of 
 a granular assembly, i.e.\ non-linear elasticity, plasticity and contact adhesion.
It is mesoscopic in spirit, i.e.\ it does not resolve
 all the details of every single contact, but is designed 
 to represent an ensemble of particles with many contacts in a bulk system.
One goal of this study is to present this rich, flexible and multi-purpose granular 
matter meso-model, which can be calibrated to realistically model ensembles of 
large numbers of particles \cite{imole_dosing_2015}.
The analytical solution for the contact dissipation is given for contact
and non-contact forces both active, but viscosity inactive, in section \ref{sec:split_COR}.
A sensible set of dimensionless parameters is defined in section \ref{sec:dimless},
before the influence of the model parameters on the overall impact behavior is 
discussed in detail, focusing on the irreversible, adhesive, elasto-plastic part 
of the model, in section \ref{sec:fa0}.

\paragraph{Analysis of the coefficient of restitution}
When the dependence of the coefficient of restitution, $e$, 
on the relative velocity between particles is analyzed, 
two sticking regimes, $e=0$, show up, as related to different sources 
of dissipation:

(i) As previously reported in the literature 
 (see e.g.\ Refs.\ \cite{Dahneke73,Brilliantov07,Thornton98,Sorace09,Jasevicius11})
 the particles stick to each other at very low impact velocity. 
 This can happen due to irreversible short-range non-contact interaction, as
 e.g.\ liquid bridges, or due to van der Waals type force for dry adhesive particles.
 The threshold velocity, below which the particles stick depends on the magnitude of 
 the non-contact adhesive force $f_a$, while for elasto-plastic adhesive particles on both 
 non-contact adhesive force and plasticity, which together control this low-velocity sticking.
 
(ii) With increasing velocity, $e$ increases and then decreases until the second sticking regime is reached, which is strongly influenced by the plastic/adhesive (and viscous) dissipation mechanisms in the hysteretic contact meso-model.
 At small impact velocity, all details of the model are of importance, while at
 higher velocities, for a sufficiently low value of the jump-in force $f_a$, 
 the contribution of the non-contact forces can be neglected.
The theoretical results are derived in a closed analytical form, and
phrased completely in terms of dimensionless
parameters (plasticity, adhesivity and initial velocity).
The ranges of impact velocities for the second sticking regime are predicted
and discussed in detail. 

(iii) For still increasing relative velocity, beyond the second sticking region, 
$e$ starts increasing again. This regime involves a change of the physical 
behavior of the system as expected, e.g.\ for non-homogeneous materials 
with micro-structure and non-flat contacts, or materials with an elastic core, 
e.g.\ asphalt (stone with bitumen layer). 
Even though this elastic limit behavior is a feature of the model,
 completely plastic behavior can be reproduced by the model too, 
 just by tuning two input parameters $k_p$  and $\phi_f$,
 as shown in appendix \ref{sec:mod}.
 This way, the low velocity collision dynamics is kept unaffected,
 but the elastic limit regime is reached only at higher impact velocities,
 or can be completely removed.
 This modification provides the high velocity sticking regime for all high velocities, 
 as expected for ideally plastic materials.
 On the other hand, the existence of a high velocity rebound, 
 as predicted the model with elastic limit regime, 
 has been observed experimentally and numerically in cold spray
 \cite{Van99, Zhang05,Schmidt06, Wu06,Wu11} and can be expected for 
 elastic core with a thin plastic shell.

\paragraph{Additional dissipative mechanisms}
For sticking situations, on the un-/re-loading branch, the particles
 oscillate around their equilibrium position until their kinetic energy is dissipated, 
 since realistic contacts are dissipative in nature.
 Since viscosity hinders analytical solutions, it was not considered before, but
 a few simulation results with viscosity are presented in Appendix \ref{sec:viseffect}. 
 With viscosity, both un- and re-loading are not elastic anymore, resembling a 
 damped oscillation and eventually leading to a static contact at finite overlap.


\paragraph{Application to multi-particle situations}
The application of the present meso-model to many-particle systems 
(bulk behavior) is the final long-term goal, 
see e.g.\ Ref.\ \cite{Luding11,Singh2014effect,Rojas2015},
as examples, where the non-contact forces were disregarded.
An interesting question that remains unanswered concerns a suitable analogy 
to the coefficient of restitution (as defined for pair collisions)
relevant in the case of bulk systems, where particles can be permanently
in contact with each other over long periods of time, and where 
impacts are not the dominant mode of interaction, but rather long lasting
contacts with slow loading-unloading cycles prevail.

One specific example for the latter situation of slow loading-unloading of 
bulk material is given in Appendix\ \ref{sec:agg}, showing qualitatively 
similar behavior as encompassed in the contact meso-model, but on a
much larger length-scale than the contact model itself, highlighting the 
dominant role of the particle structure and the (non-flat) contact area
with related plastic (irreversible) re-arrangements \cite{dominik1997physics}.

\paragraph{Outlook}
The interest of widely different communities (viz.\ particle technology, 
 granular physics, interstellar dust, asphalt, or cold-spray) in the dependence 
 of the coefficient of restitution (or deposition/impact behavior) on the impact 
 velocity is considerable.
 We hope our study helps to connect these widely different communities 
 by providing an overview and, in particular, a flexible, multi-purpose contact model, 
 valid and useful for many practically relevant situations. 
 
 The contact meso-model has to be calibrated for different materials, while
 our reference analytical results allow to verify the model implementation. 
 With this, the model can be used to predict bulk material behavior and to 
 be validated by comparison with experiments.

\section{Acknowledgment}\label{sec:acknow}
Insightful discussions with 
L. Brendel, 
M. Ghadiri, 
N. Kumar, 
S. Li, 
M. Pasha, 
H. Tanaka, 
S. Thakur 
C. Thornton, 
J. Tomas, 
O. Walton,
and T. Weinhart 
are highly appreciated.

Financial support (project number: 07CJR06) from the research program 
``Jamming and Rheology'' of the Stichting voor Fundamenteel Onderzoek der Materie (FOM), 
and of the STW-VICI project (number 10828) 
``Bridging the gap between Discrete and Continuous systems'',
both financially supported by the ``Nederlandse Organisatie voor Wetenschappelijk Onderzoek'' 
(NWO), is acknowledged.

\appendix
\numberwithin{equation}{section}

\section{Tuning of parameters to increase the plastic range}
\label{sec:mod}

We assume that the reference dimensionless plasticity depth be $\phi_f$,
 which is, e.g.\ calculated based on the maximal volume fraction related arguments 
 of a multi-particle assembly, and $k_p$ be the reference limit stiffness.
 We propose a new ${\phi_f}^{'}>{\phi_f}$, which represents the new (larger)
 dimensionless plasticity depth (arbitrary choice or calculated based on another 
 volume fraction consideration) and a new value of ${k_p}^{'}$; the choice is such
 that the tuned model resembles exactly, consistently, the reference for 
 $\delta_0< a_{12}\phi_f$, with reduced radius $a_{12}$,  and becomes plastic for 
 $a_{12}\phi_f<\delta_0<a_{12}{\phi_f}^{'}$.
At $\delta_0=a_{12}\phi_f$, Eq.\ \eqref{eq:kpdelp} reads
\begin{equation}
 k_p
  =k_1+({k_p}^{'}-k_1) { \delta^{p}_{\rm max} } / {{ \delta^{p}_{\rm max} }}^{'},
\label{eq:kpnew}
\end{equation}
since all parameters except $\phi_f$ and $k_p$ remain unchanged. 
Using Eqs.\ \eqref{eq:kpnew} and \eqref{eq:kpdel0} 
we arrive at
\begin{equation}
 \frac{(k_p-k_1)^2}{k_p \phi_f} = \frac{({k_p}'-k_1)^2}{{k_p}' {\phi_f}'},
\end{equation}
 which gives the new limit stiffness
\begin{equation}\label{eq:final_kp1}
 {k_p}^{'}=k_1+{AB}/{2}+\sqrt{(AB/2)^2+k_1AB},
\end{equation}
where $A={(k_p-k_1)^2}/{k_p}$ and $B={{\phi_f}^{'}}/{\phi_f}$.

Using Eq.\ \eqref{eq:final_kp1}, we can calculate values of the new limit 
 plastic stiffness ${k_p}^{'}$ for any given ${\phi_f}^{'}$, such that the 
 collision dynamics for lower plastic deformation $\delta_0<{\delta^p}_{0}$ 
 is intact, while the range of plastic deformation is enhanced, depending on 
 the chosen ${\phi_f}^{'}>{\phi_f}$.

\begin{figure*}
\centering
\includegraphics[scale=0.5]{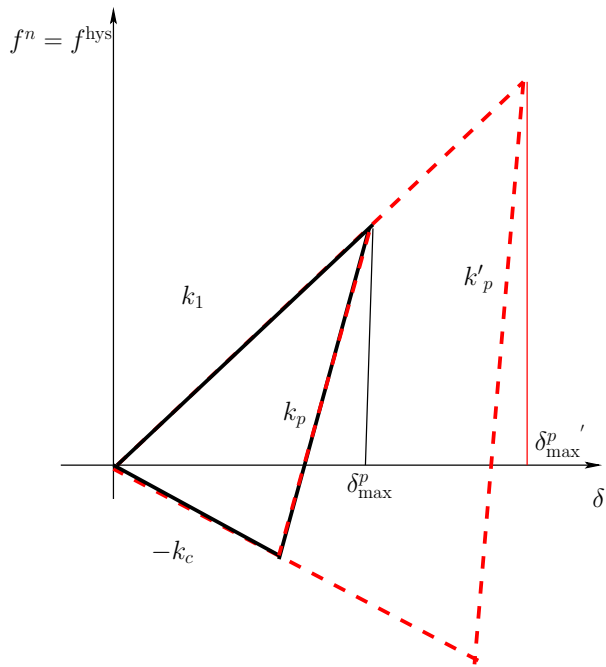}
\centering
\caption{Force-displacement law for elasto-plastic, adhesive contacts 
superimposed on the irreversible contact force law. The black solid line 
 represents the force law for reference input parameters $\phi_f$ and $k_p$, 
 while the dashed red line represents the same for a new chosen ${\phi_f}^{'}$ 
 and newly calculated ${k_p}^{'}$ resembling a wider plastic regime 
 of the particle deformation.}
\label{fig:force_adh_mod}
\end{figure*}

\section{Effect of Viscosity}\label{sec:viseffect}

Since real physical systems also can have additional dissipation modes 
that are, e.g.\ viscous in nature, we study the behavior of collisions with 
viscous damping present ($\gamma_0>0$) and compare 
it with the non-viscous case ($\gamma_0=0$). Note that any non-linear
viscous damping force can be added to the contact laws introduced 
previously, however, for the sake of simplicity we restrict ourselves
to the simplest linear viscous law as given as second term in 
Eq.\ \eqref{eq:fn_LSD}, since it can be important to choose
 the correct viscous damping term for each force law to get the physically
 correct behavior \cite{Kuwabara87,luding95,walton93,luding98c}.
In Fig.\ 11, we plot the contact force against the overlap,
and the overlap against time, during collisions for a constant value of 
$\chi=1$ and different $\beta$, for $\gamma_0=5 \times 10^{-3}$\,kg\,s$^{-1}$.

When $\beta<\beta^*$, see Fig.\ 11(a) and 
Fig.\ 11(b), the contact ends when the adhesive force 
$-k_c \delta$ goes back to zero, for both cases, with and without viscosity.
This is since the viscosity is relatively small and does not contribute
enough to the total dissipation to make the particles stick for the 
parameters used here.

For the critical adhesivity $\beta=\beta^*$, reported in 
Fig.\ 11(c), without viscosity, the overlap between the 
particles goes down to exactly zero at the end of the collision,
with all kinetic energy dissipated.
For $\gamma_0>0$, dissipation brings this marginal case into
the sticking regime and the particles stay in contact at $\delta>0$. 
This can be seen clearly in Fig.\ 11(d), where 
the particles undergo a damped oscillatory motion due to
the small residual velocity created on the re-loading branch.

For larger values $\beta>\beta^*$, the overlap at which kinetic energy is
lost completely (on the $k_c$ branch) is finite, 
for both $\gamma_0=0$ and $\gamma_0>0$, 
see Fig.\ 11(e).
In both cases, the particles stick and remain in contact.
Without viscosity, the particles keep oscillating along the slope $k_2$, 
while with viscosity the oscillation is damped and kinetic energy
vanishes. During loading and unloading the apparent slope changes with time 
due to the additional viscous force that leads to the dissipation of energy, 
as evident from the ellipsoidal converging spiral.
Waiting long enough, for some oscillation cycles, the particles stick to 
each other with a finite overlap and zero relative kinetic energy.
The difference is also visible in Fig.\ 11(f), where
for $\gamma_0=0$ the particles keep oscillating with constant amplitude, 
whereas, for $\gamma_0>0$, the particles undergo a damped oscillatory motion, 
until the velocity becomes $0$ at $\delta>0$. 
The time evolution of the overlap in 
Fig.\ 11(f) resembles that of the displacement 
evolution in Ref.\ \cite{Hatzes1991113}, where the authors studied 
sticking of particles in Saturn's rings.
\footnote{
In general, one could add a viscous law that is proportional to
$k_2-k_1$ or to a power of overlap $\delta$, such that the jump-in viscous force 
in Fig.\ 11(e) at the beginning 
of the contact is not there, however, we do not go into this detail.} 
\begin{figure*}
  \centering
    \mbox{
\subfigure[]{\includegraphics[scale=0.5,angle=-90]{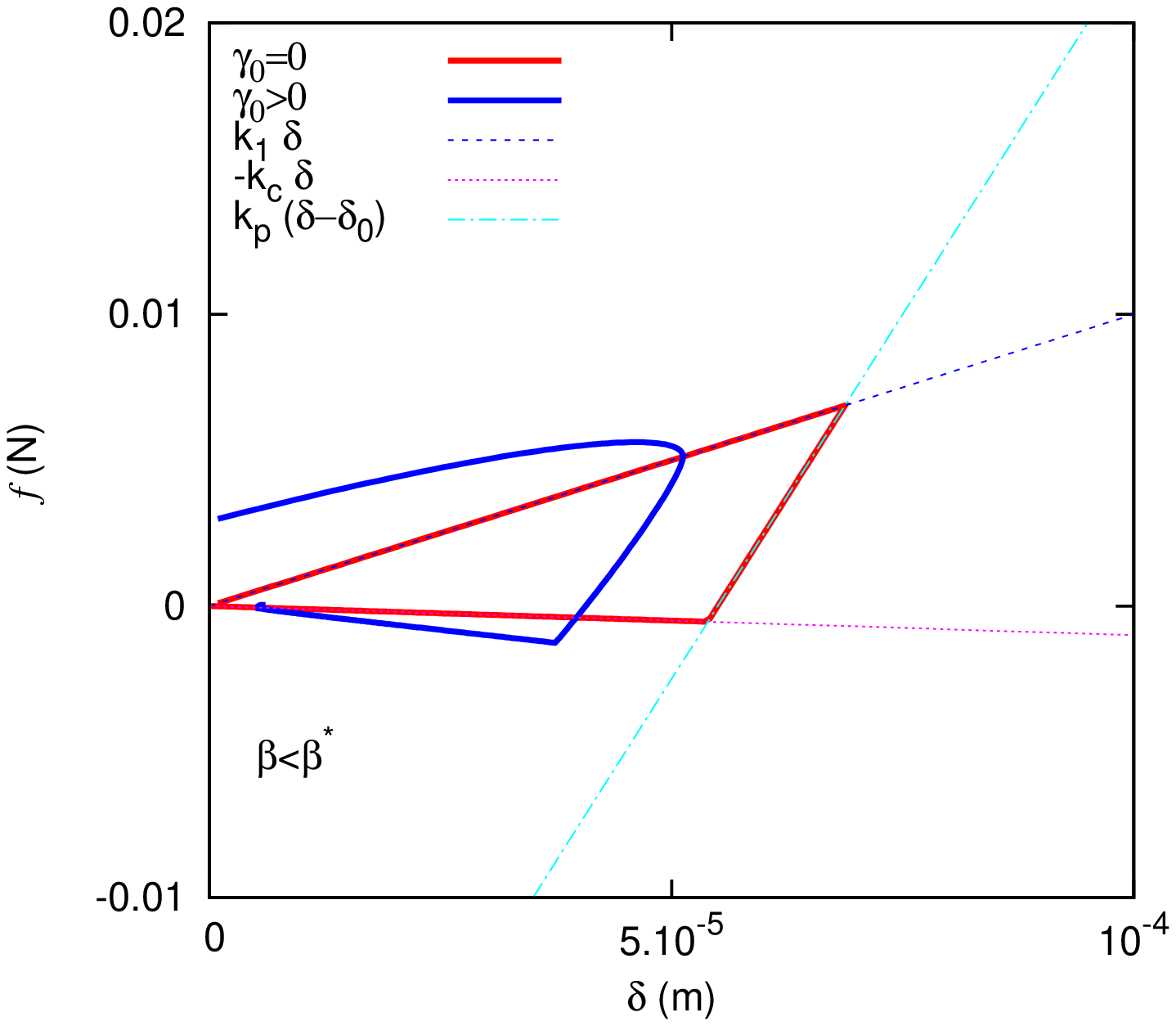}\label{fig:low_kc_force}}\quad
\subfigure[]{\includegraphics[scale=0.5,angle=-90]{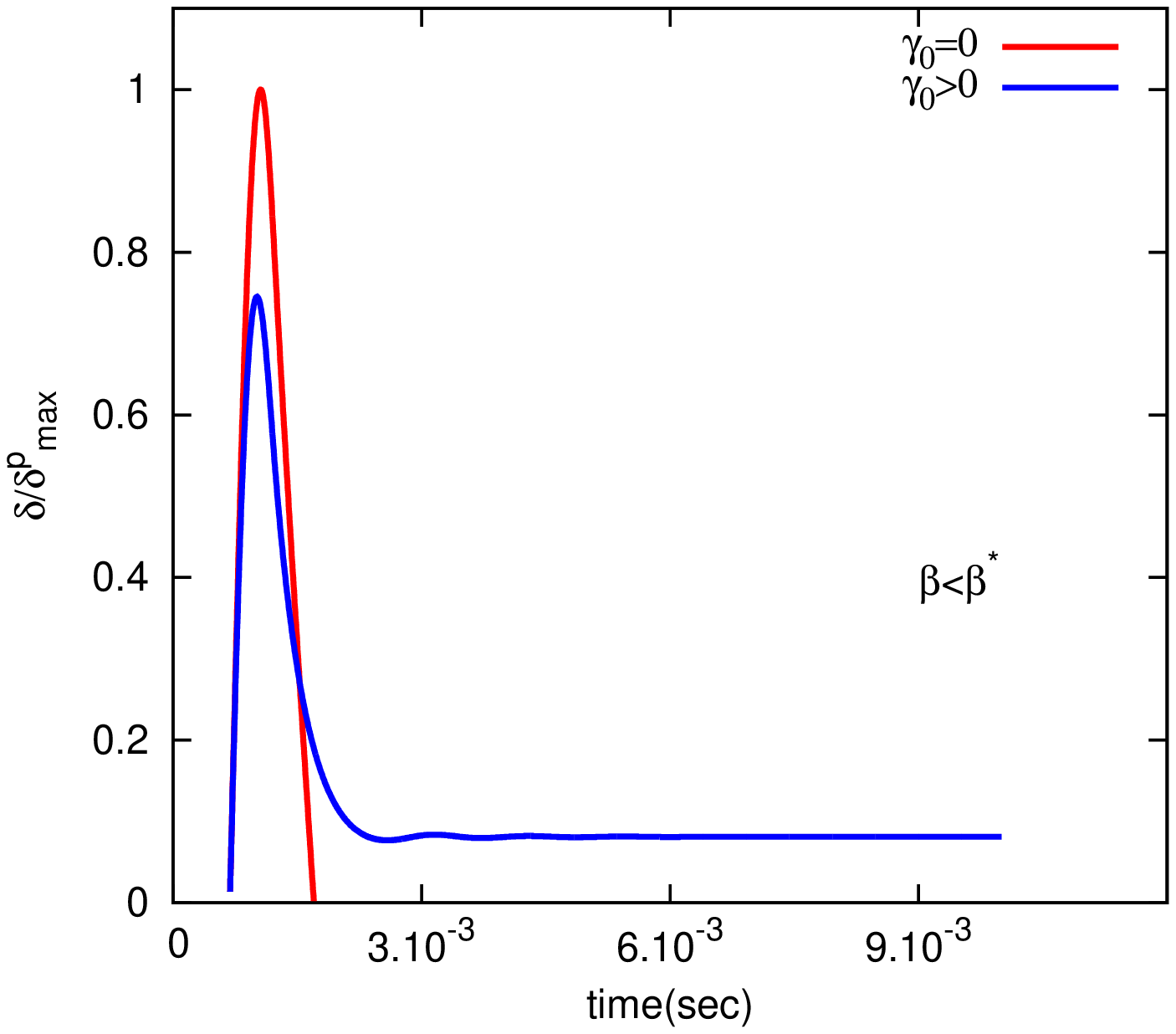}\label{fig:low_kc_del}}
}
\mbox{
\subfigure[]{\includegraphics[scale=0.5,angle=-90]{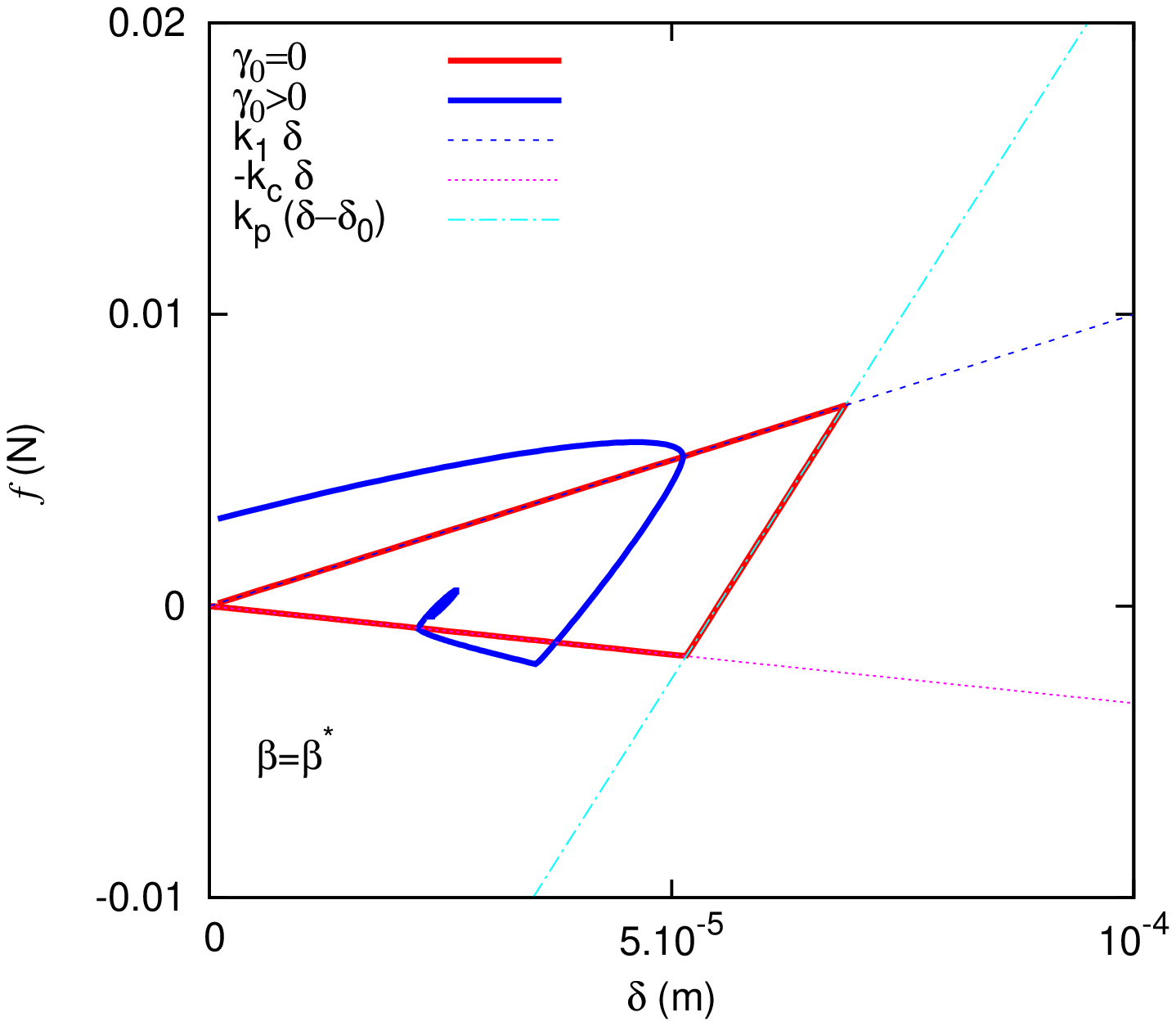}\label{fig:medium_kc_stick}}\quad
\subfigure[]{\includegraphics[scale=0.5,angle=-90]{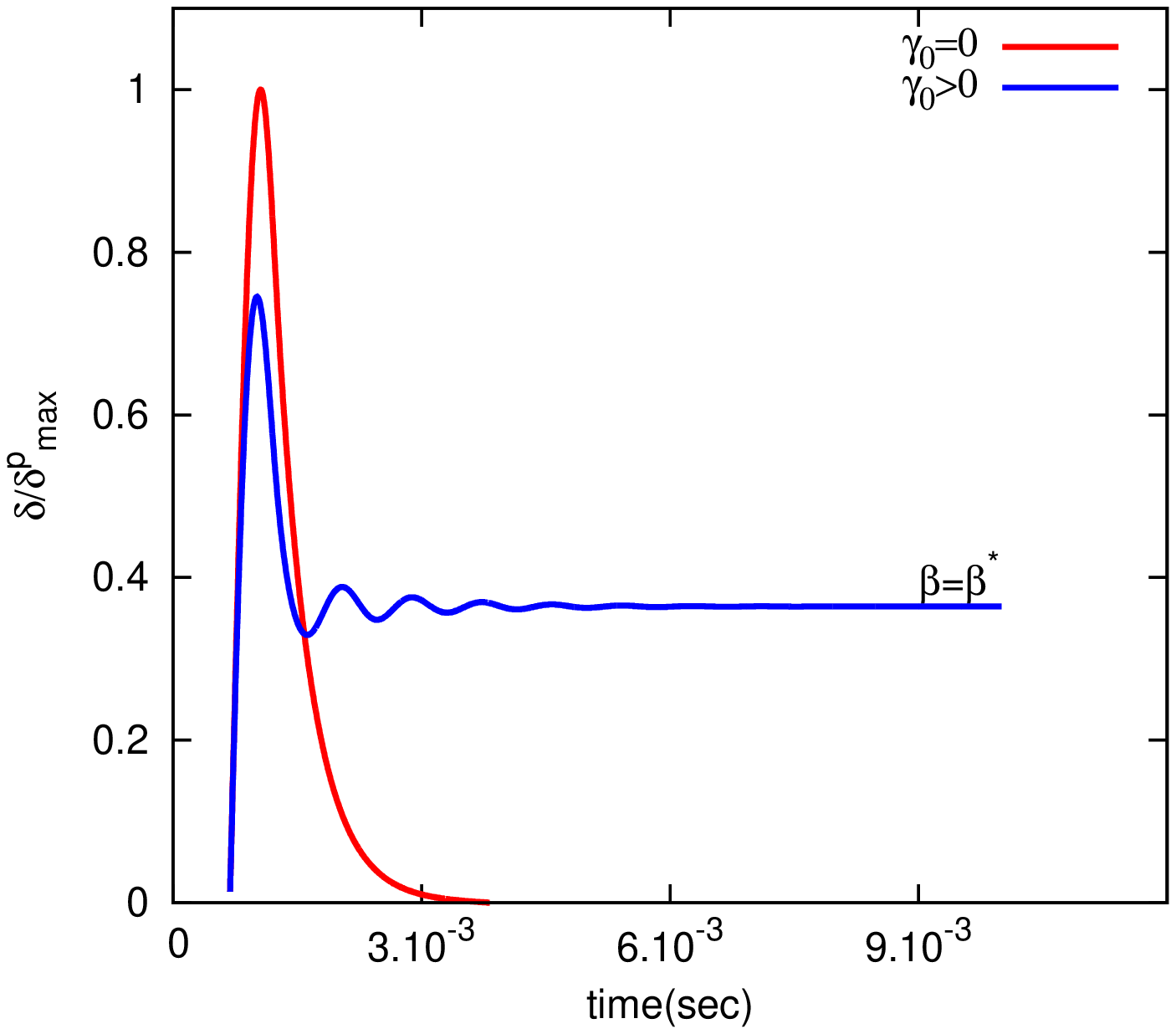}\label{fig:medium_kc_del_stick}}
}
\mbox{
\subfigure[]{\includegraphics[scale=0.5,angle=-90]{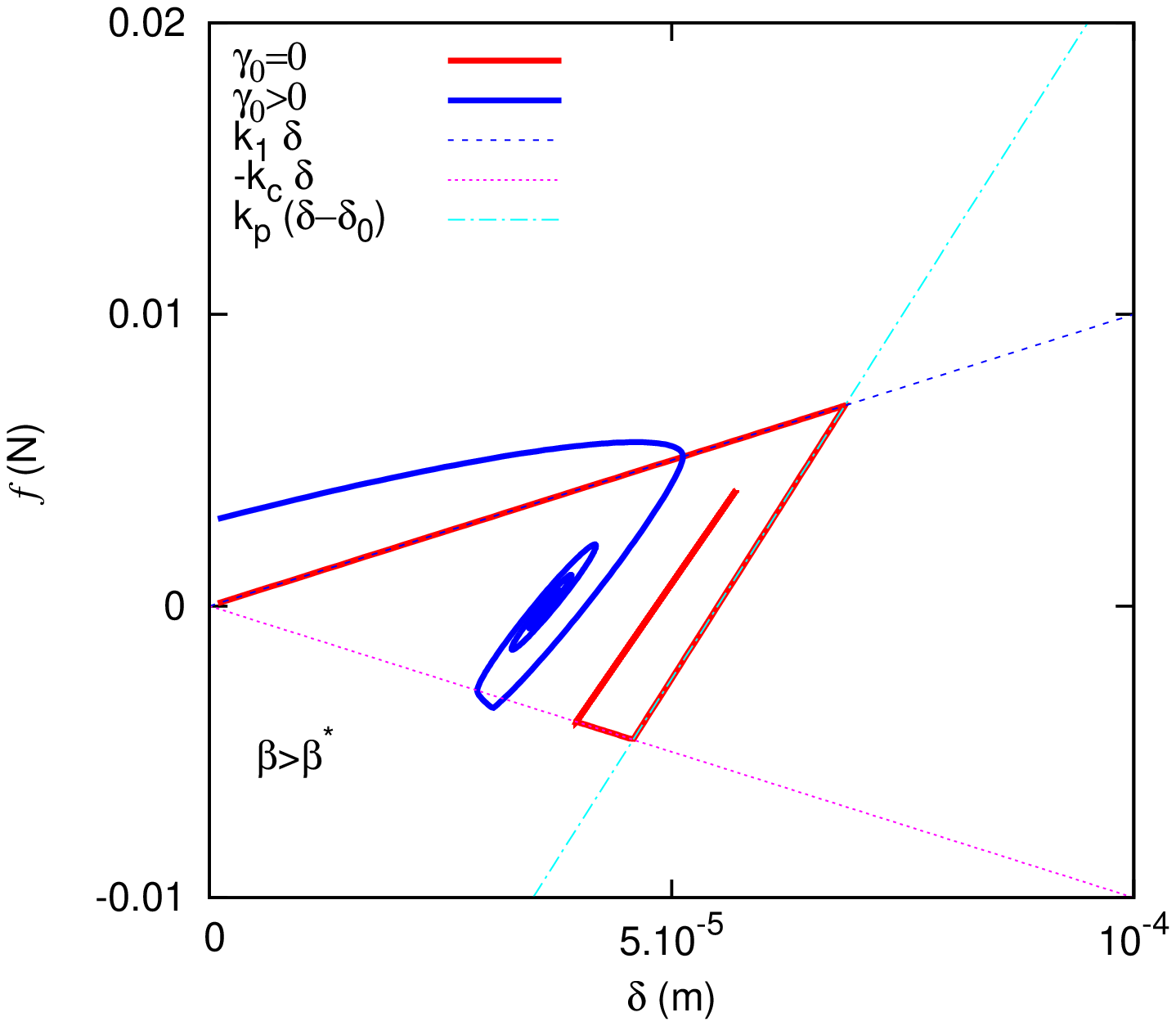}\label{fig:large_kc_stick}}\quad
\subfigure[]{\includegraphics[scale=0.5,angle=-90]{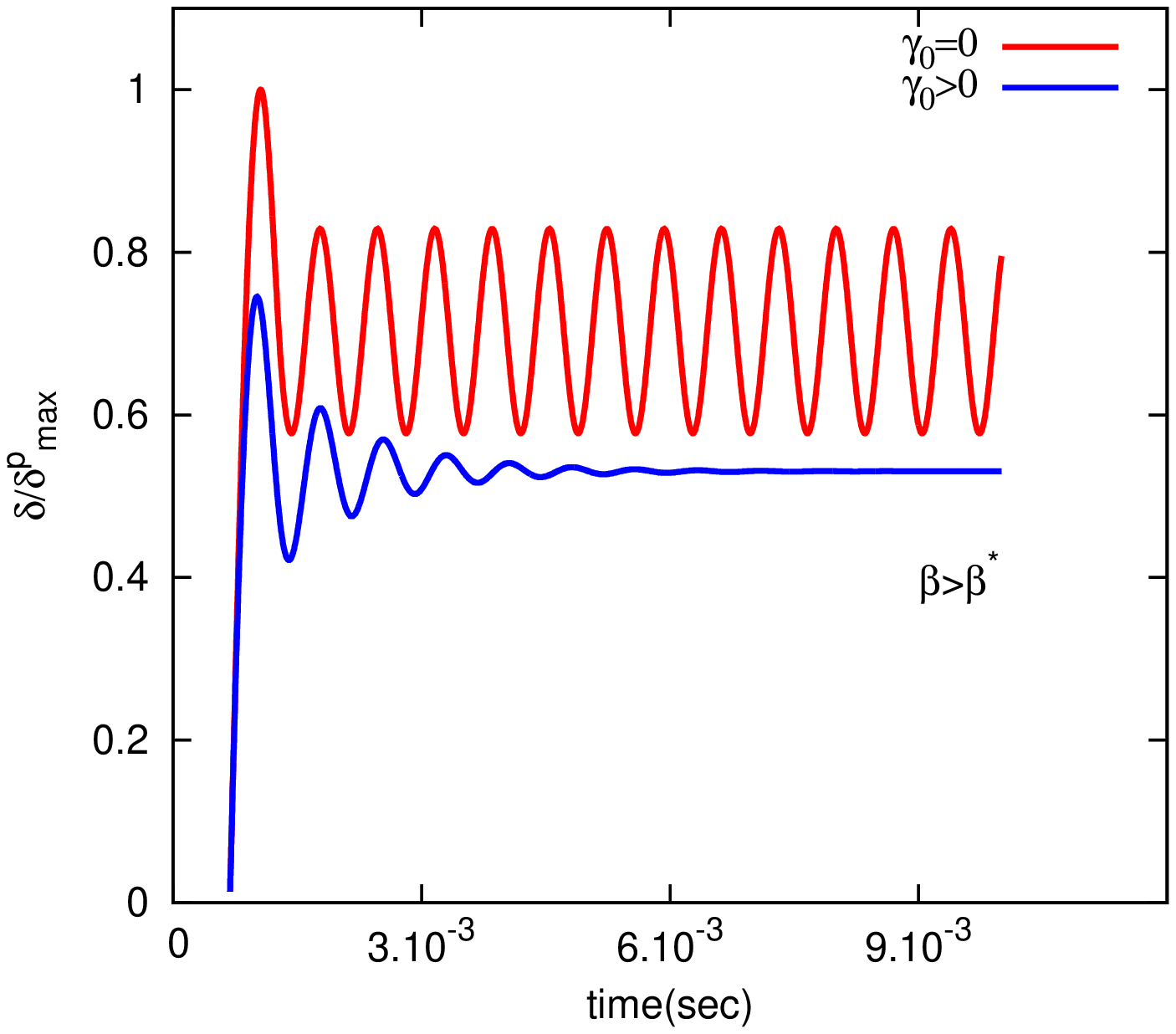}\label{fig:large_kc_del_stick}}}
\caption{
(a), (c), (e) Contact forces plotted against overlap and 
(b), (d), (f) time evolution of ${\delta}/{\delta^p_{\rm max}}$ 
for pair collisions with parameters $k_1=10^2$, $k_p=5\times 10^2$ and 
different $k_c=10$, $33.33$, and $100$, (units $\mathrm{Nm^{-1}}$), 
i.e.\ with $\eta=4$, $\beta<\beta^*$, $\beta=\beta^*$ and $\beta>\beta^*$,
for the same situations as shown in Fig.\ 9. 
The red and blue lines represent the data in the presence and absence of 
viscosity respectively, where $\gamma_0=5\times 10^{-3}$, (unit $\mathrm{Nm^{-1}sec}$).
}
\label{fig:f_del_t}
\end{figure*}

\section{Asymptotic Solutions}\label{sec:ASY}

In this subsection, we focus on the case $\chi\leq 1$, 
and study the asymptotic behavior of the coefficient of 
restitution as function of the impact velocity.

For the sake of simplicity, let us start with an elasto-plastic 
system without adhesion, i.e.\ $k_c=0$, in Eq.\ \eqref{eq:fine1} 
such that 
\begin{subequations}
\begin{equation}
e^{(1)}_n(\eta,\beta=0,\chi<1)=\sqrt{\frac{1}{1+\eta\chi}},
\end{equation}
inserting the definitions of  $\eta$, $\beta$ and $v_p$,
\begin{equation}
 e^{(1)}_n(\beta=0,v<v_p)=\sqrt{\dfrac{1}{1+\dfrac{k_p-k_1}{k_1}\dfrac{v_i}{\sqrt{{\dfrac{2k_1}{m}}}{\delta^p_{\rm max}}}}},
\end{equation}
using Eq.\ \eqref{eq:kpdelpWB}, where we defined $S=\frac{k_p-k_1}{k_1{\delta^p_{\rm max}}}$ and assuming $\omega_o=\sqrt{\frac{2k_1}{m}}$, we get
\begin{equation}\label{eq:enWB}
  e^{(1)}_n(\beta=0,v<v_p)=\sqrt{\frac{1}{1+\frac{Sv_i}{\omega_0}}}.
\end{equation}
\end{subequations}
Eq.\ \eqref{eq:enWB} is exactly the same as Eq.\ (5) in \cite{walton86}. For non-cohesive particles, and in the range
 $v<v_p$ we get exactly the same solution as Walton and Braun \cite{walton86}.

Further to study the asymptotic solution
\begin{equation}
e^{(1)}_n(\eta,\beta=0,\chi<1)=\sqrt{\frac{1}{1+\eta\chi}} 
        \approx (\eta\chi)^{-1/2} 
\label{eq:power_law_plastic}
\end{equation}
with the approximation valid for $\eta\chi \gg 1$.
\begin{figure*}
\centering
\includegraphics[scale=0.5]{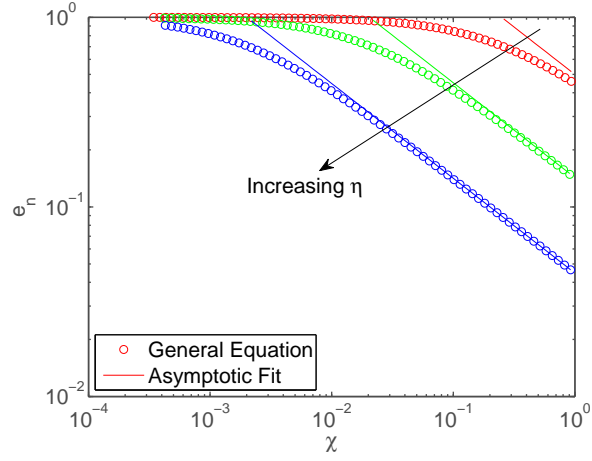}
\caption{The coefficient of restitution is plotted against the scaled initial velocity 
$\chi$ in log-log-scale for $\beta=0$ and three values of $\eta=5$, $50$, and $500$, 
with the other parameters as in Fig.\ 6. 
Red, green and blue circles denote, respectively, 
the solution of Eq.\ \eqref{eq:power_law_plastic}, while the 
solid lines represent the approximation for high scaled impact velocity
and large plasticity $\eta \gg 1$. 
}
\label{fig:cor_plas}
\end{figure*}
%
Since the scaled velocity is moderate, $\chi<1$, the condition requires
a large plasticity, i.e.,
a strong difference between the limit stiffness and the plastic
loading stiffness, $\eta \gg 1$ (or $k_p\gg k_1$). 
In Fig.\ 12, we plot the coefficient of restitution against 
the scaled initial velocity $\chi$ for three different values of $\eta=k_p/k_1$,
together with the power law prediction of Eq.\ \eqref{eq:power_law_plastic}. 
We observe, that for the smallest $\eta$ (red circle and line), 
the approximation is far from the data, while for higher $\eta$, 
the approximation works well even for rather small velocities 
$\chi \approx 0.1$.

Next, when studying the elasto-plastic adhesive contact model, $\beta>0$ and $\beta\ll 1$,
again, we restrict ourselves to values of $\eta$ such that asymptotic condition
$\eta\chi\gg 1$ is satisfied. Hence, Eq.\ \eqref{eq:fine1} can be 
approximated as
\begin{equation}
 e^{(1)}_n(\eta,\beta,\chi<1)
     \approx \sqrt{\frac{1}{\eta\chi}-\beta} ~,
\label{eq:power_law_adh_high}
\end{equation}
as long as $\eta\chi \gg \beta \ge 0$ and $\frac{1}{\eta}>\beta$ holds. 

\begin{figure*}
\centering
\includegraphics[scale=0.5]{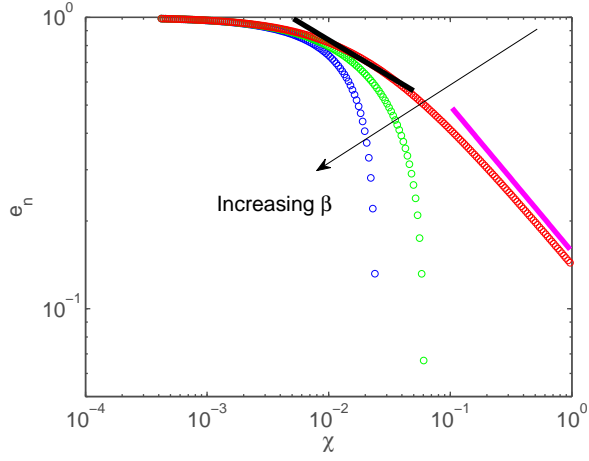}
\caption{Log-log plot of the coefficient of restitution against the 
scaled initial velocity $\chi$ for four different values of $\beta=0.01$, $0.1$, 
 and $1.0$, with $\eta=50$. 
Red, green and blue circles denote the 
respective solutions of the general equation, 
Eq.\ \eqref{eq:fine1}, solid black line represents power law $e_n\sim v^{-1/4}$, 
while magenta line denotes $e_n\sim v^{-1/2}$.
}
\label{fig:cor_coh}
\end{figure*}

In Fig.\ 13, we plot the coefficient of restitution 
against the scaled initial velocity $\chi$ for different values of $\beta$ and
superimpose the approximation, Eq.\ \eqref{eq:power_law_adh_high}. 
For small $\beta$ and large $\chi$, one observes good agreement between 
the full solution and the approximation. Differently,
for the highest values of $\beta$ the approximation is 
not valid. Due to the adhesive force, for large $\chi$, with 
increasing $\beta$, the deviation from the $\chi^{-1/2}$ power
law becomes increasingly stronger, leading to the sticking
regime, as discussed in the previous subsections. 
On the other hand, for smaller velocities, one observes
a considerably smaller power-law, resembling the well-known
$\chi^{-1/4}$ power law for plastic contacts, as indicated
by the dashed line in Fig.\ 13.

\section{Dependence on interpolation}

The choice of the interpolation rule for the unloading stiffness $k_2$
in Eq.\ \eqref{eq:k2_vel} is empirical. Therefore, for 
$\delta_{\rm max} / \delta^{p}_{\rm max} < 1$, a different choice 
could be:
\begin{equation}
k_2(\delta_{\rm max}) = k_1 (1+\eta\sqrt{\chi}).
\label{eq:diff_interp}
\end{equation}
Inserting Eq.\ \eqref{eq:diff_interp} into 
Eq.\ \eqref{eq:cor1} leads to a different expression for the
normal coefficient of restitution $e_n^{(1)}$, 
which for high values of $\eta \sqrt{\chi}$,
and for small $\beta$, reduces to
\begin{equation}\label{eq:interp-cor}
 e_n \propto \sqrt{\eta} (\chi)^{-1/4} ~.
\end{equation}
A similar power law prediction for moderate velocities has been previously 
obtained by Thornton {\em et~al.}\ in Ref.\ \cite{Thornton98}, using a non-linear 
Hertzian loading and unloading.
Fig.\ 14 shows the agreement between the power law 
approximation $\chi^{-1/4}$ and Eq.\ \eqref{eq:cor1} with the
alternative interpolation rule \eqref{eq:diff_interp}, for moderate
velocities. The choice of different interpolation laws for $k_2$
shows the flexibility of the model and requires input from 
experiments to become more realistic.
 The convexity of linear interpolation for zero cohesion is very similar to
 that of low $\beta$ in Fig.\ 9.
\begin{figure*}
\centering
\includegraphics[scale=0.5]{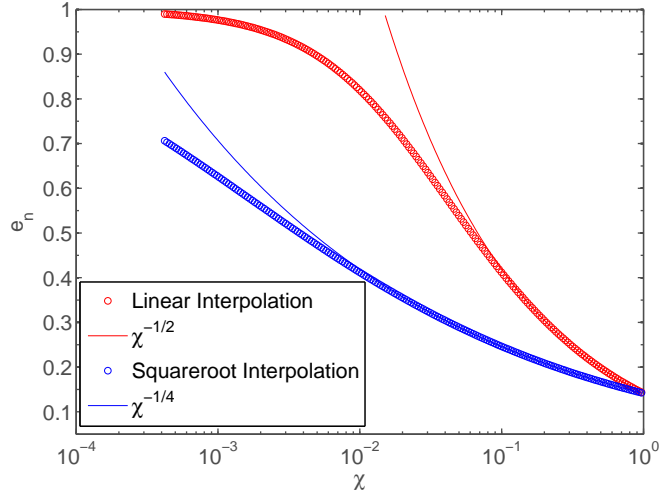}
 \caption{Semi-log plot of the coefficient of restitution as function of 
the scaled initial velocity $\chi$, using  different interpolation rules for $k_2$,
for pair collisions with $\eta=50$ and $\beta=0$. 
The symbols denote the solutions of the general equation, 
Eq.\ \eqref{eq:power_law_plastic} with linear interpolation (red circles)
or square root interpolation (blue circles), as given in 
Eq.\ \eqref{eq:diff_interp}.
The red and blue solid lines represent the approximations for high impact velocity  
$e_n \sim \chi^{-1/2}$ and $e_n \sim \chi^{-1/4}$.}
\label{fig:diff_interp}
\end{figure*}

\section{Energy Picture}
\label{sec:energy_picture}

This appendix shows the energies of two particles during contact,
where the difference between the different branches of the contact
model, namely irreversible/unstable or reversible/elastic, 
will be highlighted.

\begin{figure*}
  \centering
    \mbox{
\subfigure[]{\includegraphics[scale=0.5,angle=-90]{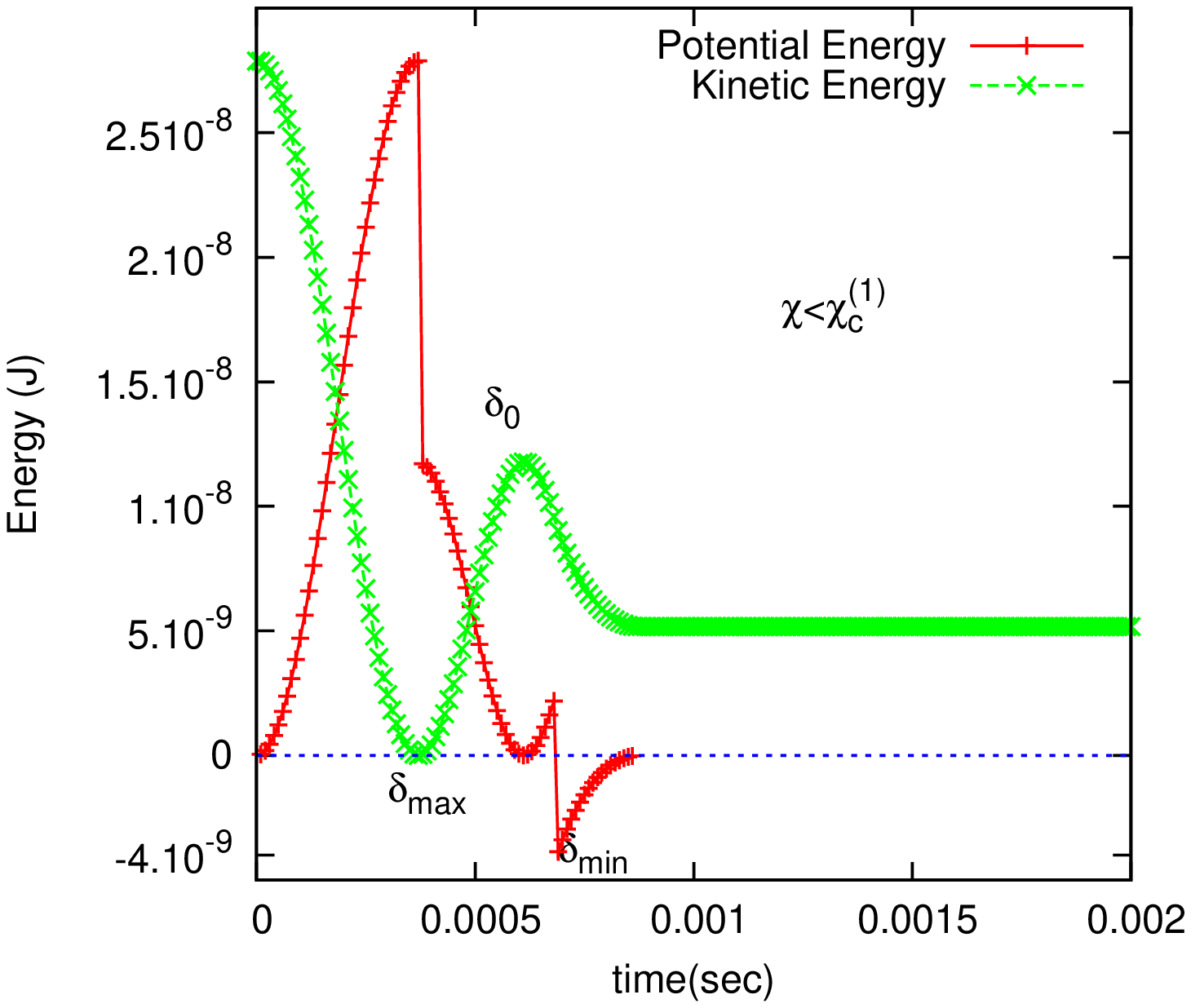}\label{fig:ene_small_ene}}\quad
\subfigure[]{\includegraphics[scale=0.5,angle=-90]{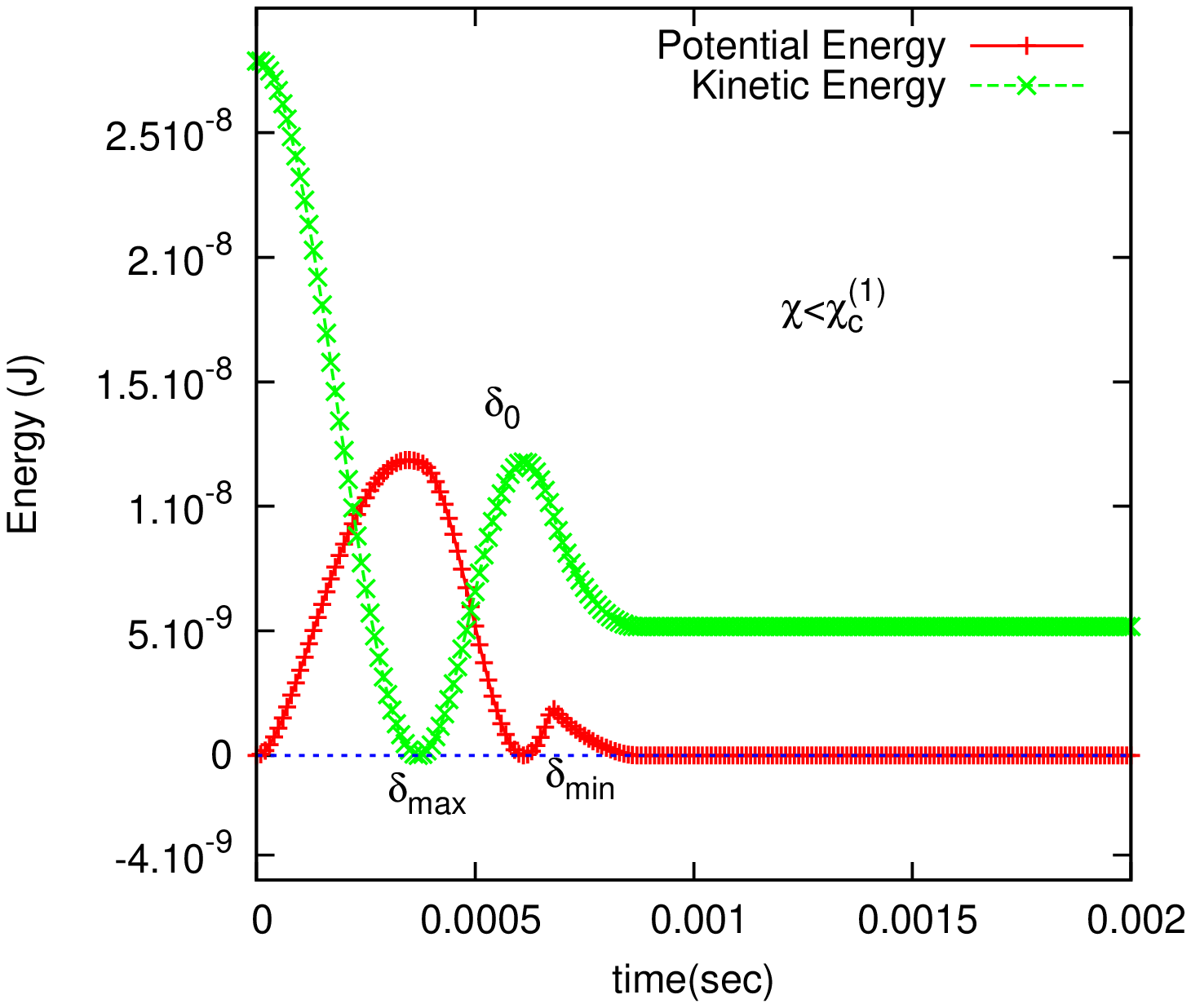}\label{fig:ene_small_f}}
}
\mbox{
\subfigure[]{\includegraphics[scale=0.5,angle=-90]{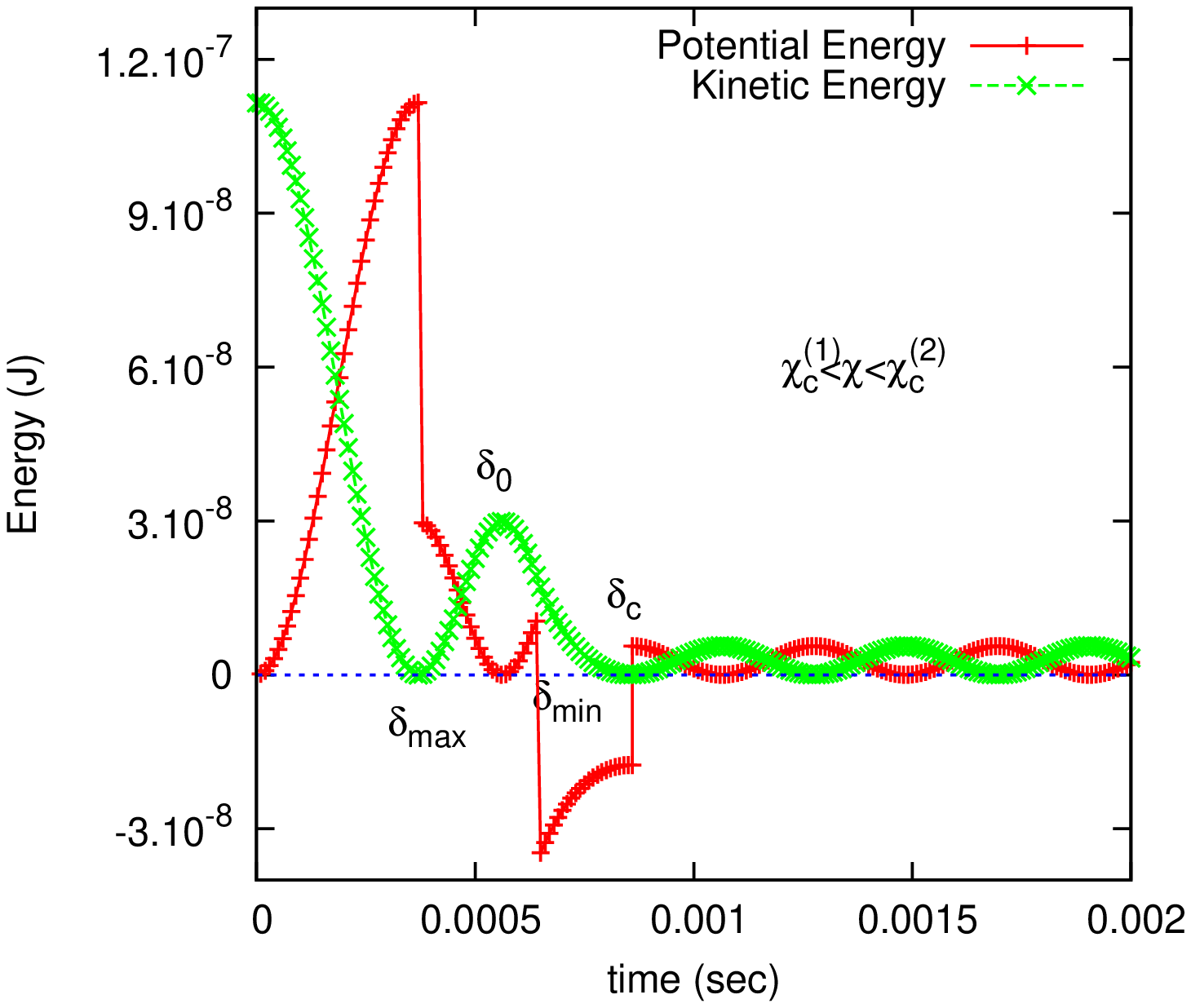}\label{fig:ene_stick_ene}}\quad
\subfigure[]{\includegraphics[scale=0.5,angle=-90]{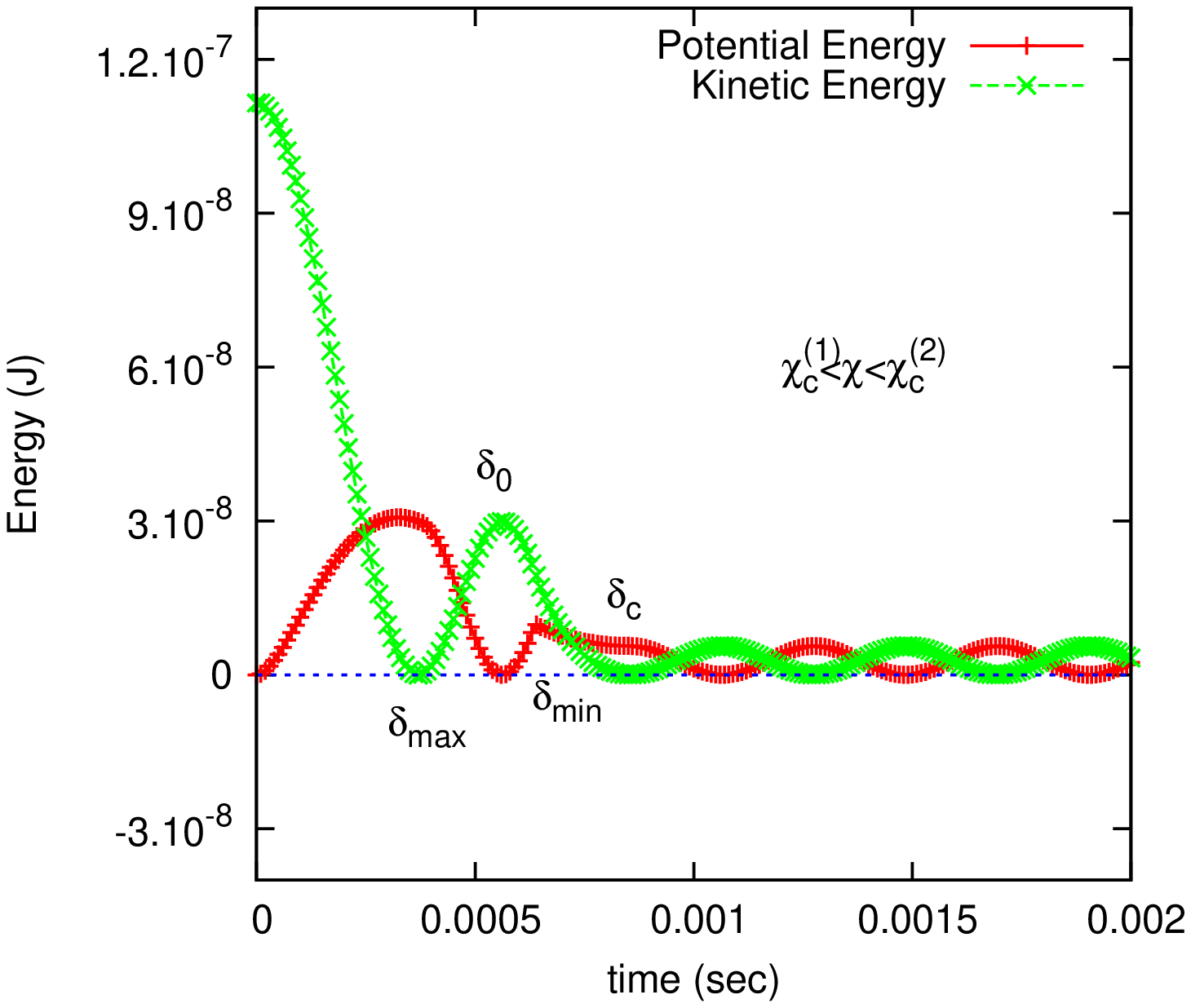}\label{fig:ene_stick_f}}
}
\caption{(a), (c) Kinetic and (irreversible, plastic, ``potential'') 
energy of the particles, and (b), (d) kinetic and available (elastic) 
potential energy (for re-loading) of the particles,
plotted against time for pair collisions with $k_1=10^2$\,Nm$^{-1}$. 
$k_p=5\times 10^2$\,Nm$^{-1}$, and $k_c=10^2$\,Nm$^{-1}$, 
i.e.\ $\eta=4$ and $\beta=1$. The initial velocity $\chi$ 
is $\chi=0.34$ (a,b) and $\chi=0.69$ (c,d), in the regimes
defined in the inset of each plot.
}\label{fig:comp_ene}
\end{figure*}

In Fig.\ 15, the time-evolution of kinetic and 
potential energy is shown; the graphs can be viewed in parallel 
to Figs.\ 7(a) and 7(b).
In Fig.\ 15(a), we plot the kinetic and potential energy of 
the particles against time for low initial velocity $\chi<\chi^{(1)}_c$, 
corresponding to Fig.\ 7(a), for which dissipation is
so weak that particles do not stick.
The kinetic energy decreases from its initial value and is converted 
to potential energy (the conversion is complete at $\delta_{\rm max}$). 
Thereafter, the potential energy drops due to the change between the loading 
and unloading slope from $k_1$ to $k_2$. The potential energy decreases to 
zero (at the force-free overlap $\delta_{\rm 0}$), where it is converted to 
(less) kinetic energy. Then the kinetic energy decreases further due to the 
acting adhesive force. 
At $\delta_{\rm min}$ the increasing potential energy drops to 
a negative value due to the change in unloading slope from $k_2$ 
to the adhesive (instable) slope $-k_c$. From there it increases from 
this minimum, negative value to zero, for $\delta=0$. From
here the kinetic energy remains constant and the potential energy 
stays at zero, since the particles are separated.

In Fig.\ 15(b), we plot the time evolution of 
kinetic and potential energy that the particles would have if 
un-/re-loading would take place at that moment, along the branch
with slope $k_2$, namely the available (elastic) potential energy. 
This energy increases from zero at $t=0$, and reaches a maximum when the 
kinetic energy becomes zero (note that it is not equal to the initial 
kinetic energy due to the plastic change of slope of $k_2$.) 
Thereafter, the available potential energy decreases to zero at the
force-free overlap $\delta_{\rm 0}$. For further unloading, the 
available potential energy first increases and then drops rapidly
on the unstable branch with slope $-k_c$.
The change in sign of the unloading slope, from $k_2$ to $-k_c$, 
is reflected in the kink in the curve at $\delta_{\rm min}$. 
Note, that comparing Figs.\ \ref{fig:ene_small_ene} and \ref{fig:ene_small_f}, 
the available potential energy always stays positive, while the total,
plastic ``potential'' energy drops to negative values after the kink at 
$\delta_{\rm min}$.

Figs.\ 15(c) and 15(d) show 
the time evolution of kinetic and potential energy (total and available, 
respectively) for an initial velocity $\chi^{(1)}_c<\chi<\chi^{(2)}_c$
in the sticking regime, see Fig.\ 7(b). 
In Fig.\ 15(c), a similar trend as that of 
Fig.\ 15(a) is observed until the potential energy 
becomes negative at $\delta_{\rm min}$. The difference to
the case of smaller impact velocity is that at this 
point, the kinetic energy is less than the magnitude of the negative 
potential energy and hence first reaches zero, i.e., the particles
stick. At this point, the (plastic) potential energy increases 
and jumps to a positive value indicating the change in sign of 
the unloading slope from $-k_c$ to $k_2$. 
Finally, it oscillates between this positive value at 
$\delta_{\rm c}$, exchanging energy with the kinetic degree of 
freedom. When the available potential energy is plotted in 
Fig.\ 15(d), a similar trend as that of 
Fig.\ 15(b) is observed up to the kink 
at $\delta_{\rm min}$. Here, the two energies have comparable values 
when they reach $\delta_{\rm min}$ and the kinetic energy decreases 
to zero with a non-zero available potential energy, 
which causes the contact to re- and un-load along $k_2$.

\section{Cyclic agglomerate compression and tension tests}
\label{sec:agg}

Goal of this appendix is to show the unloading and re-loading behavior 
of an agglomerate, i.e.\ its effective, mesoscopic force-displacement
relation, which clearly is different from the contact force law 
applied at the primary particle contacts.
We will report incomplete detachment and partly/weaker elastic response 
for re-loading after various different compressive and tensile loading
amplitudes.

The system considered here is an agglomerate (cubic) of size $L_0=0.115$, 
made of $N=1728$ primary particles of diameter $d_0=0.01$ 
(with some variation in size to avoid monodisperse artefacts), just as
in Ref.\ \cite{Luding08}.  The cubic sample was first compressed 
(pressure-sintered) with a dimensionless wall stress 
$d_0 p_s / k_p = 0.02$ to form a stable, rather dense agglomerate 
or \textquotedblleft tablet\textquotedblright . The stress is first
released to a value $2.10^{-5}$, i.e.\ $p_r/p_s=10^{-3}$ for all walls. 
Then various uni-axial, unconfined tension/compression tests are carried
out applying either further tension or compression starting from the released
 state of the sample \cite{Luding08}.
The simulation parameters are same as in Ref.\ \cite{Luding08} 
(table 2), except for the cohesion that is set here to
a rather small intensity, $k_c/k_p = 0.2$, rolling and sliding friction 
coefficients that are double as large, $\mu_r = \mu_o =0.2$, 
and viscous damping of those degrees of freedom,
$\gamma_r/\gamma = \gamma_o/\gamma = 0.1$, which also is larger
than that of the reference situation.


The force-displacement curves for the tests at different amplitudes
are shown in Figs.\ 16 and 18 for tension and
 compression tests respectively. All simulations in Figs.\ 16 and 18
 start from the same configuration, i.e. the released state mentioned above and
 is indicated by the black circle at point $(0,0)$.
These plots represent the mesoscopic contact 
model of agglomerates consisting of multiple primary particles
and their geometrical surface configurations and change in shape
during the tests.

Fig.\ 16 shows the force-displacement curve for 
an unconfined uniaxial tension test. 
 The black arrow shows the unloading/tension path, and finally arrows with different colors 
show the re-loading paths for
different deformation amplitudes, as given in the inset. 
Each of the tests, when it reaches the original strain at zero, 
is then repeated for three more cycles.  Note that repeated 
cyclic loading remains on the same branch with positive slope, displaying 
the elastic nature of the contact, while it is {\em not} completely,
perfectly detached. The contact surface is changing plastically by 
restructuring of the primary particles and surely is not flat, 
see Fig.\ 17, as one would expect for ideal, homogeneous,
plastic materials. For the largest amplitude, the behavior is not
perfectly elastic anymore, since the first plastic effects show up.
For deformations as large as 0.2 of the primary particle diameter, $d_0$, 
before re-loading (arrow with positive slope on the red curve) has mostly, but not completely lost mechanical contact.
 The complete detachment of the assembly happens for much higher amplitude, than what is expected
 from a two-particle interaction.
Note that the contact model of the primary particles is behaving 
elasto-plastically $(\phi_f=0.05)$ on the scale of only $0.05 d_0$; 
the reversible, elastic un-/re-loading is thus {\em not}
due to the primary particle contact model, since it stretches to four 
times $\phi_f d_0$ and even higher displacements. 
Finally, in order to confirm that this is not an effect of viscosity, 
qualitatively, the thick lines are simulations performed four times 
slower than those with thin lines.

\begin{figure*}
\centering
\includegraphics[scale=0.5,angle=-90]{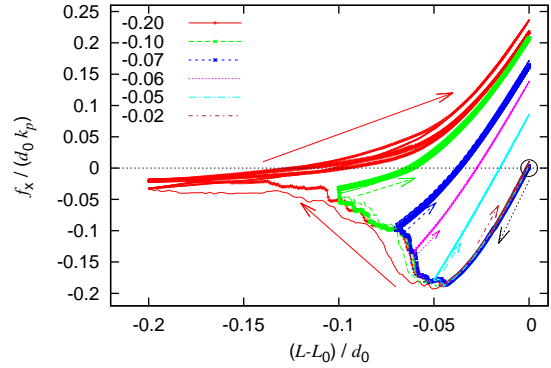}
\centering
\caption{Dimensionless force-displacement curve for an unconfined uni-axial 
tension test (negative horizontal axis),  
with the various different deformation amplitudes $D_x$ given in the inset.
The downward arrow indicates the direction of first tensile unloading,
while the upwards-right arrows indicate the change of force during 
re-loading. Except for the red curve, all these branches are reversible,
for repeated un-/re-loading.}
\label{fig:force_disp_ten}
\end{figure*}

\begin{figure*}
  \centering
    \mbox{
\subfigure[]{\includegraphics[scale=0.18]{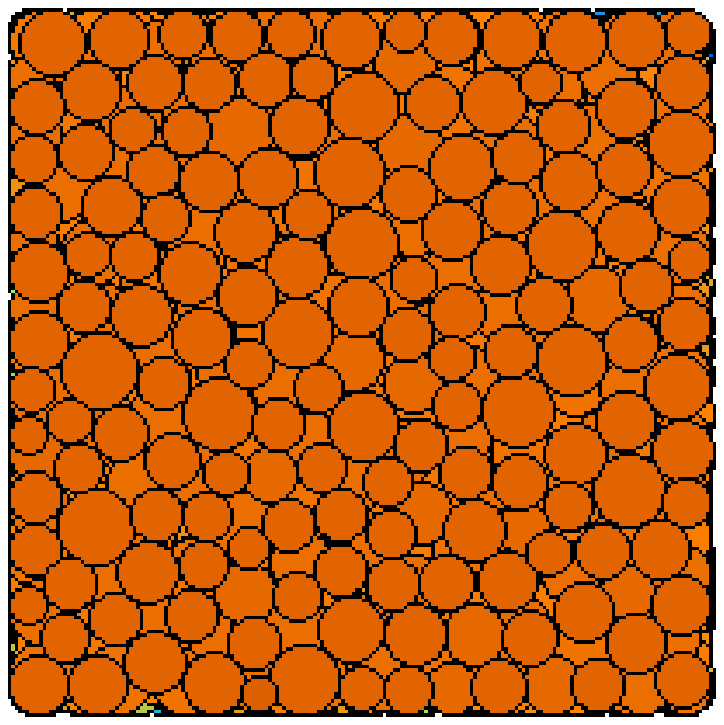}}
\subfigure[]{\includegraphics[scale=0.18]{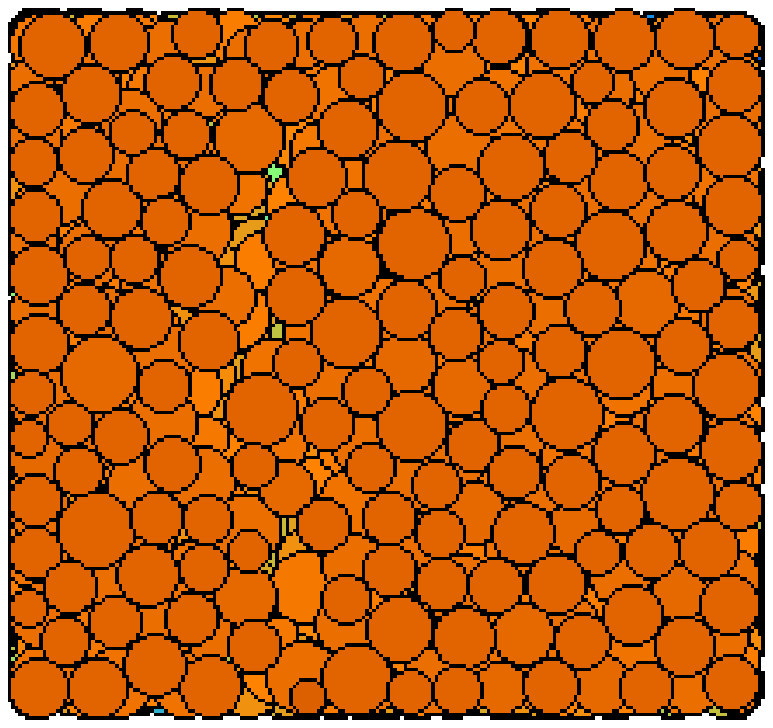}}
\subfigure[]{\includegraphics[scale=0.18]{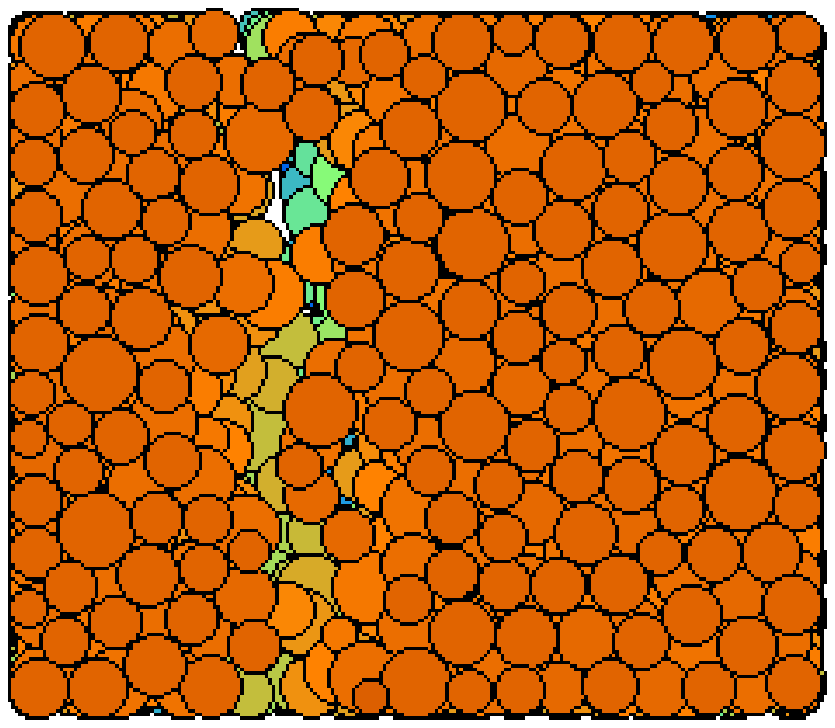}}
\subfigure[]{\includegraphics[scale=0.18]{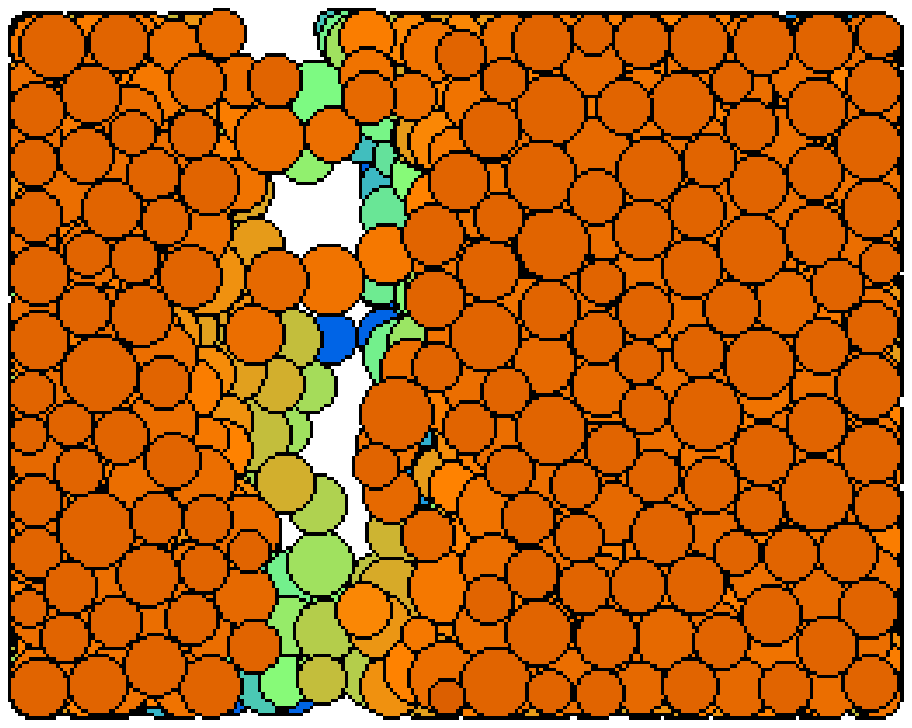}}
\subfigure[]{\includegraphics[scale=0.18]{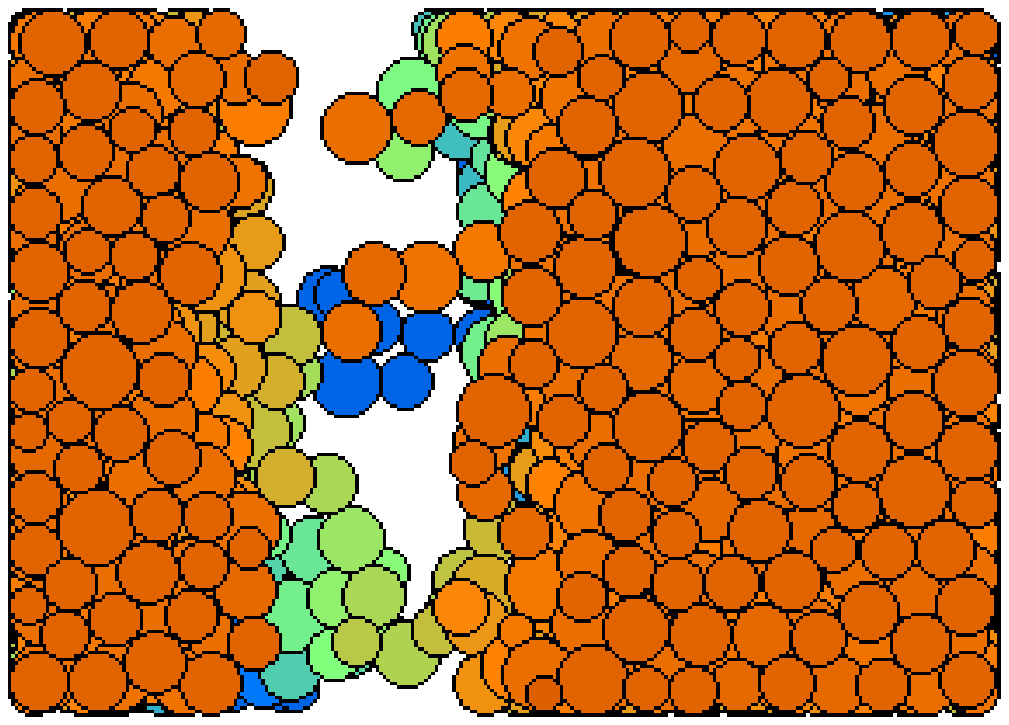}}
\subfigure[]{\includegraphics[scale=0.18]{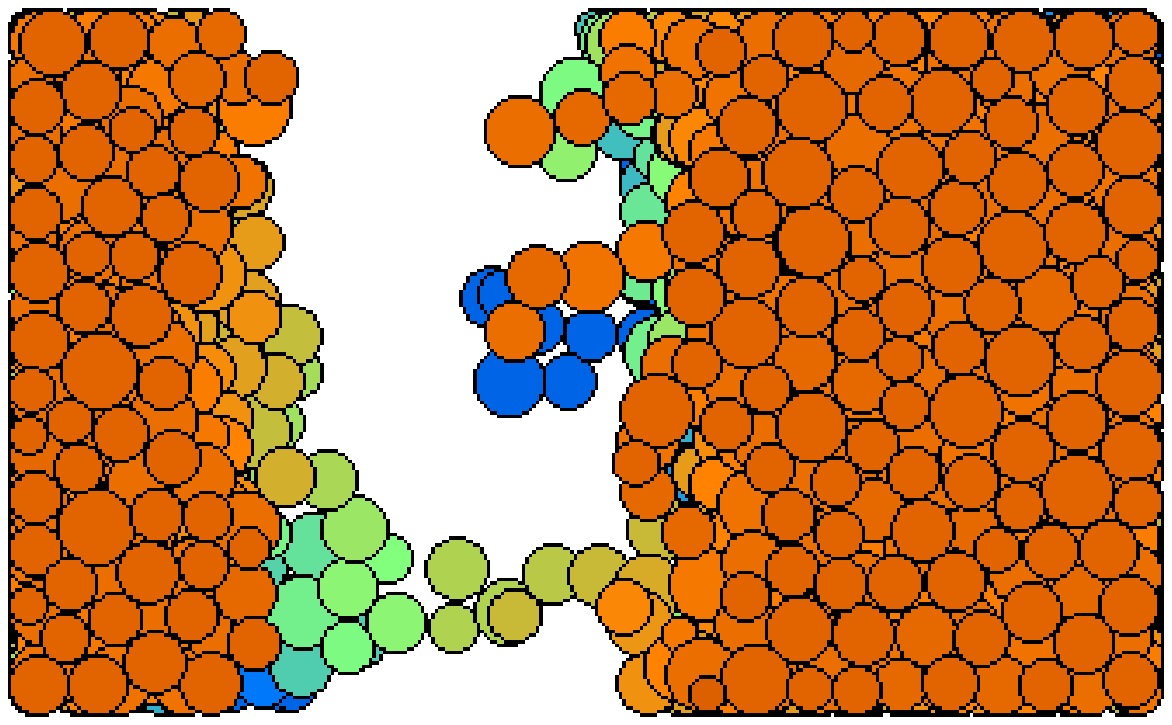}}
\subfigure[]{\includegraphics[scale=0.18]{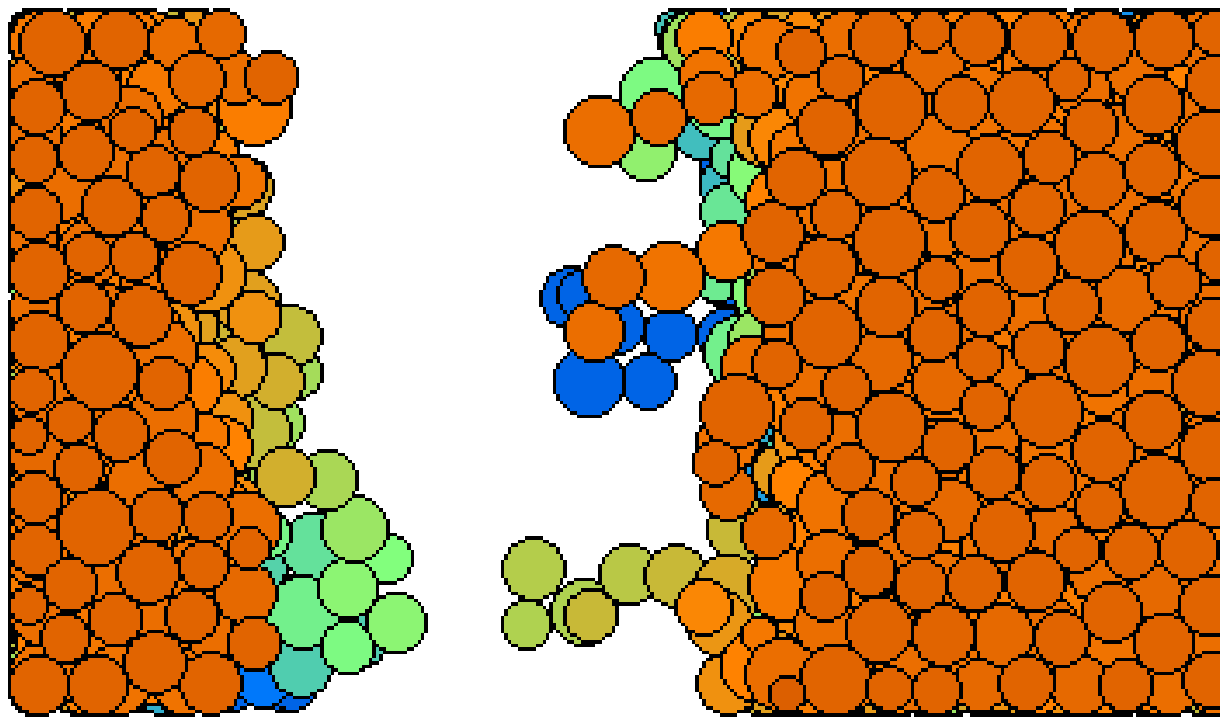}}
}
\caption{Snapshots of the tablet-sample during (large) tensile deformations
for $D_x=(L-L_0)/d_0 = 0$ (a), 0.81 (b), 1.8 (c), 3.1 (d), 4.7 (e), 
7.4 (f), and 8.6 (g).
The primary particles are colored according to their distance from the 
viewer (red, green, blue is increasing distance).
}
\label{fig:ten_snaps}
\end{figure*}

In Fig.\ 17, a few snapshots during the tensile deformation 
are presented. The first snapshot corresponds to the undisturbed sample, 
while the others are increasing tensile deformation amplitudes.
Note that these deformations are much larger than in 
Fig.\ 16. The contact is completely lost only 
at the extreme, final deformation in Fig.\ 19(g).
In Fig.\ 17, it is also visible that the contact surface 
has developed a roughness of the size of several primary particles;
the first visible gap is opened at a total deformation of $D_x \sim d_0$, 
and the contact is lost only at $D_x \sim 8 d_0$, when the last of the 
thin “threads” breaks.
The elastic, irreversible tension branch, however, 
is strongly developed only for much smaller $D_x \sim d_0/5$.

\begin{figure*}
\centering
\includegraphics*[scale=0.5,angle=-90]{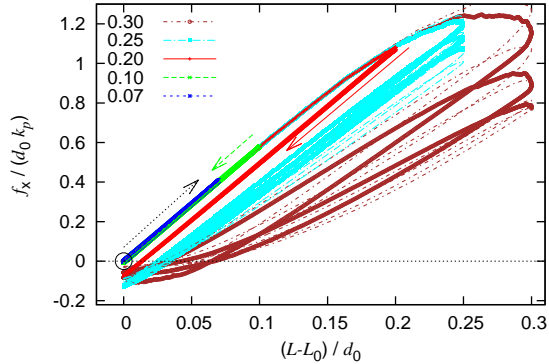}
\centering
\caption{Dimensionless force-displacement curve for the same sample as in 
Fig.\ 16, but under compressive initial loading and 
un-/re-loading. The values in the inset indicate the maximal amplitudes
$D_x$.
}
\label{fig:force_disp_comp}
\end{figure*}

Complementing the tension test above, 
Fig.\ 18 shows the behavior of the same sample
during compression cycles. The values given in the inset indicate the 
amplitude of un-/re-loading. The smallest amplitudes remain elastic 
throughout, while plastic deformation kicks in for $D_x > 0.1$ (see 
the red curve). However, the unloading and re-loading take place on
the same branch, i.e.\ a new elastic branch (e.g.\ for $D_x = 0.2$).
For even larger amplitudes, e.g.\ the yellow curve with $D_x = 0.3$,
the continuous damage/plastic destruction of the sample 
(by considerable irreversible re-arrangement during each cycle). 
Again, thick lines indicate simulations four times slower, which 
shows a small quantitative difference, but qualitative agreement even
for the largest amplitude/rate. The snapshots in Fig.\ 19
show the continuous plastic deformation of the sample at large strains.

\begin{figure*}
  \centering
\subfigure[]{\includegraphics[scale=0.18]{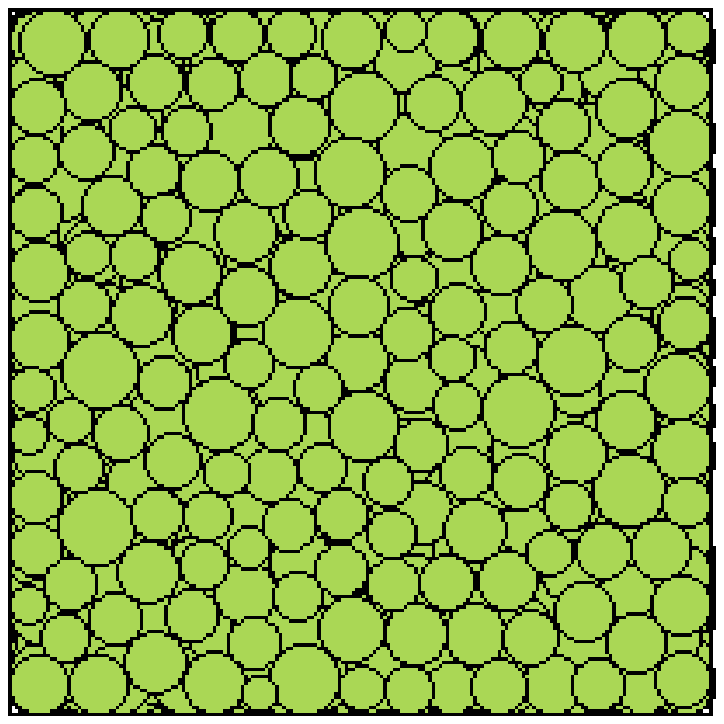}}
\subfigure[]{\includegraphics[scale=0.18]{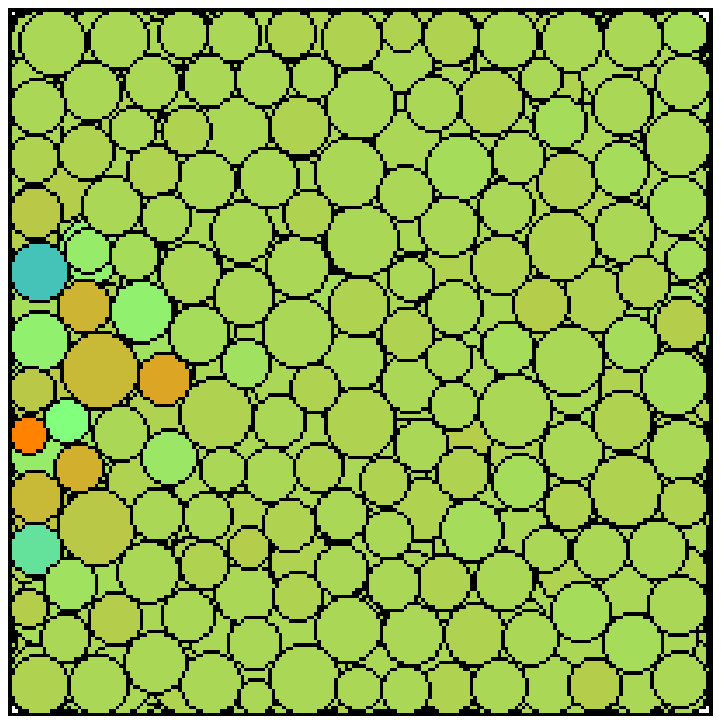}}
\subfigure[]{\includegraphics[scale=0.18]{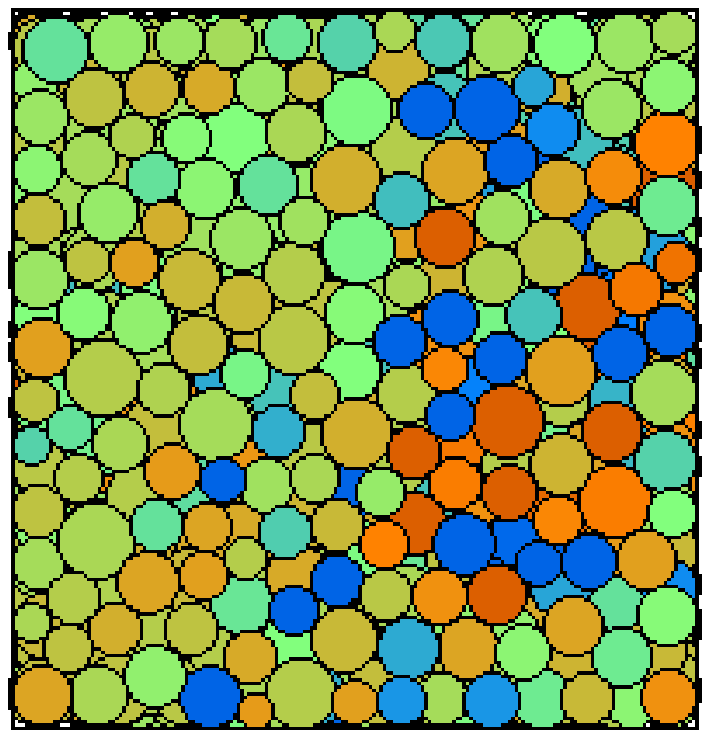}}
\subfigure[]{\includegraphics[scale=0.18]{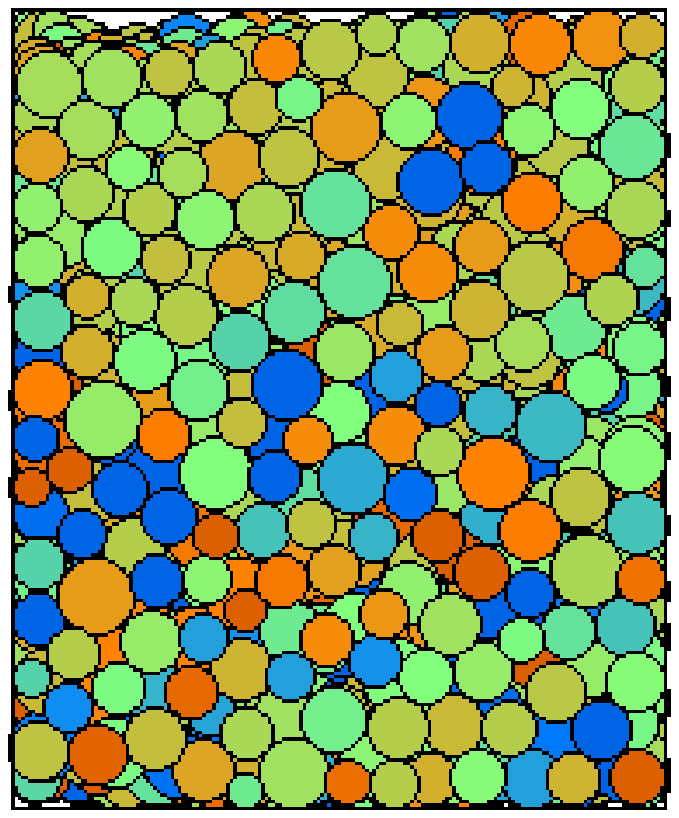}}
\subfigure[]{\includegraphics[scale=0.18]{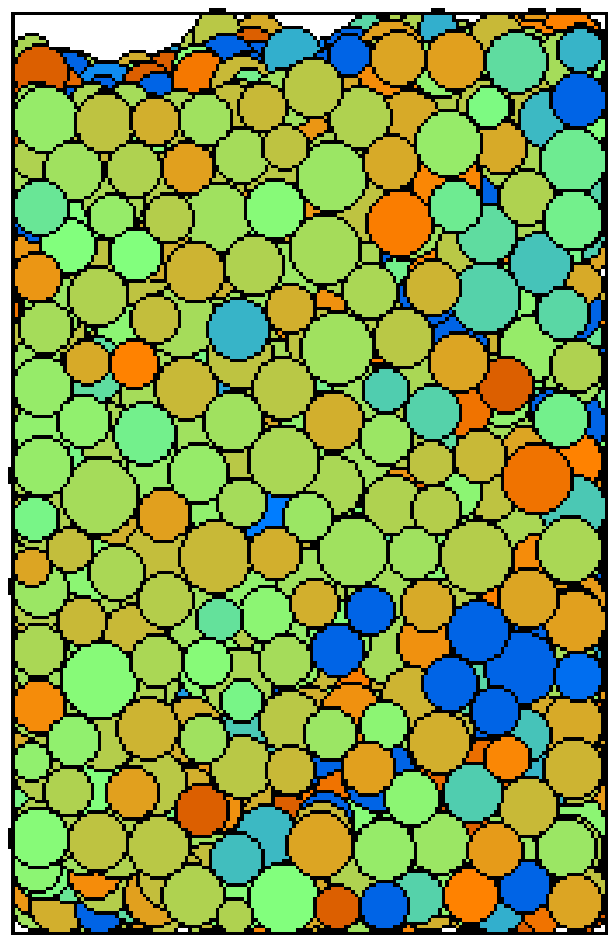}}

\caption{Snapshots of the tablet sample during compression at 
(large) deformations $D_x = (L-L_0)/d_0 = 0$ (a), 0.01 (b), 
0.3 (c), 0.8 (d), and 1.8 (e).
The color code is small stress (green) and compressive/tensile larger 
stress (red/blue) averaged/isotropically per primary particle.
}
\label{fig:comp_snaps}
\end{figure*}

\bibliographystyle{ieeetr}

\bibliography{PowTech}


\end{document}